\documentclass{unifal}

\usepackage[english]{babel} 
\usepackage[num, bibjustif]{abntex2cite} 
\citebrackets{[}{]} 

\usepackage{amsmath}
\usepackage{bm}
\usepackage{graphicx}
\usepackage[compat=1.1.0]{tikz-feynman}

\usepackage[dvipsnames]{xcolor}
\usepackage{caption}
\usepackage{subcaption}
\usepackage{enumitem} 
\usepackage{setspace}

\title{Systematic Study of Neutrino-Nucleus Interactions for the DUNE Experiment}
\author{Luiz Felipe Brandão Magalhães Rodrigues}
\local{Poços de Caldas/MG}
\date{2026}

\orientador{Prof. Dr. Fernando Gonçalves Gardim}
\coorientador{Prof. Dr. Gustavo do Amaral Valdiviesso}
\tipo{Dissertação apresentada como parte dos requisitos para obtenção do título de Mestre em Física pela Universidade Federal de Alfenas.\\Área de concentração: Física.}

\defesa{09 de Outubro de 2026}
\bancaum{Prof. Dr. Fernando Gonçalves Gardim}{Universidade Federal de Alfenas}
\bancadois{Prof. Dr. André Fabiano Steklain Lisbôa}{Universidade Tecnológica Federal do Paraná}
\bancatres{Prof. Dr. Anibal Thiago Bezerra}{Universidade Federal de Alfenas}

\fcautor{Rodrigues, Luiz Felipe Brandão Magalhães.}
\fctipo{Dissertação (Mestrado em Física)}
\fcilustrado{il.}
\fccatalogacao{1. Neutrino Oscillation. 2. DUNE. 3. GENIE Event Generator. 4. Final State Interactions. 5. Cross-section. I. Gardim, Fernando Gonçalves, orient. II. Valdiviesso, Gustavo do Amaral, coorient. III. Título.}

\palchaves{Oscilação de Neutrinos; DUNE; GENIE; Interações de Estado Final; Seção de Choque; Divulgação Científica.}
\keywords{Neutrino Oscillation; DUNE; GENIE Event Generator; Final State Interactions; Cross-section; Scientific Outreach.}

\usepackage{floatrow}
\usepackage[dvipsnames]{xcolor}

\definecolor{verde}{rgb}{0.25,0.5,0.35}
\definecolor{jpurple}{rgb}{0.5,0,0.35}
\definecolor{darkgreen}{rgb}{0.0, 0.2, 0.13}

\usepackage{listings}
\newcommand{\estiloJava}{
\lstset{
    language=Java,
    basicstyle=\ttfamily\scriptsize,
    keywordstyle=\color{jpurple}\bfseries,
    stringstyle=\color{red},
    commentstyle=\color{verde},
    morecomment=[s][\color{blue}]{/**}{*/},
    extendedchars=true,
    showspaces=false,
    showstringspaces=false,
    numbers=left,
    numberstyle=\tiny,
    breaklines=true,
    backgroundcolor=\color{RoyalPurple!5},
    breakautoindent=true,
    captionpos=b,
    xleftmargin=0pt,
    tabsize=2
}}

\tikzfeynmanset{warn luatex=false}

\begin{document}

\maketitle

\thanks{I would like to express my deep gratitude to my family, Katia,  Armando, Ana Katherine, Wesley and Hermione, and my lovely girlfriend, Sarah. I thank you for being my solid foundation. Your love, understanding during times of absence due to work, and constant encouragement were indispensable in helping me complete this stage.

To my childhood friends, Gustavo, Isaque, Alexandre and Luis, and my friends at the Federal University of Alfenas (UNIFAL-MG), Thamires, Lauane, João Gabriel, Lucas, Sabrina, Thais, Rodrigo, Fabiana and all the others. I thank you for all the enriching discussions, the fellowship, and for making the research environment more pleasant and collaborative.

To my advisors, Prof. Dr. Fernando Gonçalves Gardim and Prof. Dr. Gustavo do Amaral Valdiviesso, for all the patience, trust, and guidance throughout this journey. Your excellent mentorship was fundamental not only to the development and rigor of this research but also to my growth as a researcher.

I would like to acknowledge the National Laboratory for Scientific Computing (LNCC/MCTI, Brazil) for providing HPC resources of the SDumont supercomputer, which have contributed to the research results reported within this dissertation.

Finally, this work was supported by the National Council for Scientific and Technological Development – Brazil (CNPq).
}

\epigrafe{``I would rather have questions that can't be answered than answers that can't be questioned.''\\
Richard P. Feynman}

\newpage
\thispagestyle{empty}
\begin{center}
    \Large \textbf{ABSTRACT}
\end{center}
\vspace{0.5cm}

\noindent The recent generation of high-precision neutrino oscillation experiments, such as the Deep Underground Neutrino Experiment (DUNE), relies on the accurate reconstruction of incident neutrino energy to determine fundamental parameters like Charge-Parity (CP) violation and the neutrino mass hierarchy. The use of Liquid Argon ($^{40}\text{Ar}$), introduces significant theoretical complexities. This dissertation provides a comprehensive investigation into neutrino-nucleus scattering phenomenologies, connecting fundamental electroweak theory with computational simulations using the GENIE Monte Carlo event generator. The study evaluates the limitations of traditional frameworks, such as the Relativistic Fermi Gas (RFG) model, specifically addressing the Axial Mass parameter and its impact on the Charged Current Quasi-Elastic (CCQE) channel. By comparing baseline RFG predictions against experimental data from MicroBooNE, this research quantifies the necessity of advanced many-body nuclear frameworks. The implementation of the Local Fermi Gas (LFG) model, Random Phase Approximation (RPA) screening, and the SuperScaling Approximation (SuSAv2) are systematically analyzed to resolve kinematic biases. Additionally, the impact of modernized phenomenological constraints, including the updated $M_A^{RES} = 0.96$ GeV for resonant pion production and the 2021 Bodek-Yang parameter updates for Deep Inelastic Scattering (DIS), is evaluated. Furthermore, this research integrates university extension initiatives by developing and presenting macroscopic 3D-printed educational models, consisting of a Linear Accelerator (LINAC) analog and a scattering experiment, and finally presenting it into the festival ``Mais Ciencia Por Favor".

\vspace{1cm}
\noindent \textbf{Keywords:} Neutrino Interactions; DUNE; GENIE Event Generator; Final State Interactions; Cross-section; Scientific Outreach.
\clearpage

\newpage
\thispagestyle{empty}
\begin{center}
    \Large \textbf{RESUMO}
\end{center}
\vspace{0.5cm}

\noindent A recente geração de experimentos de oscilação de neutrinos de alta precisão, como o Deep Underground Neutrino Experiment (DUNE), depende da reconstrução precisa da energia do neutrino incidente para determinar parâmetros fundamentais, como a violação de Carga-Paridade (CP) e a hierarquia de massa dos neutrinos. O uso de Argônio Líquido ($^{40}\text{Ar}$) introduz complexidades teóricas significativas. Esta dissertação fornece uma investigação abrangente sobre as fenomenologias do espalhamento neutrino-núcleo, conectando a teoria eletrofraca fundamental com simulações computacionais usando o gerador de eventos Monte Carlo GENIE. O estudo avalia as limitações dos modelos tradicionais, como o modelo do Gás de Fermi Relativístico (RFG), abordando especificamente o parâmetro de Massa Axial e seu impacto no canal Espalhamento Quase-Elástico de Corrente Carregada (CCQE). Ao comparar as previsões de linha de base do RFG com dados experimentais do MicroBooNE, esta pesquisa quantifica a necessidade de modelos nucleares avançados de muitos corpos. A implementação do modelo do Gás de Fermi Local (LFG), a blindagem por Aproximação de Fase Aleatória (RPA) e a Aproximação de SuperEscalamento (SuSAv2) são analisadas sistematicamente para resolver viéses cinemáticos. Adicionalmente, avalia-se o impacto de restrições fenomenológicas modernizadas, incluindo o $M_A^{\text{RES}} = 0{,}96\text{ GeV}$ atualizado para a produção ressonante de píons e as atualizações dos parâmetros de Bodek-Yang de 2021 para o Espalhamento Inelástico Profundo (DIS). Além disso, esta pesquisa integra iniciativas de extensão universitária ao desenvolver e apresentar modelos educacionais macroscópicos impressos em 3D, consistindo em um análogo de Acelerador Linear (LINAC) e um experimento de espalhamento, e, finalmente, apresentando-os no festival ``Mais Ciência Por Favor''.

\vspace{1cm}
\noindent \textbf{Palavras-chave:} Interações de Neutrinos; DUNE; Gerador de Eventos GENIE; Interações de Estado Final; Seção de Choque; Letramento Científico.

\clearpage
\pagestyle{empty}           

\tableofcontents

\clearpage
\pagestyle{fancy}          
\setcounter{page}{1}       

\chapter{Introduction}

The field of neutrino physics has definitively entered an era of high precision \cite{NuSTEC:2018}. Following the historic discovery of neutrino oscillations, a phenomenon demonstrating that neutrinos possess non-zero mass and mix among different flavor states, the global scientific community has shifted its focus towards determining the remaining unknown parameters of the Standard Model's lepton sector, which include the precise measurement of the neutrino mass hierarchy and the observation of Charge-Parity (CP) symmetry violation in the leptonic sector~\cite{neutrinoHistory}. The Deep Underground Neutrino Experiment (DUNE), currently under construction, represents the flagship of this global effort. To achieve its unprecedented physics goals, DUNE will utilize Liquid Argon Time Projection Chambers (LArTPCs), a detector technology that offers exceptional spatial resolution and precise capabilities for reconstructing the complex tracks of final-state particles \cite{P5, U11}.

Despite the technological evolution of the detectors, a critical theoretical challenge persists: in order to extract the oscillation parameters, physicists must accurately reconstruct the energy of the incident neutrino. Because DUNE operates with a wide-band neutrino beam, the initial energy of any single interacting neutrino is unknown and must be inferred entirely from the measured kinematics of the outgoing particles \cite{DUNE_TDR_Vol2}. This reconstruction is complicated by the fact that neutrinos in the DUNE detectors do not interact with isolated free nucleons, instead, they interact with nucleons that are bound within the nuclei, specifically Argon-40 ($^{40}$Ar). Consequently, the reconstruction of the incident neutrino energy is affected and distorted by multiple nuclear effects \cite{P4, P12}.

The theoretical and experimental difficulties introduced by the nuclear environment can be categorized into initial-state and final-state effects. In the initial state, protons and neutrons, also called nucleons, are not at rest, they possess an intrinsic momentum due to Fermi motion and are subject to nuclear binding energies \cite{NuSTEC:2018}. This initial dynamic state distorts the outgoing kinematics of the scattered lepton and hadrons. Also, a significant fraction of neutrinos interact with correlated pairs of nucleons through Meson Exchange Currents (MEC). The theoretical models describing these multi-nucleon processes are currently constrained by experimental data, thereby representing a major source of uncertainty in the interaction cross-section \cite{P2, P4}. Equally challenging are the Final State Interactions (FSI). When a neutrino interaction produces hadrons at the primary weak interaction, these particles are produced within the nuclear potential. Before they can emerge from the Argon nucleus and be detected by the LArTPC, they must traverse the nuclear medium. During this transit, hadrons can undergo elastic rescattering, charge exchange, or complete absorption. These FSI drastically alter the final topology of the event observed in the detector~\cite{U12}. For instance, a pion produced in a resonant interaction may be completely absorbed by the nucleus, effectively disappearing from the observable final state. This causes the event to imitate a quasi-elastic topology and leads to a severe underestimation of the total deposited energy, directly biasing the reconstruction of the neutrino energy \cite{P7, P8}.

Modern neutrino experiments rely on Monte Carlo event generators \cite{NuSTEC:2018}. The Generative Neutrino Interaction Evaluation (GENIE) framework is the standard simulation tool adopted by the DUNE collaboration to model the complete neutrino-nucleus interaction pipeline, from the primary vertex to the intra-nuclear hadronic cascades \cite{P14, U7}. However, recent high-precision studies have indicated significant discrepancies between GENIE predictions and actual experimental data derived from electron-nucleus and neutrino scattering experiments \cite{P1, P3}. While recent theoretical advancements have been integrated into the generator, their validation for the DUNE neutrino flux on Argon targets requires investigation.

The central motivation of this Master's research is the need to validate, refine, and quantify the uncertainties of the neutrino-nucleus interaction models. By utilizing the GENIE event generator, this work aims to evaluate the uncertainties associated with neutrino-Argon scattering. Specifically, the research will perform a comparative study of different models available within GENIE, focusing on the differences between the established baseline models and the most recent theoretical implementations \cite{P2}. A major component of this evaluation will be quantifying the impact of FSI on the energy reconstruction, with emphasis on interaction channels involving pion production \cite{P8}. Ultimately, by identifying the most sensitive kinematic variables and evaluating the biases in energy reconstruction, this research aims to contribute directly to the reduction of systematic errors in DUNE's oscillation analyses, aligning with the efforts of the experiment's cross-section working groups \cite{U17}. This dissertation is organized to provide a logical progression from fundamental particle physics to advanced computational simulations. Following this introduction, Sec. 2 provides a historical and theoretical overview of the neutrino; Sec. 3 is a review of particle physics, detailing the Standard Model, relativistic kinematics, and the mathematical derivation of cross-sections via Fermi's Golden Rule; Sec. 4 is an introduction to the Deep Underground Neutrino Experiment, describing its facilities, the Liquid Argon TPC technology, and the dominant neutrino interaction channels at the few-GeV energy scale; Sec. 5 is the theoretical modeling of the nucleus, exploring the Fermi Gas and Relativistic Fermi Gas models that form the background of interaction simulations; Sec. 6 gives the details of the architecture and operational mechanics of Neutrino Event Generators, specifically focusing on the GENIE framework and its treatment of Monte Carlo sampling; Sec. 7, 8 and 9 show the results and the discussion about the systematic study performed in the axial mass parameter, the nuclear models and the Deep Inelastic Scattering update, respectively; Sec. 10 shows the highlights of the university extension and scientific outreach activities developed during this program, demonstrating the translation of abstract particle physics concepts into accessible educational models; finally, Sec. 12 concludes the discussion and present the next steps and future work.

\newpage
\chapter{The Neutrino}
In December 1930, the physicist Wolfgang Pauli proposed the existence of a new particle in a letter addressed to the participants of a nuclear conference in Tübingen, Germany \cite{neutrinoHistory}. At that time, the theoretical model of beta decay ($\beta$-decay) faced a significant challenge, as its predictions did not coincide with experimental data. The prevailing model considered a two-body decay process:
\begin{equation}
   (A,Z) \longrightarrow (A, Z + 1) + e^{-}.
   \label{eq:wrongbetaD}
\end{equation}
    
\noindent According to the laws of energy and momentum conservation for such a process, the emitted electron should possess a single, discrete kinetic energy. However, experimental observations contradicted this prediction, revealing that the electron's energy was distributed over a continuous spectrum \cite{neutrinoHistory}. To resolve this discrepancy, Pauli postulated a new neutral particle, which he called a ``neutron," as described in his letter:

\begin{quote}
\itshape
Dear Radioactive Ladies and Gentlemen,
\par
[...] the continuous $\beta$-spectrum would then become understandable under the assumption that in $\beta$-decay a neutron is emitted in addition to the electron, in such a way that the sum of the energies of the neutron and the electron is constant.
\end{quote}

\noindent Thus, according to Pauli's proposal, $\beta$-decay was, in fact, a three-body process:
\begin{equation}
   (A,Z) \longrightarrow (A, Z + 1) + e^{-} + n .
   \label{eq:betaD}
\end{equation}

\noindent This model successfully explained the continuous energy spectrum. Subsequently, the particle proposed by Pauli was renamed ``neutrino" by Enrico Fermi, and it was assigned its corresponding Greek letter symbol, $\nu$ \cite{MestradoGustavo}.

It is now well established that neutrinos exist in three distinct types, known as flavors: the electron neutrino ($\nu_e$), the muon 
neutrino ($\nu_\mu$), and the tau neutrino ($\nu_\tau$). In the Standard Model of particle physics, these are described as massless spin 1/2 particles. However, recent experiments, such as the one conducted at the Sudbury Neutrino Observatory (SNO), have conclusively demonstrated a phenomenon known as neutrinooscillation. This process, in which neutrinos change from one flavor to another, requires that they have mass, a property not included in the original Standard Model \cite{MestradoGustavo}. While the exact values remain unknown, various experiments have established stringent upper limits on the neutrino masses. The latest results from global analyses of experimental data are regularly updated and compiled, for instance, by the Nu-fit collaboration \cite{nu-fit-data}.

\newpage
\chapter{Particle Physics Review}
A description of quantum particle interactions requires a foundational review of particle physics. Key concepts from quantum mechanics and statistical physics are essential to construct the mathematical framework that governs fundamental particle interactions.

\section{The Standard Model}

\begin{figure}
    \centering
    \includegraphics[width=1\linewidth]{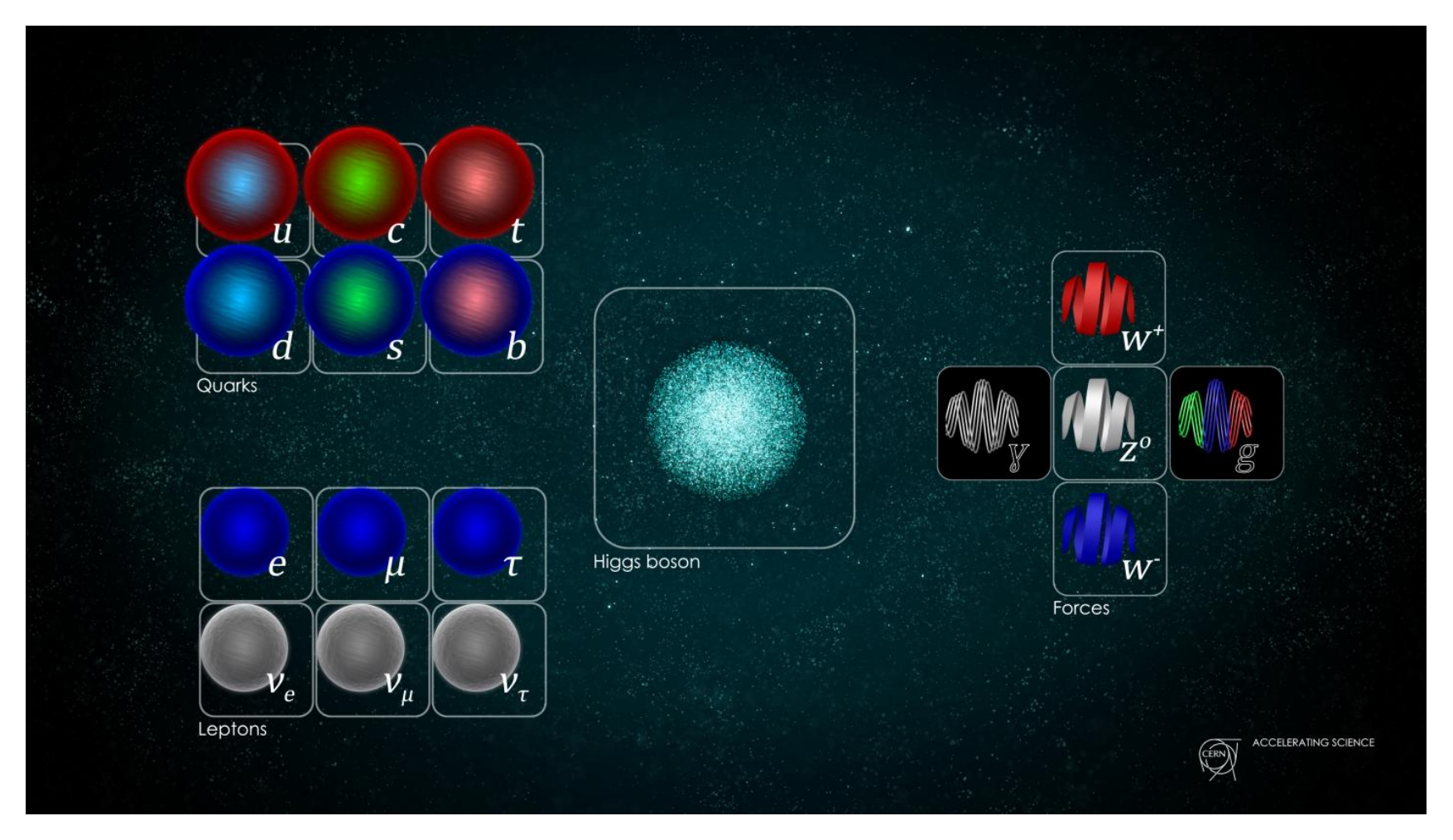}
    \caption{Schematic representation of the Standard Model of Particle Physics produced by CERN. The diagram organizes elementary particles into three main sectors: on the left, the matter-constituting fermions, subdivided into quarks ($u, d, c, s, t, b$) in the upper panel and leptons ($e, \mu, \tau$ and their corresponding neutrinos) in the lower panel. Positioned in the center is the Higgs boson, responsible for the mass-generation mechanism. On the right are the gauge bosons (forces), which mediate the fundamental interactions: the photon ($\gamma$), weak bosons ($W^+, W^-, Z^0$), and the gluon ($g$) \cite{cern_sm}.}
    \label{fig:standardModel}
\end{figure}

 By definition, fundamental particles are the elementary constituents of matter and radiation. To classify these particles, physicists developed the Standard Model (SM) of Particle Physics, as illustrated in Figure \ref{fig:standardModel}. The matter particles, known as fermions, are organized into three generations. The upper-left section of Figure \ref{fig:standardModel} displays the quarks. These are categorized into six types, known as flavors: up (u), down (d), charm (c), strange (s), top (t), and bottom (b). Quarks are the constituents of composite particles such as the protons and neutrons that form atomic nuclei \cite{cern_sm}. The lower-left section of Fig. \ref{fig:standardModel} represents the leptons. This group includes three charged leptons, the electron ($e$), muon ($\mu$), and tau ($\tau$) and their neutral counterparts, the neutrinos. Neutrinos also come in three flavors, each corresponding to a charged lepton: the electron neutrino ($\nu_e$), muon neutrino ($\nu_\mu$), and tau neutrino ($\nu_\tau$). The right-hand side of the figure shows the bosons, spin-1 particles that mediate the fundamental forces \cite{cern_sm}. The Standard Model accounts for three of the four fundamental forces: the strong force, mediated by the gluon (g); the electromagnetic force, mediated by the photon ($\gamma$); and the weak force, mediated by the W and Z bosons. Finally, the SM includes the Higgs boson, a spin-0 particle associated with the Higgs field, which gives mass to the W and Z bosons, quarks, and charged leptons \cite{cern_sm}. The mathematical foundation of the Standard Model is Quantum Field Theory (QFT), which treats each fundamental particle as an excitation of a corresponding quantum field. 

\section{Relativity in Particle Physics}
Proposed in 1905 by Albert Einstein, Special Relativity revolutionized physics by establishing fundamental principles such as the constancy of the speed of light for all inertial observers and the equivalence between mass and energy. This theoretical framework is indispensable for describing the subatomic world, as particles in accelerators and cosmic ray interactions routinely reach energies where their speeds approach that of light. Consequently, relativistic kinematics govern all particle interactions.

Special Relativity postulates the union of three-dimensional space and time into a single space known as Minkowski Spacetime. In this formalism, physical events are described by four-vectors, whose contravariant form is given by \cite{griffiths2020introduction}:
\begin{equation}
    A^{\mu} = (A^0, A^1, A^2, A^3) = (A^0, \bm{A}).
    \label{eq:contrafourvector}
\end{equation}

\noindent The geometry of Minkowski spacetime is defined by a scalar product, which is formulated using the metric tensor, $g_{\mu\nu}$ \cite{griffiths2020introduction}:
\begin{equation}
    A \cdot B = A^{\mu}B_{\mu} = g_{\mu\nu}A^{\mu}B^{\nu} = A^0B^0 - \boldsymbol{A} \cdot \boldsymbol{B},
    \label{eq:scalarfourvec}
\end{equation}

\noindent For the particle physics convention used in this work, the metric tensor is:
\begin{equation}
    g_{\mu\nu} = \begin{pmatrix}
         1 &  0 &  0 &  0 \\
         0 & -1 &  0 &  0 \\
         0 &  0 & -1 &  0 \\
         0 &  0 &  0 & -1 \\
    \end{pmatrix}.
    \label{eq:metrictensor}
\end{equation}

\noindent A key feature of this scalar product is that it forms a Lorentz invariant a quantity whose value remains the same for all inertial observers. The inner product of the four-momentum with itself is one of the most important invariants in physics. Applying the definition from Eq. (\ref{eq:scalarfourvec}) to the four-momentum, $p^\mu = (E/c, \boldsymbol{p})$, yields:
\begin{equation}
    p^2 = p^{\mu}p_{\mu} = \left(\frac{E}{c}\right)^2 - |\boldsymbol{p}|^2 = (m_0c)^2.
    \label{eq:invariantmass}
\end{equation}

\noindent Rearranging this invariant quantity yields the energy-momentum relation:
\begin{equation}
    E^2 = (m_0c^2)^2 + (pc)^2.
    \label{eq:energymomentum}
\end{equation}

\noindent This relation encapsulates the principle of mass-energy equivalence, allowing for the creation of new massive particles from kinetic energy in the high-energy collisions studied in experimental particle physics.

\section{Natural Units}
In High-Energy Physics, the standard International System of Units (SI) proves to be cumbersome due to the microscopic scales of length and time, and the relativistic speeds involved. Consequently, it is standard practice to adopt the system of \textit{Natural Units}, which simplifies the mathematical formalism by normalizing fundamental constants of nature to unity. This convention is defined by setting the reduced Planck constant ($\hbar$) and the speed of light in a vacuum ($c$) equal to one \cite{NaturalUnits}:

\begin{equation}
    \hbar = k_B = c = 1.
\end{equation}

\noindent By setting $c=1$, the distinction, in therms of units, between space ($L$) and time ($T$) is removed, treating them as equal, like required by Special Relativity. Similarly, setting $\hbar=1$ relates energy to frequency and momentum to wavenumber, bridging the concepts of classical mechanics and quantum theory. 

\subsection{Dimensional Analysis}
A direct consequence of this convention is that all physical quantities can be expressed in terms of a single dimension, typically chosen to be energy ($E$). In particle physics, the standard unit of energy is the electron-volt (eV), with relevant scales often in the range of MeV ($10^6$ eV) or GeV ($10^9$ eV).

Using dimensional analysis with $[c] = [L][T]^{-1} = 1$ and $[\hbar] = [E][T] = 1$, the dimensions of mass ($M$), length ($L$), and time ($T$) relate to energy as follows:

\begin{equation}
    [L] = [T] = [E]^{-1}, \quad [M] = [E].
\end{equation}

\noindent Therefore, mass, momentum, and temperature are all measured in units of energy (e.g., GeV), while length and time are measured in inverse energy (e.g., $\text{GeV}^{-1}$). This leads to significant simplifications in fundamental equations. For instance, the relativistic energy-momentum relation becomes:
\begin{equation}
    E^2 = p^2 + m^2.
\end{equation}

\noindent Table \ref{tab:natural_units} summarizes the dimensions of common physical quantities in this system.

\begin{table}
    \centering
    \caption{Dimensions of physical quantities in Natural Units, expressed as powers of energy $[E]$ (Author).}
    \label{tab:natural_units}
    \begin{tabular}{|| l | c | c ||}
        \hline
        \textbf{Quantity} & \textbf{Symbol} & \textbf{Dimension} $[E]^n$ \\ [0.5 ex]
        \hline\hline
        Mass & $m$ & $1$ \\
        \hline
        Energy & $E$ & $1$ \\
        \hline
        Momentum & $p$ & $1$ \\
        \hline
        Length & $L$ & $-1$ \\
        \hline
        Time & $t$ & $-1$ \\
        \hline
        cross-section & $\sigma$ & $-2$ \\
        \hline
    \end{tabular}
\end{table}

\subsection{Conversion to SI Units}

While theoretical derivations are performed in natural units, experimental results must often be compared with macroscopic measurements. To convert quantities back to the SI system (or convenient mixed units like fm and mb), one utilizes the conversion factors derived from the values of $\hbar$ and $c$. The most useful conversion constant in nuclear and particle physics is the product $\hbar c$: 

\begin{equation}
    \hbar c = 197.326 980 4 \text{ MeV} \text{fm}.
\end{equation}

\noindent For example, to convert a cross-section calculated in natural units (where it has dimensions of $\text{GeV}^{-2}$) into units of area (such as millibarns, mb), one applies the conversion factor $(\hbar c)^2$:

\begin{equation}
    1 \text{ GeV}^{-2} = \frac{(\hbar c)^2}{(1 \text{ GeV})^2} \approx (0.197 \text{ fm})^2 \approx 0.389 \text{ mb},
\end{equation}
\noindent where $1 \text{ fm} = 10^{-15} \text{ m}$ and $1 \text{ mb} = 10^{-27} \text{ cm}^2$.

\section{Cross-section: The Bridge Between Theory and Experiment}
\begin{figure}
    \centering
    \includegraphics[width=0.5\linewidth]{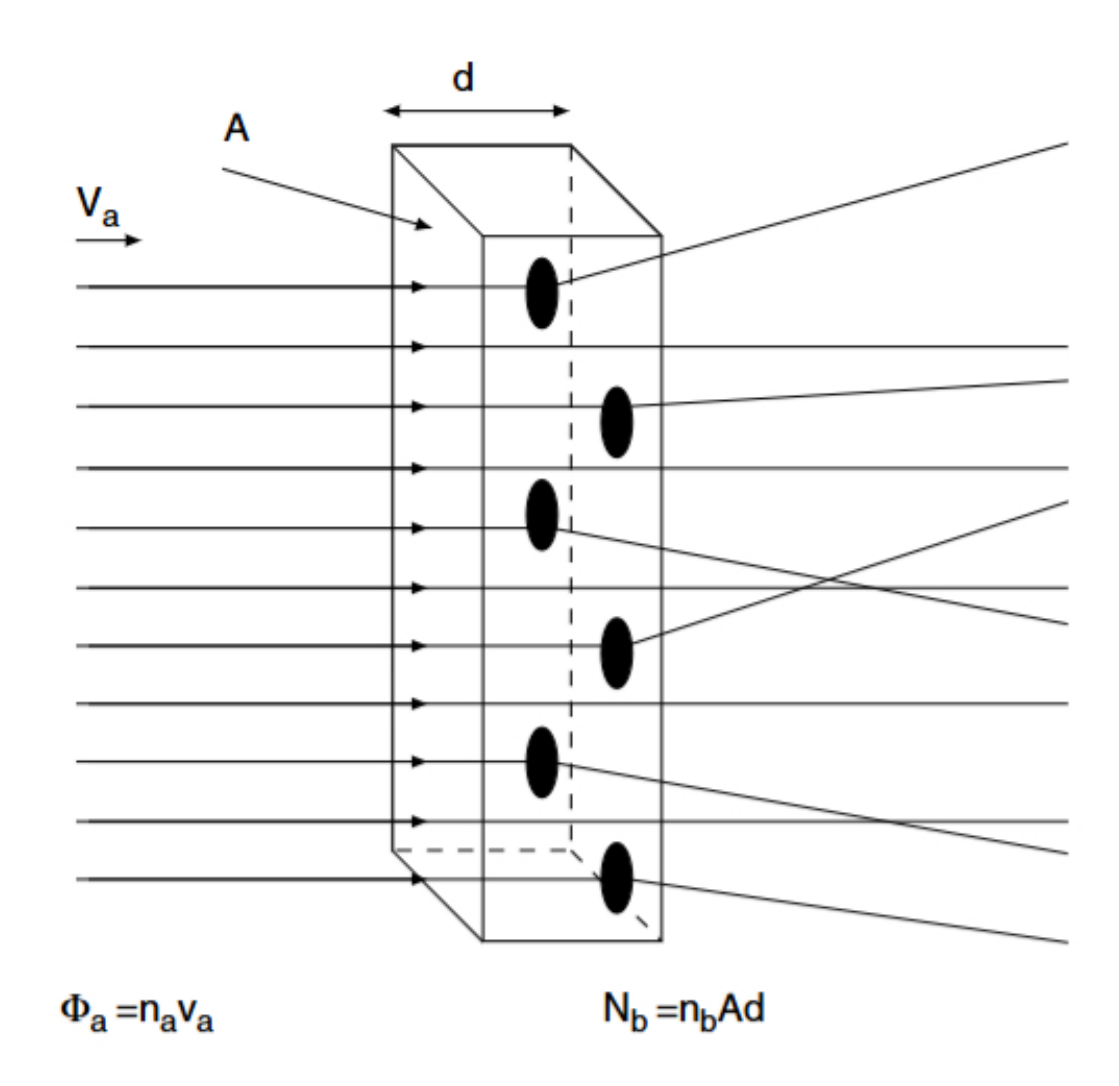}
    \caption{Schematic representation of a macroscopic scattering process used to derive the concept of an interaction cross-section. An incident beam of particles with number density $n_a$ and velocity $v_a$ produces a flux $\Phi_a = n_a v_a$. The beam impinges perpendicularly on a target block of cross-sectional area $A$ and thickness $d$. The target contains a total of $N_b = n_b A d$ scattering centers, where $n_b$ is the target number density. In this geometric interpretation, the cross-section $\sigma$ represents the effective area that each target particle presents to the incident flux, determining the overall probability of a scattering event \cite{crossection}.}
    \label{fig:cross-section}
\end{figure}

In particle physics, theoretical models are validated through comparison with experimental measurements. One physical observable that serves as the quantitative bridge between these two domains is the cross-section. Conceptually denoted by $\sigma$, represents the effective area that a target particle presents to a projectile for a specific interaction to occur. It is a measure of the intrinsic probability of that interaction, and for this reason, it is expressed in units of area. Figure \ref{fig:cross-section} illustrates this concept.

The cross-section is a measured quantity derived from counting the number of interactions ($N_{\text{events}}$) produced by a beam of particles incident on a target. The rate ($R$) of these events depends on the intrinsic probability of the interaction ($\sigma$), the number of scattering centers ($N_b$) and the flux ($\phi_a$), which is the number of projectiles hitting the target per unit area and per unit time \cite{crossection}.
\begin{equation}
    R=\phi_a N_b \sigma_b.
    \label{eq:Rate}
\end{equation}

\noindent Since the flux and the scattering centers are properties of the experimental setup, it is defined a single parameter called luminosity ($\mathcal{L}=\phi_a N_b$), which these factors are encapsulated in, with units of cm$^{-2}$s$^{-1}$. So according to eq. (\ref{eq:Rate}) the relation is:

\begin{equation}
    R = \frac{dN_{\text{events}}}{dt} =\sigma\phi_a N_b,
    \label{eq:diferencialRate}
\end{equation}

\begin{equation}
    dR= \mathcal{L} \cdot \frac{d\sigma}{d\Omega} d\Omega .
    \label{eq:diferencialLuminosity}
\end{equation}

\noindent In practice, experimentalists use this relation to extract the cross-section from the data. Eq. (\ref{eq:diferencialLuminosity}) is expressed in terms of the differential cross-section.

From a theoretical point of view, the cross-section is a quantity derived from the dynamics of the interaction as described by QFT. Theorists use the fundamental principles of the Standard Model, for instance Fermi's Golden Rule as we'll see below, to compute $|\mathcal{M}|^2$ and predict how $d\sigma/d\Omega$ should behave as a function of energy and angle \cite{crossection}.

The connection between theory and experiment is established by comparing the measured cross-section with the predicted cross-section. A precise agreement between the two sides validates our theoretical understanding, while any significant discrepancy could be a sign of new, undiscovered physics.

\section{Electroweak Interaction in the Standard Model}

Since this work focuses on neutrino interactions, it is necessary to first understand how the neutrino couples to other particles within the Standard Model. These couplings are categorized into two primary types: Charged Current (CC) and Neutral Current (NC) interactions. Understanding these mechanisms is fundamental to identifying which interaction channels are relevant to this study.

\subsection{Charged Current (CC) Interactions}

Charged Current interactions are mediated by the $W$ bosons, which carry electric charge ($W^+$ or $W^-$). The characteristic of CC interactions is the exchange of electric charge between the fermions and the boson, resulting in a change of the flavor.

Mathematically, the interaction is described by the Lagrangian density coupling the weak charged current ($J^\mu_{CC}$) to the $W$ boson  \cite{electroweak}:
\begin{equation}
    \mathcal{L}_{CC} = -\frac{g}{\sqrt{2}} \left( J^\mu_{CC} W^+_\mu + J^{\mu\dagger}_{CC} W^-_\mu \right),
\end{equation}
\noindent where $g$ is the weak coupling constant. The charged current has a purely chiral $V-A$ (Vector minus Axial) structure, involving only left-handed fermions (or right-handed antifermions).

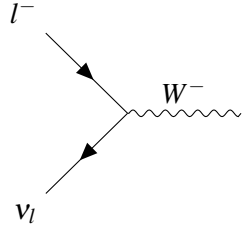
\begin{figure}[!ht]
\begin{center}
    \begin{tikzpicture}
    \begin{feynman}
        \vertex (a);
        \vertex [above left=of a] (b) {$l^-$};
        \vertex [below left=of a] (c) {$\nu_l$};
        \vertex [right=of a] (d);

        \diagram* {
        (b) -- [fermion] (a) -- [fermion] (c);
        (a) -- [boson, edge label=$W^-$] (d);
        };
    \end{feynman}
\end{tikzpicture}
\end{center}

\caption{Fundamental Feynman diagram representing a charged-current (CC) neutrino scattering process. The interaction is mediated by the exchange of a $W^-$ boson. At the top-left vertex, an incoming negative charged lepton ($l^-$). At the down-left an outgoing lepton-neutrino ($\nu_l$). At the middle, the negative $W^-$ boson. It represents a lepton emitting a $W$ boson, changing the momentum direction and turning into a lepton-neutrino. This basic topology forms the mechanism for the CC interactions (Author).}
\label{fig:Wdiagram}
\end{figure}

\noindent For leptons, this process involves a charged lepton ($l^\pm$) exchanging a $W$ boson and transforming into its corresponding neutrino flavor ($\nu_l$), or vice-versa, ensuring the conservation of lepton number. The fundamental vertex is illustrated in the Feynman Diagram in Figure \ref{fig:Wdiagram}.

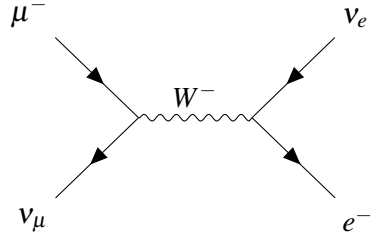
\begin{figure}[!ht]
\begin{center}
    \begin{tikzpicture}
    \begin{feynman}
        \vertex (a);
        \vertex [above left=of a] (b) {$\mu^-$};
        \vertex [below left=of a] (c) {$\nu_\mu$};
        \vertex [right=of a] (d);
        \vertex [above right=of d] (e) {$\nu_e$};
        \vertex [below right=of d] (f) {$e^-$};

        \diagram* {
        (b) -- [fermion] (a) -- [fermion] (c);
        (a) -- [boson, edge label=$W^-$] (d);
        (e) -- [fermion] (d) -- [fermion] (f);
        };
    \end{feynman}
\end{tikzpicture}
\end{center}

\caption{Feynman diagram representing the $\mu^- + \nu_e \to e^- + \nu_\mu$ neutrino process. The interaction is mediated by the exchange of a $W^-$ boson. At the top-left vertex, an incoming muon ($\mu^-$). At the down-left an outgoing muon-neutrino ($\nu_{\mu}$). At the middle, the negative $W^-$ boson. At top-right vertex, an incoming electron-neutrino ($\nu_e$) and at the down-right, the outgoing electron $e^-$ (Author).}
\label{fig:Wdiagram_example}
\end{figure}

\noindent As an example, the inverse muon decay ($\nu_\mu + e^- \to \mu^- + \nu_e$) or the reaction $\mu^- + \nu_e \to e^- + \nu_\mu$ can be expressed by the diagram in Figure \ref{fig:Wdiagram_example} \cite{griffiths2020introduction}.

\subsection{Neutral Current (NC) Interactions}
Neutral Current interactions are mediated by the $Z^0$ boson, which carries no electric charge. Unlike CC interactions, the NC processes do not change the flavor of the participating fermions.

The corresponding Lagrangian for the neutral current interaction is given by \cite{electroweak}:
\begin{equation}
    \mathcal{L}_{NC} = -\frac{g}{2\cos\theta_W} J^\mu_{NC} Z_\mu,
\end{equation}
where $\theta_W$ is the Weinberg angle (or weak mixing angle) and $J^\mu_{NC}$ is the weak neutral current.

The fundamental vertex representation is represented by Figure \ref{fig:Zdiagram}:

\begin{figure}[!ht]
\begin{center}
    \begin{tikzpicture}
    \begin{feynman}
        \vertex (a);
        \vertex [above left=of a] (b) {$l$};
        \vertex [below left=of a] (c) {$l$};
        \vertex [right=of a] (d);

        \diagram* {
        (b) -- [fermion] (a) -- [fermion] (c);
        (a) -- [boson, edge label=$Z^0$] (d);
        };
    \end{feynman}
\end{tikzpicture}
\end{center}

\caption{Fundamental Feynman diagram representing a neutral-current (NC) neutrino scattering process. The interaction is mediated by the exchange of a $Z$ boson. At the top-left vertex, an incoming lepton ($l$). At the down-left an outgoing lepton ($l$). At the middle, the $Z$ boson. It represents a lepton emitting a $Z$ boson and changing the momentum direction. This basic topology forms the mechanism for the NC interactions (Author).}
\label{fig:Zdiagram}
\end{figure}
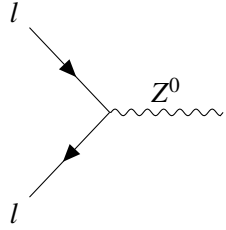

A classic example is the elastic neutrino-electron scattering ($\nu_\mu + e^- \to \nu_\mu + e^-$), represented by the diagram in Figure \ref{fig:Zdiagram_example} \cite{griffiths2020introduction}:

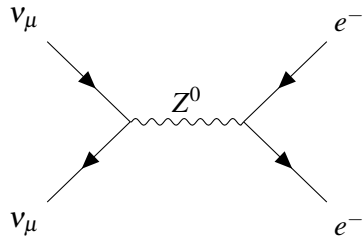
\begin{figure}[!ht]
\begin{center}
    \begin{tikzpicture}
    \begin{feynman}
        \vertex (a);
        \vertex [above left=of a] (b) {$\nu_\mu$};
        \vertex [below left=of a] (c) {$\nu_\mu$};
        \vertex [right=of a] (d);
        \vertex [above right=of d] (e) {$e^-$};
        \vertex [below right=of d] (f) {$e^-$};

        \diagram* {
        (b) -- [fermion] (a) -- [fermion] (c);
        (a) -- [boson, edge label=$Z^0$] (d);
        (e) -- [fermion] (d) -- [fermion] (f);
        };
    \end{feynman}
\end{tikzpicture}
\end{center}

\caption{Feynman diagram representing the $\nu_\mu + e^- \to \nu_\mu + e^-$ neutrino process. The interaction is mediated by the exchange of a $Z$ boson. At the top-left vertex, an incoming muon-neutrino ($\nu_\mu^-$). At the down-left an outgoing muon-neutrino ($\nu_{\mu}$). At the middle, the $Z$ boson. At top-right vertex, an incoming electron ($e^-$) and at the down-right, the outgoing electron $e^-$ (Author).}
\label{fig:Zdiagram_example}
\end{figure}

\section{Mandelstam Variables}
In the analysis of scattering processes involving particles interactions, for instance two initial and two final state particles ($1 + 2 \to 3 + 4$), it is essential to utilize a kinematic formalism that remains independent of the reference frame. Historically, scattering amplitudes were often described using frame-dependent variables, such as the laboratory energy or the scattering angle. However, in 1958, Stanley Mandelstam introduced a set of three Lorentz-invariant variables now known as Mandelstam variables ($s, t, u$) to formulate dispersion relations for pion-nucleon scattering in a covariant manner \cite{mandelstam1958determination}. These variables encode the energy, momentum transfer, and angular dependence of the interaction in a unified framework essential for the SM phenomenology.

\subsection{Definitions and Lorentz Invariance}

Consider a general two-body scattering process represented by the four-momenta $p_1 + p_2 \to p_3 + p_4$, where $p_i = (E_i, \vec{p}_i)$ denotes the four-momentum of particle $i$, and the energy-mass relation, given by Eq. (\ref{eq:energymomentum}), implies $p_i^2 = m_i^2$. The Mandelstam variables are defined as the square of the sums and differences of these four-momenta \cite{peskin2018introduction}:

\begin{align}
    s &= (p_1 + p_2)^2 = (p_3 + p_4)^2, \label{eq:mandel_s} \\
    t &= (p_1 - p_3)^2 = (p_2 - p_4)^2, \label{eq:mandel_t} \\
    u &= (p_1 - p_4)^2 = (p_2 - p_3)^2. \label{eq:mandel_u}
\end{align}

The utility of these variables stems from their property as Lorentz scalars. In Special Relativity, the scalar product of two four-vectors $A^\mu$ and $B^\mu$, given by $A \cdot B = g_{\mu\nu}A^\mu B^\nu$, is invariant under Lorentz transformations, as discussed in the Relativity Review section. Since the four-momentum $p^\mu$ transforms as a vector, any linear combination of momenta (e.g., $q^\mu = p_1^\mu - p_3^\mu$) is also a four-vector. Consequently, the square of such a vector, $q^2 = q_\mu q^\mu$, is a Lorentz scalar. Therefore, the values of $s$, $t$, and $u$ are identical in all inertial reference frames, whether in the Laboratory frame or the Center of Mass (CM) frame. Furthermore, these three variables are not independent. Conservation of four-momentum ($\sum p_{in} = \sum p_{out}$) leads to the sum rule \cite{peskin2018introduction}:
\begin{equation}
    s + t + u = \sum_{i=1}^{4} m_i^2,
\end{equation}
where $m_i$ are the masses of the participating particles.

\subsection{Physical Interpretation and Interaction Channels}

The variables $s$, $t$, and $u$ correspond to three distinct topological configurations in Feynman diagrams, referred to as channels. Each channel represents a different physical process depending on which variable corresponds to the square of the momentum carried by the intermediate propagator.

\begin{enumerate}
    \item \textbf{s-channel:} In this configuration, particles 1 and 2 annihilate or merge to form an intermediate state (resonance) which subsequently decays into particles 3 and 4. The variable $s$ represents the square of the total energy in the Center of Mass frame ($s = E_{CM}^2$).
    
    \item \textbf{t-channel:} This channel describes a scattering process where a mediator boson is exchanged between the line $1 \to 3$ and the line $2 \to 4$. The variable $t$ represents the square of the four-momentum transfer. In the context of neutrino-nucleus scattering discussed in this work, the relevant kinematic variable is often $Q^2$, defined as $Q^2 \equiv -t$, which serves as a measure of the resolution scale of the probe.
    
    \item \textbf{u-channel:} This channel corresponds to an exchange where particle 1 turns into particle 4, while particle 2 turns into particle 3. It is topologically similar to the t-channel but involves the exchange of the outgoing particles (crossed channel), often relevant for backward scattering processes.
\end{enumerate}

\begin{figure}
    \centering
    \begin{tikzpicture}
        \begin{feynman}
            \vertex (a1) {1};
            \vertex [below=2cm of a1] (a2) {2};
            \vertex [right=1.5cm of a1] (b1);
            \vertex [below=1cm of b1] (center_s);
            \vertex [right=1.5cm of center_s] (out_s);
            \vertex [above right=1.0cm of out_s] (a3) {3};
            \vertex [below right=1.0cm of out_s] (a4) {4};
            
            \diagram* {
                (a1) -- [fermion] (center_s),
                (a2) -- [fermion] (center_s),
                (center_s) -- [boson, edge label=$s$] (out_s),
                (out_s) -- [fermion] (a3),
                (out_s) -- [fermion] (a4),
            };
        \end{feynman}
        \node [below=0.5cm of center_s] {(a) s-channel};
    \end{tikzpicture}
    \hspace{0.5cm}
    \begin{tikzpicture}
        \begin{feynman}
            \vertex (t1) {1};
            \vertex [below=2cm of t1] (t2) {2};
            \vertex [right=2cm of t1] (t3) {3};
            \vertex [right=2cm of t2] (t4) {4};
            \vertex [right=1cm of t1] (vm1);
            \vertex [right=1cm of t2] (vm2);
            
            \diagram* {
                (t1) -- [fermion] (vm1) -- [fermion] (t3),
                (t2) -- [fermion] (vm2) -- [fermion] (t4),
                (vm1) -- [boson, edge label=$t$] (vm2),
            };
        \end{feynman}
         \node [below=0.5cm of vm2] {(b) t-channel};
    \end{tikzpicture}
    \hspace{0.5cm}
    \begin{tikzpicture}
        \begin{feynman}
            \vertex (u1) {1};
            \vertex [below=2cm of u1] (u2) {2};
            \vertex [right=2cm of u1] (u3) {3};
            \vertex [right=2cm of u2] (u4) {4};
            \vertex [right=1cm of u1] (vu1);
            \vertex [right=1cm of u2] (vu2);
            
            \diagram* {
                (u1) -- [fermion] (vu1) -- [fermion] (u4),
                (u2) -- [fermion] (vu2) -- [fermion] (u3),
                (vu1) -- [boson, edge label=$u$] (vu2),
            };
        \end{feynman}
        \node [below=0.5cm of vu2] {(c) u-channel};
    \end{tikzpicture}
    \label{fig:mandelstam_channels}
    \caption{Feynman diagrams representing the three fundamental scattering topologies described by the Mandelstam variables. (a) The $s$-channel, where the initial particles 1 and 2 annihilate to form an intermediate boson that subsequently decays into the final state particles 3 and 4. (b) The $t$-channel, characterizing a direct scattering process mediated by the exchange of a boson between the incoming and outgoing particle lines ($1 \to 3$ and $2 \to 4$). (c) The $u$-channel, representing a crossed-exchange process where the initial particle 1 transforms into particle 4, while particle 2 transforms into particle 3. These invariant kinematic channels are essential for evaluating scattering amplitudes and corresponding cross-sections within the Standard Model formalism (Author, inspired by ref. \cite{peskin2018introduction}).}
\end{figure}
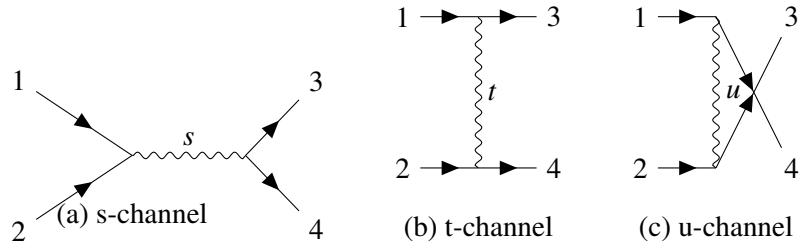

\noindent These topologies are illustrated in the diagrams in Figure 7.

\section{Fermi's Golden Rule}
One of the most important formalisms for describing interactions in particle physics is Fermi's Golden Rule, which provides a method to calculate the reaction rate, defined as the probability of a transition occurring per unit of time.

The calculation begins by determining the transition amplitude for a given process. This is derived from the matrix element of the interaction Hamiltonian, $\mathcal{H}_{\text{int}}$, between the initial and final quantum states \cite{FermiGoldenRule}:

\begin{equation}
    \mathcal{M}_{fi} = \langle\psi_f|\mathcal{H}_{\text{int}}|\psi_i\rangle = \mathcal{M}\sqrt{2E_a2E_b...2E_n}.
    \label{eq:MatrixElement}
\end{equation}

\noindent This matrix element can be calculated in practice using Feynman diagrams and their corresponding rules. According to Fermi's Golden Rule, expressed here in natural units ($\hbar=1$), the transition rate is proportional to the square of this matrix element \cite{crossection}:

\begin{equation}
    \Gamma = 2\pi |\mathcal{M}_{fi}|^2 \rho(E_f),
    \label{eq:FermiGoldenRule}
\end{equation}

\noindent here, $\rho(E_f)$ is the density of final states. To relate the transition rate in Eq. (\ref{eq:FermiGoldenRule}) to the experimentally measurable cross-section, the specific process of interest must first be defined. For the purposes of this work, the cross-section for a generic 2-to-2 scattering process ($A+B \to C+D$) will be derived, as it forms the basis for subsequent sections. The first step is to define the density of states as the number of available final quantum states ($dN_f$) per unit of energy \cite{FermiGoldenRule}:

\begin{equation}
    \rho(E_f) = \frac{dN_f}{dE_f}.
    \label{eq:stateDensity}
\end{equation}

\noindent Writing Eq.~(\ref{eq:FermiGoldenRule}) in its differential form and incorporating the definition from Eq.~(\ref{eq:stateDensity}) leads to the following expression for the differential transition rate, where energy conservation is enforced by a Dirac delta function:

\begin{equation}
    d\Gamma = 2\pi |\mathcal{M}_{fi}|^2 \delta(E_i - E_f) \, dN_f.
\end{equation}

\noindent In this expression, the total final energy is $E_f=E_C+E_D$, the total initial energy is $E_i=E_A+E_B$, and the differential number of final states, $dN_f$, which also enforces momentum conservation, is given by \cite{FermiGoldenRule}:

\begin{equation}
    dN_f = (2\pi)^3 \prod_{j=1}^{N_f}\frac{d^3\boldsymbol{p}_j}{(2\pi)^3}\delta^3\left(\boldsymbol{p}_{\text{i}}-\sum_{j=1}^{N_f}\boldsymbol{p}_j\right).
    \label{eq:numberofstates}
\end{equation}

\noindent By substituting the expression for the final state phase space, the total transition rate $\Gamma$ can be written as an integral over the momentum of the final state particles.

\begin{equation}
    \Gamma = (2\pi)^4 \int |\mathcal{M}_{fi}|^2 \delta(E_i-E_f) \delta^3(\boldsymbol{p}_i-\boldsymbol{p}_f) \frac{d^3\boldsymbol{p}_C}{(2\pi)^3} \frac{d^3\boldsymbol{p}_D}{(2\pi)^3}.
\end{equation}

\noindent Using Eq. (\ref{eq:MatrixElement}) we can write the rate expression in the form of the Lorentz-invariant amplitude $\mathcal{M}$:
\begin{equation}
        \Gamma = \frac{(2\pi)^{-2}}{4E_AE_B} \int |\mathcal{M}|^2 \delta(E_A + E_B - E_C - E_D) \delta^3(\boldsymbol{p}_A + \boldsymbol{p}_B - \boldsymbol{p}_C - \boldsymbol{p}_D) \frac{d^3\boldsymbol{p}_C}{2E_C} \frac{d^3\boldsymbol{p}_D}{2E_D}.
        \label{eq:RateLIPS}
\end{equation}

\noindent The most convenient frame choice is the center-of-mass (CM) frame where $\boldsymbol{p}_A=-\boldsymbol{p}_B=\boldsymbol{p}_i$, $\boldsymbol{p}_C=-\boldsymbol{p}_D=\boldsymbol{p}_f$ and for the energy $\sqrt{s}=(E_A+E_B)$. So the transition rate takes the form:
\begin{equation}
        \Gamma = \frac{(2\pi)^{-2}}{4\boldsymbol{p}_i \sqrt{s}} \int |\mathcal{M}|^2 \delta(\sqrt{s}- E_C - E_D) \delta^3(\boldsymbol{p}_C + \boldsymbol{p}_D) \frac{d^3\boldsymbol{p}_C}{2E_C} \frac{d^3\boldsymbol{p}_D}{2E_D}.
        \label{eq:FinalRate}
\end{equation}

According to Eq. (\ref{eq:Rate}) we relate the rate and the cross-section:
\begin{equation}
    rate=\phi_a N_b \sigma=\phi_a n_b V\sigma=(v_a + v_b)n_an_bV\sigma,
\end{equation}

\noindent where $n_a$ and $n_b$ are densities. Normalizing the wavefunctions so that each particle occupies a single volume ($V$): $n_a=n_b=1/V$, we get the following relation:
\begin{equation}
    \sigma=\frac{\Gamma}{(v_a + v_b)}.
    \label{RateCrossRelation}
\end{equation}

\noindent Using Eq. (\ref{eq:FinalRate}) into Eq. (\ref{RateCrossRelation}), we get:
\begin{equation}
    \sigma=\frac{(2\pi)^{-2}}{4\boldsymbol{p}_i \sqrt{s}} \int |\mathcal{M}|^2 \delta(\sqrt{s}- E_C - E_D) \delta^3(\boldsymbol{p}_C + \boldsymbol{p}_D) \frac{d^3\boldsymbol{p}_C}{2E_C} \frac{d^3\boldsymbol{p}_D}{2E_D}.
\end{equation}

\noindent After some manipulations, we get the final expression for the differential cross-section of the 2-body scattering:
\begin{equation}
    \left(\frac{d\sigma}{d\Omega}\right)_{\text{CM}} = \frac{1}{64\pi^2 s} \frac{|\boldsymbol{p}_f|}{|\boldsymbol{p}_i|} |\mathcal{M}|^2,
    \label{eq:theo_cross_section}
\end{equation}

\noindent where $s$ is the Mandelstam variable representing the square of the CM energy, and $|\boldsymbol{p}_i|$ and $|\boldsymbol{p}_f|$ are the magnitudes of the initial and final state momenta, respectively. By integrating this expression over all solid angles and other variables, the total cross-section $\sigma$ can be obtained.

\newpage
\chapter[DUNE]{Deep Underground Neutrino Experiment}
The Deep Underground Neutrino Experiment (DUNE) is a next-generation international project designed to measure neutrino oscillation parameters with unprecedented precision. This research is developed as part of the Brazilian contribution to the collaboration (DUNE-BR), focusing specifically on the phenomenology of neutrino-nucleus interactions relevant to Liquid Argon Time Projection Chambers. Given that DUNE will operate with a wide-band beam spanning the transition region from hundreds of MeV to several GeV, the experimental sensitivity relies on the correct modeling of various interaction mechanisms. Therefore, it is essential to detail the specific interaction channels and topologies that characterize the neutrino scattering physics at DUNE energies.

\section{The DUNE Experimental Facilities}
\begin{figure}
    \centering
    \includegraphics[width=0.75\linewidth]{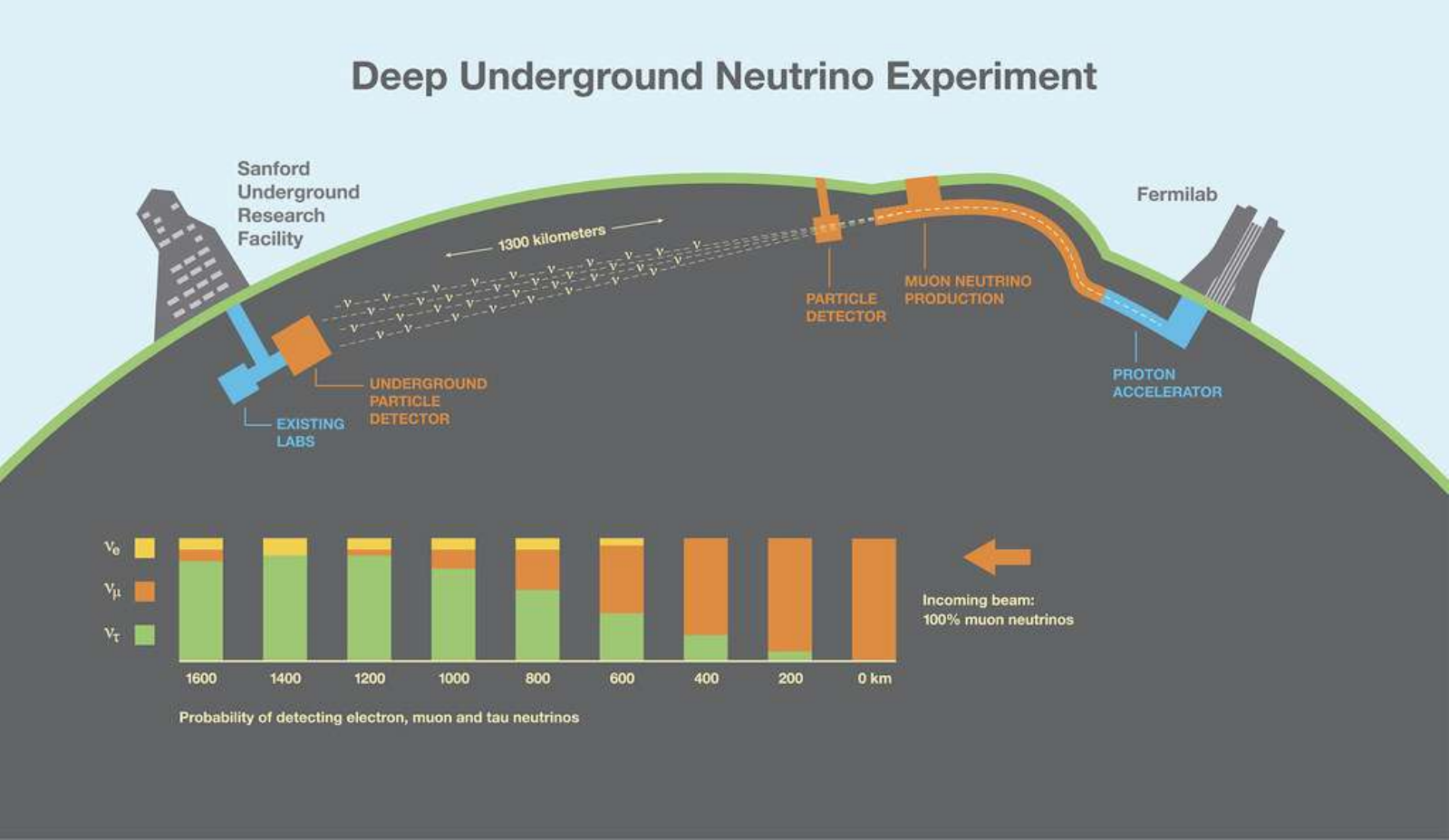}
    \caption{Schematic representation of the DUNE project. On the right, the Fermilab complex (Illinois) is shown, which houses the proton accelerator and generates the initial beam, composed primarily of muon neutrinos ($\nu_\mu$). The beam travels along a 1300 km baseline through the Earth's crust to reach the SURF in South Dakota, where the far underground detector is located. The lower section of the diagram illustrates the neutrino oscillation phenomenon, detailing the detection probabilities and flavor transitions ($\nu_e, \nu_\mu, \nu_\tau$) as a function of the distance traveled from the source~\cite{fnal_dune_layout}.}
    \label{fig:dune_map}
\end{figure}

The experimental setup utilizes the Long-Baseline Neutrino Facility (LBNF) to generate a high-intensity wide-band neutrino beam at Fermi National Accelerator Laboratory (Fermilab) in Illinois, directed towards the Sanford Underground Research Facility (SURF) in South Dakota as illustrated in Figure \ref{fig:dune_map}. The interaction physics and oscillation parameters are probed by comparing data from a high-resolution Near Detector (ND) complex at the source and a massive Far Detector (FD) situated 1300 km downstream \cite{DUNE_TDR_Vol2}.

\subsection{Neutrino Beamline Production (LBNF)}
The neutrino beam is generated using the accelerator complex at Fermilab. Protons are accelerated to energies between 60 and 120 GeV and impinged upon a graphite target. The collision produces a cascade of secondary hadrons, primarily pions ($\pi^\pm$) and kaons ($K^\pm$) \cite{LBNF_CDR_Vol3}. These charged mesons are focused by a system of three magnetic horns. By selecting the polarity of the current in the horns, the experiment can focus positive mesons (which decay into neutrinos, $\nu_\mu$) or negative mesons (which decay into antineutrinos, $\bar{\nu}_\mu$), allowing DUNE to operate in both neutrino and antineutrino modes. The focused mesons travel through a decay pipe, producing the neutrino beam \cite{LBNF_CDR_Vol3}. The information about the beamline production are listed in Table \ref{tab:LBNF_info}.

\begin{table}
    \centering
    \caption{Reference operating parameters for the primary proton beam of the LBNF. The table presents the expected beam power and the annual integrated flux, quantified as Protons on Target (POT), for three distinct proton momentum configurations: 60, 80, and 120 GeV \cite{LBNF_CDR_Vol3}.}
    \label{tab:LBNF_info}
    \begin{tabular}{|| c | c | c ||} 
         \hline
         Proton Momentum (GeV) & Expected Beam Power (MW) & Expected POT/year \\ [0.5ex] 
         \hline\hline
         120 & 1.2 & $1.2 \times10^{21}$ \\
         \hline
         80 & 1.07 & $1.47 \times10^{21}$ \\
         \hline
         60 & 1.03 & $1.89 \times10^{21}$ \\
         \hline
    \end{tabular}
\end{table}

\begin{figure}
    \centering
    \includegraphics[width=1\textwidth]{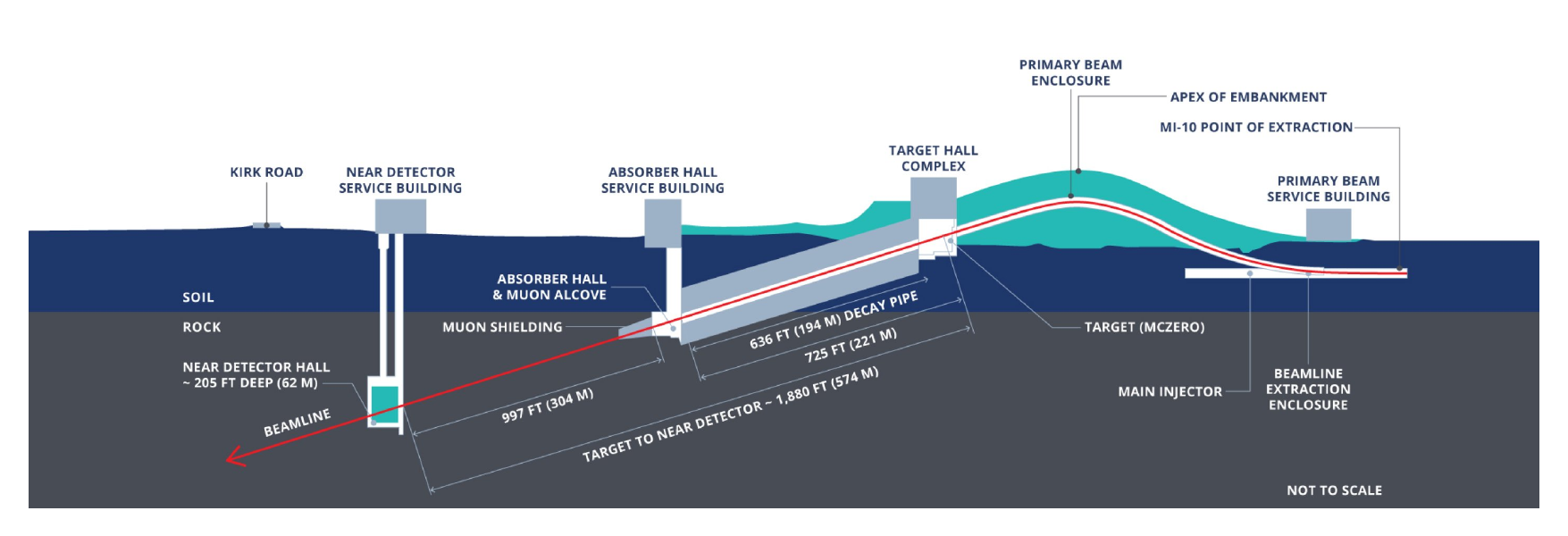}
    \caption{Schematic view of the LBNF neutrino beamline, showing the sequence of neutrino production stages, beginning with the primary proton beam transport to the Target Hall. Following the target interaction, secondary mesons travel through a 194-meter (636 ft) helium-filled Decay Pipe, where they decay into neutrinos. The remaining hadrons are stopped in the Absorber Hall, while the neutrino beam continues through the rock to the Near Detector Hall, located approximately 574 meters (1880 ft) downstream from the target at a depth of 62 meters \cite{fnal_dune_layout}.}
    \label{fig:lbnf_beamline}
\end{figure}

\subsection{The Near Detector Complex}
\begin{figure}
    \centering
    \includegraphics[width=0.8\textwidth]{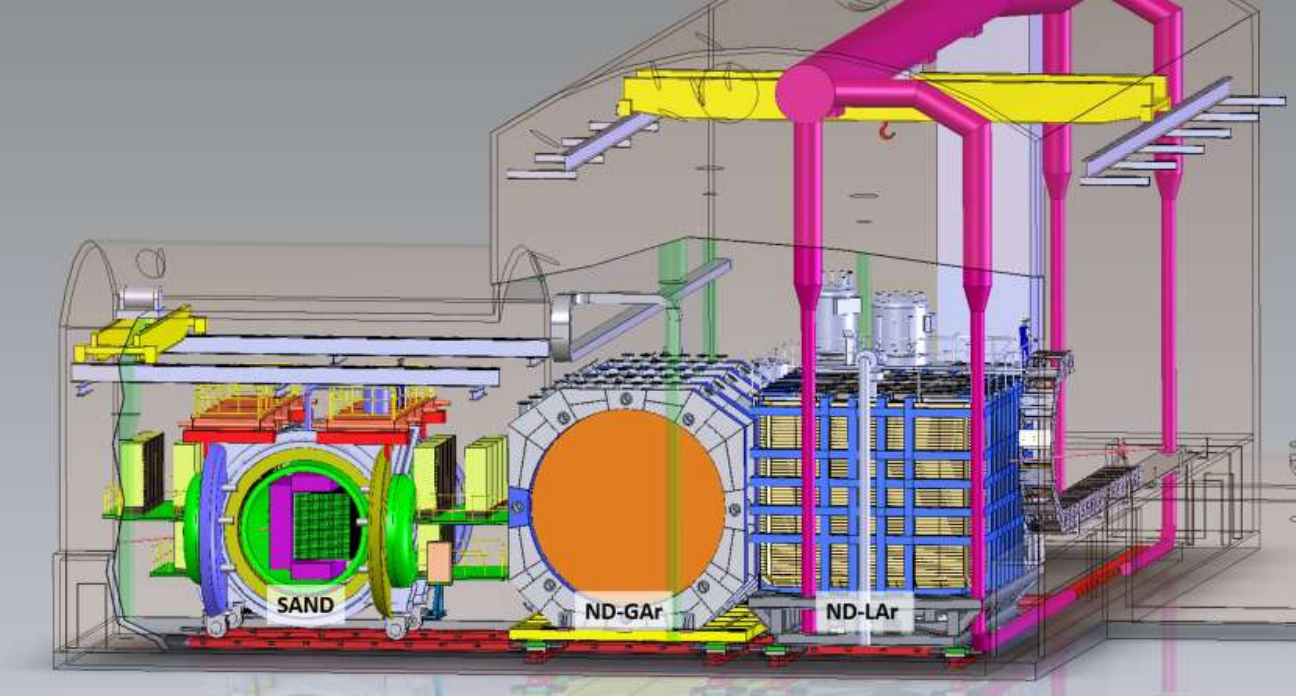}
    \caption{Three-dimensional rendering of the DUNE Near Detector complex situated in the underground hall. The system consists of three primary components arranged along the beamline (from right to left): the modular Liquid Argon TPC (ND-LAr), the Gas Argon TPC (ND-GAr), and the System for on-Axis Neutrino Detection (SAND). The rails visible beneath the ND-LAr and ND-GAr modules allow these detectors to move off-axis to sample different neutrino energy spectra, while SAND remains fixed on-axis \cite{DUNE_ND_CDR}.}
    \label{fig:near_detector}
\end{figure}

Located approximately 574 meters downstream from the target, the Near Detector (ND) complex (Figure \ref{fig:near_detector}) serves a dual purpose: to constrain the systematic uncertainties associated with the neutrino flux and to measure neutrino-nucleus interaction cross-sections with high precision \cite{DUNE_ND_CDR}. The ND is critical for characterizing the beam spectrum.

\subsection{The Far Detector at SURF}
\begin{figure}
    \centering
    \includegraphics[width=0.6\textwidth]{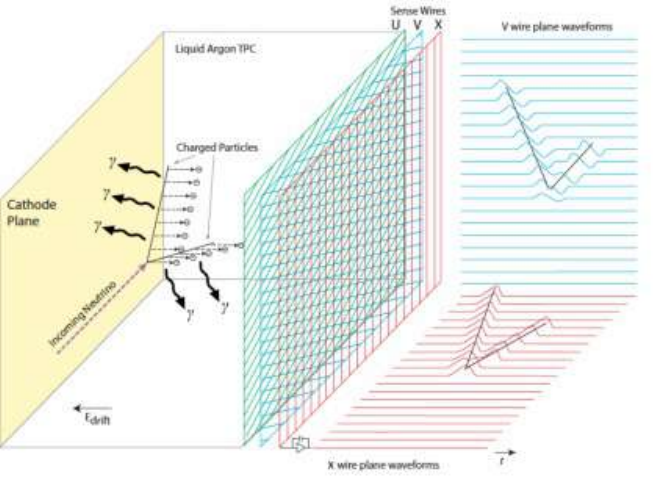}
    \caption{Illustration of the Liquid Argon Time Projection Chamber (LArTPC) working principle. A neutrino interaction vertex produces charged particles that traverse the liquid argon volume, leaving behind tracks of ionization electrons and emitting scintillation light ($\gamma$). Under the influence of a uniform electric field, the ionization electrons drift from the cathode towards the anode wire planes. The readout system comprises three wire planes with different angular orientations (U, V, and X), allowing for 3D spatial reconstruction. As electrons pass through the induction planes, characteristic waveforms are digitized to visualize the event topology \cite{sacerdoti2021vdtpc}.}
    \label{fig:LArTPC}
\end{figure}

The Far Detector (FD) is located at the Sanford Underground Research Facility (SURF) in Lead, South Dakota, at a depth of 1480 meters (4850 feet) underground. This overburden provides shielding equivalent to 4300 meters of water, drastically reducing the cosmic ray background \cite{DUNE_TDR_Vol2}. It consists of four separate cryostat modules, each containing 17 kilotons of liquid argon \cite{DUNE_TDR_Vol2}. The modules utilize Liquid Argon Time Projection Chamber (LArTPC) technology, which offers exceptional calorimetric capabilities and 3D spatial reconstruction of particle tracks. The interaction of neutrinos with argon atoms produces ionization electrons and scintillation light. The electrons drift under a strong electric field towards anode planes, allowing for the precise imaging of the event topology \cite{sacerdoti2021vdtpc}, as represented in Figure \ref{fig:LArTPC}.

\section{Neutrino Interaction Channels at DUNE Energies}
\begin{figure}
    \centering
    \includegraphics[width=0.75\linewidth]{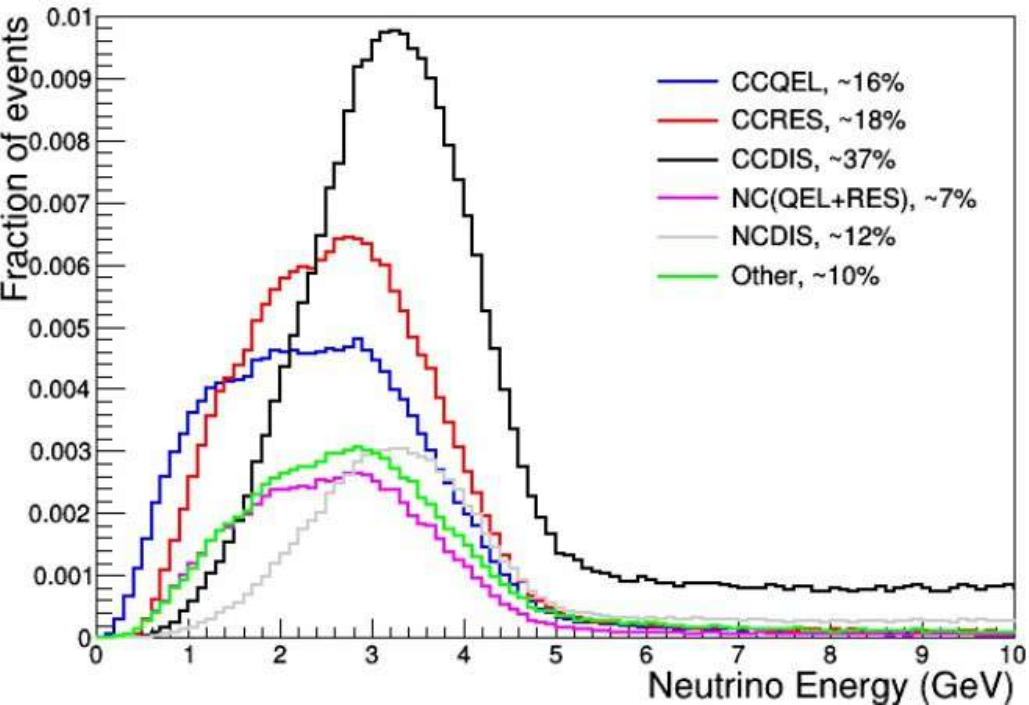}
    \caption{Fraction of neutrino events classified by interaction mechanism versus neutrino energy. The plot illustrates the relative contributions of Charged Current (CC) and Neutral Current (NC) channels within the DUNE flux energy range. The legend indicates the integrated percentage of each channel in the total sample, highlighting the statistical prevalence of Deep Inelastic Scattering (CCDIS) over Resonant (CCRES) and Quasi-Elastic (CCQE) modes in the peak region \cite{DUNE_TDR_Vol2}.}
    \label{fig:InteractionChannels}
\end{figure}

Neutrino-nucleus interactions are broadly categorized based on the energy transfer to the hadronic system and the resulting final state topology. DUNE will utilize the Long-Baseline Neutrino Facility (LBNF) wide-band beam, which peaks at approximately $2.5$ GeV \cite{LBNF_CDR_Vol3}. This energy range places DUNE in a kinematic transition region where no single interaction mechanism dominates. Instead, the cross-section is a complex convolution of Quasi-Elastic (QEL) scattering, Resonant (RES) pion production, and Deep Inelastic Scattering (DIS) as shown in Figure \ref{fig:InteractionChannels}. Understanding these channels and their distinct topologies in a Liquid Argon Time Projection Chamber (LArTPC) is critical for accurate energy reconstruction and oscillation \cite{DUNE_TDR_Vol2}. 

\subsection{Charged Current Quasi-Elastic (CCQEL)}
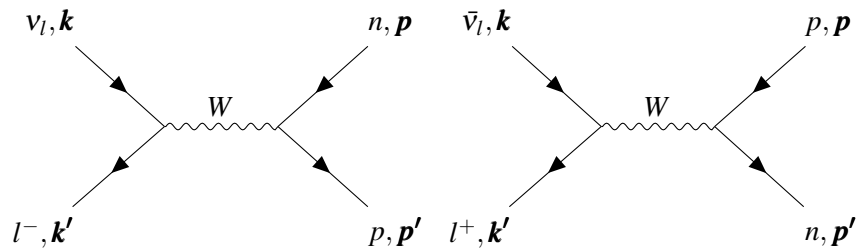
\begin{figure}[!ht]
\begin{center}
    \begin{tikzpicture}
    \begin{feynman}
        \vertex (a);
        \vertex [above left=of a] (b) {$\nu_l,\boldsymbol{k}$};
        \vertex [below left=of a] (c) {$l^-, \boldsymbol{k'}$};
        \vertex [right=of a] (d);
        \vertex [above right=of d] (e) {$n, \boldsymbol{p}$};
        \vertex [below right=of d] (f) {$p, \boldsymbol{p'}$};

        \diagram* {
        (b) -- [fermion] (a) -- [fermion] (c);
        (a) -- [boson, edge label=$W$] (d);
        (e) -- [fermion] (d) -- [fermion] (f);
        };
    \end{feynman}
\end{tikzpicture}
\begin{tikzpicture}
    \begin{feynman}
        \vertex (a);
        \vertex [above left=of a] (b) {$\bar{\nu_l},\boldsymbol{k}$};
        \vertex [below left=of a] (c) {$l^+,\boldsymbol{k'}$};
        \vertex [right=of a] (d);
        \vertex [above right=of d] (e) {$p, \boldsymbol{p}$};
        \vertex [below right=of d] (f) {$n, \boldsymbol{p'}$};

        \diagram* {
        (b) -- [fermion] (a) -- [fermion] (c);
        (a) -- [boson, edge label=$W$] (d);
        (e) -- [fermion] (d) -- [fermion] (f);
        };
    \end{feynman}
\end{tikzpicture}
\end{center}

\caption{Feynman diagrams representing Charged Current Quasi-Elastic (CCQE) interactions. The left diagram illustrates a neutrino ($\nu_l$) scattering off a neutron ($n$) via the exchange of a virtual $W$ boson, producing a negatively charged lepton ($l^-$) and a proton ($p$). The right diagram describes the antineutrino ($\bar{\nu}_l$) interaction with a proton ($p$), yielding a positively charged antilepton ($l^+$) and a neutron ($n$). These fundamental processes are the dominant interaction channels in the lower-energy regime of the DUNE neutrino flux (Author).}
\label{fig:CCQEdiagram}
\end{figure}

The Charged Current Quasi-Elastic (CCQEL) interaction is one of the dominant process at lower energies ($E_\nu < 1$ GeV) and remains a significant component of the signal at DUNE energies. There are two processes for CCQEL interactions, where the neutrino converts a neutron into a proton, or vice-versa for antineutrinos:

\begin{equation}
    \nu_l + n \to l^- + p,
\end{equation}
\begin{equation}
     \bar{\nu_l} + p \to l^+ + n.
\end{equation}

\noindent which can also be represented as the diagrams of the interactions in Figure \ref{fig:CCQEdiagram}.

The $\boldsymbol{k}$ and $\boldsymbol{k'}$ are respectively the incoming four-momenta of the incoming neutrino (anti-neutrino) and the four-momenta of the outgoing lepton. Also $\boldsymbol{p}$ and $\boldsymbol{p'}$ are respectively the four-momenta of the incoming nucleon and the outgoing nucleon. 

In the specific case of Charged Current interactions, the process is termed ``quasi"-elastic rather than purely elastic for two primary reasons: first, the weak interaction induces a flavor change in the nucleon, converting a neutron into a proton. Since the neutron mass exceeds the proton mass ($M_n > M_p$), the kinematics are slightly modified compared to a scattering where the target mass is invariant. Second, for heavy nuclear targets such as Argon ($^{40}\text{Ar}$), nuclear effects such as Fermi motion occur, consequently the interaction is not perfectly elastic in the frame of the nucleus, as energy is dissipated to liberate the nucleon and account for the mass difference.

The theoretical description of the differential cross-section for this process on free nucleons relies on the formalism developed by Llewellyn Smith~\cite{LlewellynSmith:1972zz}. The calculation starts by defining the squared matrix element of the interaction, which is derived from the quantum field theory \cite{Athar:2020kqz}.

\begin{equation}
    |\mathcal{M}|^2 = \frac{G^2_F}{2} \cos^2{\theta_C} L^{\mu \nu} H_{\mu \nu},
    \label{eq:llewellyn_Matrix_El}
\end{equation}

\noindent where $G^2_F$ is the Fermi constant, $\theta_C$ is the Cabibbo angle, $L^{\mu \nu}$ is the leptonic tensor and $H_{\mu \nu}$ is the hadronic tensor. The tensor terms are obtained by QFT, and represent respectively the electroweak coupling, taking the neutrino and the charged lepton, which is fully calculable within the Standard Model perturbative theory, and the the hadronic tensor encapsulates the complex internal structure and dynamics of the target nucleon, containing the information regarding the distribution of quarks and gluons, as well as non-perturbative QCD effects such as nuclear form factors and binding energy corrections \cite{Halzen:1984mc}. The contraction $L^{\mu\nu}H_{\mu\nu}$ quantifies the probability amplitude for the lepton to transfer energy and momentum to the hadronic system via the exchange of a vector boson~\cite{Halzen:1984mc}.

So the leptonic tensor can be expressed as \cite{Athar:2020kqz}:

\begin{equation}
    L^{\mu \nu} = 8[k^{\mu}k'^{\nu} + k'^{\mu}k^{\nu} - g^{\mu \nu} k \cdot k' \pm i\epsilon^{\mu \nu \alpha \beta} k'^{\alpha} k^{\beta} ],
    \label{eq:leptonic_tensor}
\end{equation}

\noindent where $\pm$ indicate a neutrino (+) or a anti-neutrino (-) process, and the $\epsilon^{\mu \nu \alpha \beta}$ is the Levi-Civita pseudo-tensor in four dimensions, which represents the parity violation.

The hadronic tensor is a construction of various terms \cite{Athar:2020kqz}.

The contraction of the leptonic tensor $L^{\mu \nu}$ with the hadronic tensor $H_{\mu \nu}$ yields:

\begin{equation}
    |\mathcal{M}|^2 = \frac{G^2_F}{2} \cos^2{\theta_C}N(q^2),
    \label{eq:llewellyn_Matrix_El_contraction}
\end{equation}

\noindent where $N(q^2) = L^{\mu \nu}H_{\mu \nu}$. The diferencial cross-section is derived from equation (\ref{eq:theo_cross_section}) considering $|\boldsymbol{p_f}| \approx |\boldsymbol{p_i}|$ since it is a QEL process, and transforming the expression to the lab frame \cite{Athar:2020kqz}:

\begin{equation}
    \frac{d \sigma}{d q^2}=\frac{|\mathcal{M}|^2}{8 \pi M^2 E^2_{\nu (\bar{\nu})}},
    \label{eq:intermediate_cross-section}
\end{equation}

\noindent where $E^2_{\nu (\bar{\nu})}$ is the neutrino (anti-neutrino) energy. Calculating the contraction of the tensors, the expression of the differential cross-section for CCQEL process in terms of the mandelstam variables give us the Llewellyn Smith model \cite{LlewellynSmith:1972zz}:

\begin{equation}
    \frac{d \sigma}{d q^2}=\frac{G^2_F M^2 \cos^2(\theta_C)}{8 \pi E^2_{\nu}}\left[A(q^2) \mp B(q^2)\frac{(s-u)}{M^2} + C(q^2)\frac{(s-u)^2}{M^4}\right],
    \label{eq:final_cross-section}
\end{equation}

\noindent where the negative (positive) sign before $B(q^2)$ refers to neutrino (anti-neutrino) scattering. The terms $A(q^2)$, $B(q^2)$ and $C(q^2)$ contain the form factors and are given as \cite{Athar:2020kqz}:

\begin{align}
    A(q^2) &= \frac{m^2 - q^2}{4M^2} \Bigg[ \left( 4 - \frac{q^2}{M^2} \right) g_1^2(q^2) - \left( 4 + \frac{q^2}{M^2} \right) f_1^2(q^2) - \frac{q^2}{M^2} \left( 1 + \frac{q^2}{4M^2} \right) f_2^2(q^2) \nonumber \\
    &\quad - \frac{4q^2}{M^2} f_1(q^2) f_2(q^2) - \frac{m_l^2}{M^2} \left( \left( f_1(q^2) + f_2(q^2) \right)^2 + \left( g_1(q^2) + 2g_3(q^2) \right)^2 \right) \nonumber \\
    &\quad + \left( \frac{q^2}{M^2} - 4 \right) g_3^2(q^2) \Bigg], \label{eq:10.80} \\[10pt]
    B(q^2) &= - \frac{q^2}{M^2} g_1(q^2) \left[ f_1(q^2) + f_2(q^2) \right], \label{eq:10.81} \\[10pt]
    C(q^2) &= \frac{1}{4} \left[ g_1^2(q^2) + (f_1(q^2))^2 - \frac{q^2}{4M^2} (f_2(q^2))^2 \right]. 
    \label{eq:form_factors_CCQEL}
\end{align}

\noindent The form factors $f_i(q^2)$ and $g_i(q^2)$ parametrize the internal structure of the nucleon, reflecting its composite nature bound by strong interactions.

\subsection{Charged Current Resonant Pion Production (CCRES)}
As the neutrino energy increases, the interaction can excite the target nucleon into a baryonic resonance. Usually a resonant scattering is represented as:

\begin{equation}
    \nu_l + N \to l^- + R,
\end{equation}

\noindent where $N$ is the initial state nucleon, $R$ is the resonance, which have many types of decay, producing different final state particles. The most important for DUNE energies is the $\Delta(1232)$ resonance, which subsequently decays by emitting a pion:

\begin{equation}
    \nu_l + N \to l^- + \Delta \to l^- + N' + \pi.
\end{equation}

\noindent The study of the $\Delta(1232)$ resonance is important to DUNE because as the experiment's neutrino beam flux peaks in the $2.5$ GeV, resonant production becomes one of the dominant interaction mechanism, demonstrated in the Figure (\ref{fig:InteractionChannels}). A precise understanding of this channel is needed to understand the background estimation of the signal. The decay of the neutral resonance ($\Delta \to N\pi^0$) produces neutral pions that subsequently decay into photons; these photons generate electromagnetic showers in the detector that can be indistinguishable from the electron signals characteristic of $\nu_e$ appearance, thereby constituting a major background for oscillation analyses.

The Rein-Sehgal model \cite{rein1981neutrino} provides the framework for calculating the CCRES scattering diferentinal cross-section. Anchored in the relativistic quark model proposed by Feynman, Kislinger, and Ravndal (FKR), the formalism treats the interaction as a two-step process: the excitation of a target nucleon into a heavier baryon resonance via the absorption of a virtual intermediate vector boson ($W^\pm$), followed by the subsequent decay of this resonance into a nucleon and a meson. In this approach, the differential cross-section with respect to the momentum transfer squared $Q^2$ is derived by calculating the hadronic current matrix elements between the initial nucleon state and the final resonant state. The resulting cross-section is expressed as a linear combination of partial helicity cross-sections, denoted as $\sigma_L$, $\sigma_R$, and $\sigma_S$, which correspond to the absorption of left-handed, right-handed, and scalar $W$-bosons, respectively. The kinematic dependence of the cross-section arises from the coefficients of these helicity terms, which differ for neutrino (Equation (\ref{eq:RS_CrossSection})) and antineutrino (Equation (\ref{eq:RS_CrossSection_Antineutrino})) scattering due to the chiral nature of the weak interaction. 

\begin{equation}
    \frac{d^2\sigma}{dQ^2 dW} = \frac{G_F^2 \cos^2\theta_c}{4\pi^2} \frac{Q^2}{E_{\nu}^2} \frac{W}{M} \kappa(Q^2, W) \left[ u^2 \sigma_L + v^2 \sigma_R + 2uv \sigma_S \right] D(W),
    \label{eq:RS_CrossSection}
\end{equation}

\begin{equation}
    \frac{d^2\sigma^{\bar{\nu}}}{dQ^2 dW} = \frac{G_F^2 \cos^2\theta_c}{4\pi^2} \frac{Q^2}{E_{\bar{\nu}}^2} \frac{W}{M} \kappa(Q^2, W) \left[ u^2 \sigma_R + v^2 \sigma_L + 2uv \sigma_S \right] D(W).
    \label{eq:RS_CrossSection_Antineutrino}
\end{equation}

\noindent Where $W$ is the invariant mass of the hadronic system, $M$ is the mass of the target nucleon, $\kappa(Q^2, W)$ is the kinematic flux factor represented by Equation (\ref{eq:kinematic_flux}), $u$ and $v$ are the kinematic coefficients defined in Eq. \ref{eq:RS_uv_coefficients}. Also the partial helicity cross-sections $\sigma_L, \sigma_R, \sigma_S$ are represented respectively by Equations (\ref{eq:sigma_L}), (\ref{eq:sigma_R}) and (\ref{eq:sigma_S}).

\begin{equation}
    \kappa(Q^2, W) = \frac{W^2-M^2}{2M},
    \label{eq:kinematic_flux}
\end{equation}

\begin{equation}
    u = \frac{1}{2} \left( 1 + \frac{E_\nu - E_l}{|\mathbf{q}|} \right), \quad v = \frac{1}{2} \left( 1 - \frac{E_\nu - E_l}{|\mathbf{q}|} \right),
    \label{eq:RS_uv_coefficients}
\end{equation}

\begin{align}
    \sigma_L(W, Q^2) &= \frac{\pi}{\kappa} \frac{W}{M} \left( |f_{-1}|^2 + |f_{-3}|^2 \right) \label{eq:sigma_L}, \\[10pt]
    \sigma_R(W, Q^2) &= \frac{\pi}{\kappa} \frac{W}{M} \left( |f_{+1}|^2 + |f_{+3}|^2 \right) \label{eq:sigma_R} ,\\[10pt]
    \sigma_S(W, Q^2) &= \frac{\pi}{\kappa} \frac{W}{M} \frac{Q^2}{|\mathbf{q}|^2} \left( |f_{0+}|^2 + |f_{0-}|^2 \right) \label{eq:sigma_S}.
\end{align}

\noindent Finally, $D(W)$ is the Breit-Wigner distribution function, shown in Fig. \ref{eq:RS_BreitWigner}, it is added to Equations (\ref{eq:RS_CrossSection}) and (\ref{eq:RS_CrossSection}) to account for the finite lifetime of the excited states.

\begin{equation}
    D(W) = \frac{1}{\pi} \frac{\Gamma / 2}{(W - M_R)^2 + (\Gamma / 2)^2}.
    \label{eq:RS_BreitWigner}
\end{equation}

\subsection{Charged Current Deep Inelastic Scattering (CCDIS)}

At energies above $2-3$ GeV, the neutrino has sufficient momentum transfer $Q^2$ to resolve the individual quark constituents of the nucleon. This leads to the fragmentation of the nucleon into a hadronic shower containing multiple mesons (pions, kaons) and baryons:

\begin{equation}
    \nu_l + N \to l^- + X,
\end{equation}
\noindent where $X$ represents the system of hadrons. According to Figure \ref{fig:InteractionChannels} CCDIS interactions are the most significant portion of the DUNE flux. In this regime, the cross-section is treated using the factorization theorem, which separates the interaction from the internal structure of the hadron. This approach relies on the Quark-Parton Model (QPM), where the nucleon is viewed not as a static entity, but as a collection of quasi-free, point-like constituents called partons (quarks and gluons). At high momentum transfer ($Q^2$), the neutrino scatters elastically off an individual parton carrying a fraction of the nucleon's total momentum \cite{Halzen:1984mc}.

One of the most important frameworks to calculate CCDIS cross-sections are the Parton Distribution Functions (PDFs), denoted as $f_i(x, Q^2)$. These functions represent the probability density of finding a parton of flavor $i$ with momentum fraction $x$ at a given energy scale $Q^2$. Since the binding dynamics of partons are governed by non-perturbative QCD and cannot be calculated from first principles, PDFs must be determined experimentally through global fits to scattering data \cite{Halzen:1984mc}. They are fundamental to the calculations of DIS, acting as the bridge between the elementary lepton-quark interaction and the physical nucleon target \cite{CTEQ2002}. However, the DUNE neutrino flux peaks in the few-GeV energy range, placing a significant portion of the interactions in a transition region between resonant scattering and pure DIS, often referred to as shallow inelastic scattering (SIS). In this domain, standard perturbative QCD (pQCD) calculations become unreliable due to significant non-perturbative effects. To correct this, the GENIE neutrino event generator \cite{Andreopoulos2010} employs the Bodek-Yang effective leading-order model \cite{bodek2002modeling}. This phenomenological approach extends the applicability of the parton model to lower $Q^2$ values. 

The inclusive charged current differential cross-section for neutrino-nucleon scattering is expressed in terms of the nucleon structure functions $F_1$, $F_2$, and $F_3$ as:

\begin{equation}
    \frac{d^2\sigma^{\nu(\bar{\nu})}}{dx dy} = \frac{G_F^2 M E_\nu}{\pi} \left[ xy^2 F_1(x,Q^2) + \left(1 - y - \frac{Mxy}{2E_\nu}\right) F_2(x,Q^2) \pm xy \left(1 - \frac{y}{2}\right) F_3(x,Q^2) \right],
    \label{eq:CCDIS_xsec}
\end{equation}

\noindent where the positive (negative) sign in the $F_3$ term applies to neutrino (antineutrino) scattering. $x$ and $y$ are the dimensionless kinematic variables governing the scattering process. The Bjorken scaling variable, defined as $x = Q^2 / (2M\nu)$, represents the fraction of the nucleon's longitudinal momentum carried by the struck quark in the infinite momentum frame. The variable $y = \nu / E_\nu$ is the inelasticity, which corresponds to the fraction of the incident neutrino energy transferred to the hadronic system in the laboratory rest frame. The expressions for $F_1(x, Q^2)$, $F_2(x, Q^2)$ and $F_3(x, Q^2)$ are respectively expressed in Equations (\ref{eq:F2_BY}), (\ref{eq:xF3_BY}) and (\ref{eq:F1_BY}).

\begin{equation}
    F_2(x, Q^2) = \sum_{q} \left[ Q_q(\xi_w, Q^2) + \bar{Q}_q(\xi_w, Q^2) \right],
    \label{eq:F2_BY}
\end{equation}

\begin{equation}
    x F_3(x, Q^2) = \sum_{q} \left[ Q_q(\xi_w, Q^2) - \bar{Q}_q(\xi_w, Q^2) \right],
    \label{eq:xF3_BY}
\end{equation}

\begin{equation}
    2x F_1(x, Q^2) = F_2(x, Q^2) \frac{1 + \frac{4M^2 x^2}{Q^2}}{1 + R(x, Q^2)},
    \label{eq:F1_BY}
\end{equation}

\noindent where the modified quark distribution functions $Q_q(\xi_w, Q^2)$ are derived from the standard Leading Order PDFs $q(\xi_w, Q^2)$:

\begin{equation}
    Q_q(\xi_w, Q^2) = q(\xi_w, Q^2) \cdot K(Q^2).
\end{equation}

\noindent and the K-factors:
\begin{equation}
    K_{\text{valence}}(Q^2) = \left[ 1 - G_D^2(Q^2) \right] \left( \frac{Q^2}{Q^2 + C_v} \right), \quad K_{\text{sea}}(Q^2) = \frac{Q^2}{Q^2 + C_s},
\end{equation}

\noindent $R(x, Q^2)$ is a function described and discussed in in the model's paper \cite{bodek2002modeling}.

The most recent update to the model \cite{Bodek2021} decouples some contributions defining separate $K$-factors for the Vector and Axial components. The total structure function is now decomposed as:

\begin{equation}
    F_{2}(x, Q^2) = F_{2}^V(\xi_w, Q^2) + F_{2}^A(\xi_w, Q^2),
\end{equation}

\noindent with distinct low-$Q^2$ scaling corrections:
\begin{align}
    F_{2}^V &\propto \frac{Q^2}{Q^2 + C_V} \times K_{valence}^V(Q^2), \\
    F_{2}^A &\propto \frac{Q^2}{Q^2 + C_A} \times K_{valence}^A(Q^2).
\end{align}

\noindent This correction important for DUNE as it models better the axial current contribution.

\section{Dominance of Charged Current Interactions}

While both Charged Current (CC) and Neutral Current (NC) interactions occur in DUNE, the CC channel is of primary importance for the oscillation physics goals of DUNE. This precedence arises from two main factors: the experimental capability for flavor identification and the relative magnitude of the cross-sections.

Experimentally, the distinction between neutrino flavors is only possible via CC interactions. The flavor of the incoming neutrino is determined by identifying the corresponding charged lepton produced in the final state. In contrast, NC interactions produce an outgoing neutrino that escapes undetected. The observable signature consists solely of the hadronic recoil energy, making it impossible to distinguish the flavor of the incident neutrino.

Mathematically, the Charged Current cross-section is larger than the Neutral Current cross-section at DUNE energies. Although both interactions are described by the contraction of a leptonic tensor ($L_{\mu\nu}$) and a hadronic tensor ($H^{\mu\nu}$), the couplings of the fermions to the mediator bosons differ significantly. In the Charged Current case, the $W^\pm$ boson couples only to left-handed chiral states. For Neutral Currents, the coupling to the $Z^0$ boson is modulated by the weak mixing angle (Weinberg angle), $\theta_W$. The ratio of the total cross-sections for deep inelastic scattering (which dominates the DUNE flux) can be approximated by the Paschos-Wolfenstein relation \cite{Halzen:1984mc}:

\begin{equation}
    R_\nu = \frac{\sigma_{NC}(\nu N)}{\sigma_{CC}(\nu N)} \approx \frac{1}{2} - \sin^2\theta_W + \frac{20}{27}\sin^4\theta_W.
\end{equation}

\noindent Given the experimental value of $\sin^2\theta_W \approx 0.23$ \cite{nu-fit-data}, this ratio yields $R_\nu \approx 0.3$. This implies that the inclusive cross-section for Neutral Current interactions is roughly one-third of that for Charged Current interactions. Therefore, the statistical abundance establish Charged Current interactions as the critical signal channel for the experiment.

\section{Systematic Uncertainties in Neutrino-Nucleus Interaction}
The sensitivity of neutrino oscillation experiments, such as DUNE, is limited by the precision with neutrino-nucleus interactions can be modeled in complex targets. As detailed in the NuSTEC White Paper \cite{NuSTEC:2018}, the transition from elementary targets to heavy nuclei introduces significant systematic uncertainties, which arise from the limitations of nuclear models in describing the nuclear environment.

\subsection{Quasi-Elastic (CCQEL) Uncertainties}
For QEL processes the primary source of theoretical uncertainties lies in the Axial Mass Parameter. While measurements on deuterium give an axial mass parameter of $M_A \approx 1.0$ GeV, modern cross-section data on heavy nuclei require an inflated effective value ($M_A \approx 1.35$ GeV) to achieve agreement within the Relativistic Fermi Gas (RFG) framework \cite{NuSTEC:2018}. The NuSTEC review highlights that this discrepancy likely masks unmodeled nuclear effects, specifically Meson Exchange Currents (MEC) and multinucleon interactions.

\subsection{Resonance Production (RES) Uncertainties}
In the resonance production (RES) regime, where the cross-section is dominated by the excitation of baryon resonances, most notably the $\Delta(1232)$ \cite{NuSTEC:2018}, uncertainties come from nuclear medium effects. The Rein--Sehgal model, which is used to theoretically describe this channel, relies on axial form factors constrained by bubble chamber data from the 1970s and 1980s, introducing significant uncertainties \cite{Kitagaki1983}. Furthermore, the theoretical modeling of the resonant amplitudes and the non-resonant background remain a major source of model dependence \cite{NuSTEC:2018}.

\subsection{Deep and Shallow Inelastic Scattering (SIS/DIS)}
The Deep Inelastic Scattering (DIS) regime is well-described by the Quark-Parton Model at high momentum transfer, however the DUNE flux also populates the ``Shallow Inelastic Scattering" (SIS) transition region \cite{NuSTEC:2018}. Here, perturbative QCD calculations break down due to low $Q^2$ and non-perturbative effects \cite{NuSTEC:2018}. In this regime the phenomenological approaches, such as the Bodek-Yang model, address this transition by employing $K$-factors and scaling variables to extrapolate Parton Distribution Functions (PDFs) \cite{bodek2002modeling}, introducing systematic uncertainties into the cross-section predictions \cite{NuSTEC:2018}.

\newpage
\chapter[$\nu$-A Scattering]{Neutrino Nucleus Scattering}

To accurately describe the interactions of neutrinos within the massive detectors utilized by modern oscillation experiments, it is necessary to move beyond the mathematical models of neutrino nucleus scattering. In practice, incident neutrinos interact with nucleons that are deeply bound within complex nuclear targets. This dense nuclear environment introduces significant physical complexities: the target nucleons are subject to strong binding potentials, possess intrinsic initial momenta, and are governed by quantum mechanical exclusion principles. Consequently, to evaluate precise interaction cross-sections and predict the observable final-state kinematics, it is essential to adopt a theoretical framework capable of modeling the many-body dynamics of the target nucleus. This section explores the standard nuclear models used to mathematically approximate the nucleus, beginning with the foundational Fermi Gas model, which establishes the basic phase-space mechanics required for the subsequent relativistic formulations.

\section{Fermi Gas Model}

The Fermi Gas model provides a description of the nucleus by treating it as a gas of non-interacting fermions confined within a spherical volume. In this framework, protons and neutrons are considered as two independent gases. According to the Pauli exclusion principle, no two identical fermions can occupy the same quantum state. At absolute zero temperature ($T=0$), the nucleons fill all available momentum states from zero up to a maximum value, defined as the Fermi momentum, $k_F$. Mathematically, the probability of finding a nucleon with a specific initial momentum $\mathbf{k}$ is governed by the momentum distribution function, $n(\mathbf{k})$, which is modeled as a Heaviside step function (or theta function) \cite{Athar:2020kqz}:

\begin{equation}
    n(\mathbf{k}) = \Theta(k_F - |\mathbf{k}|) = 
    \begin{cases} 
      1, & \text{if } |\mathbf{k}| \le k_F \\
      0, & \text{if } |\mathbf{k}| > k_F 
   \end{cases}.
   \label{eq:FGprobability}
\end{equation}

\noindent This step function is the direct mathematical manifestation of the Pauli Exclusion Principle for a degenerate fermion gas. It dictates that all states below the Fermi surface are completely occupied, and all states above it are strictly empty. To derive the relationship between the Fermi momentum and the nuclear density, we evaluate the number of available states in phase space. Accounting the Pauli exclusion principle, the volume occupied by a single quantum state in phase space is given by $(2\pi)^3$. The total number of states is obtained by integrating the spherical volume element in momentum space, $4\pi p^2 dp$, from zero to $p_F$. Finally, multiplying this integral by the nuclear volume $V$ and the spin degeneracy factor $g$, which accounts for the two possible spin projections of a spin $1/2$ nucleon, yields the mathematical relation between the particle number and the Fermi momentum \cite{Athar:2020kqz}:

\begin{equation}
    N = \frac{g V}{(2\pi)^3} \int_{0}^{p_F} 4\pi p^2 dp = \frac{2 V}{(2\pi)^3} \left( \frac{4\pi p_F^3}{3} \right) = \frac{V}{3\pi^2} p_F^3.
\end{equation}

\noindent By defining the number density of the specific nucleon as $\rho_n = N/V$, the Fermi momentum $k_F$ is directly related to the density:

\begin{equation}
    k_F = (3\pi^2 \rho_n)^{1/3}.
    \label{eq:fermi_momentum}
\end{equation}

Empirical measurements from electron scattering demonstrate that the density of nuclear matter is approximately constant across medium and heavy nuclei, with $\rho \approx 0.17$ nucleons/fm$^3$ \cite{Hofstadter1956}. For a symmetric nucleus ($N \approx Z \approx A/2$), the neutron and proton densities are approximately $\rho/2$. which means the neutron density is: 

\begin{equation}
    \rho_n = \frac{\rho}{2} = 0.085 \text{ fm}^{-3}.
\end{equation}

Substituting this value into the equation \eqref{eq:fermi_momentum} in units of inverse femtometers ($\text{fm}^{-1}$):

\begin{equation}
    k_F = (3\pi^2 \times 0.085)^{1/3} \approx (2.516)^{1/3} \approx 1.36 \text{ fm}^{-1}.
\end{equation}

\noindent In high-energy and nuclear physics, it is customary to express momentum in units of $\text{MeV/c}$. To convert $k_F$ into physical momentum, we multiply by the fundamental constant $\hbar c$, which has an approximate value of $197.3 \text{ MeV}\cdot\text{fm}$ \cite{PDG2022}:

\begin{equation}
    p_F = \hbar k_F = (1.36 \text{ fm}^{-1}) \times (197.3 \text{ MeV}\cdot\text{fm}) \approx 268 \text{ MeV/c}.
\end{equation}

Because the incoming neutrino interacts with nucleons distributed throughout the entire nuclear volume, it is usually used an effective or average Fermi momentum. For intermediate-mass nuclei, averaging over the nuclear volume yields a slightly lower effective Fermi momentum of $k_F \approx 250 \text{ MeV/c}$, which is the standard value typically adopted in simulations, based on historical fits to quasielastic electron scattering data~\cite{Moniz1971}. The highest energy state occupied by a nucleon in this ground-state configuration is the Fermi energy, $E_F$. In the standard non-relativistic approximation, this is:

\begin{equation}
    E_F = \frac{k_F^2}{2m_N} \approx 33 \text{ MeV}
    \label{eq:fermiEnergy}
\end{equation}

\noindent where $m_N$ is the nucleon mass. However, in the context of high-energy neutrino scattering experiments such as DUNE, where momentum transfers routinely exceed the nucleon mass, the nucleons are ejected with relativistic velocities.

\subsection{Limitations of Fermi Gas Model}

While the non-relativistic Fermi Gas model provides an intuitive baseline for describing nuclear dynamics, it possesses severe limitations when applied to modern high-energy neutrino experiments. These problems originate both from its kinematic approximations and its oversimplified view of nuclear structure. Firstly, the non-relativistic framework fundamentally fails at the higher energy scales. The DUNE neutrino flux peaks at approximately $2.5 \text{ GeV}$, in this kinematic regime, the non-relativistic kinetic energy approximation (Equation \ref{eq:fermiEnergy}) is no longer valid \cite{SmithMoniz1972}. Secondly, the core assumption of the FG model, that nucleons behave independent and non-interacting gas of fermions,  is physically incomplete. As detailed in \cite{NuSTEC:2018}, this independent-particle picture ignores interactions between nucleons within the nuclear medium. Specifically, it fails to account for:

\begin{itemize}
    \item \textbf{Short-Range Correlations (SRC):} Strong repulsive forces at short distances cause nucleons to form temporary, highly correlated pairs. These SRC pairs possess momenta that significantly exceed the Fermi momentum, creating a high-momentum tail in the nuclear momentum distribution. The simple Fermi gas leads to an underestimation of high-$Q^2$ scattering events.
    \item \textbf{Constant Binding Energy:} The model assumes a single, average binding energy for all nucleons. In reality, nucleons reside in discrete shells with distinct separation energies. The probability of finding a nucleon with a specific momentum and a specific binding energy is more accurately described by a two-dimensional Spectral Function (SF) \cite{Benhar2005}, rather than the distribution of the FG model.
\end{itemize}

\noindent Consequently, while the Fermi Gas provides a first-order approximation for cross-sections, its inability to describe relativistic recoil and nucleon-nucleon correlations limits its predictive for precision measurements.

\section{Relativistic Fermi Gas Model}

The transition from low-energy nuclear phenomena to the high-energy regime necessitates a fundamental upgrade to the kinematic framework describing the target nucleus. The momentum transfer ($|\mathbf{q}|$) and energy transfer ($\omega$) from the incident neutrino are frequently on the order of, or significantly larger than, the nucleon rest mass. Consequently, the non-relativistic approximation for the kinetic energy of the nucleon becomes entirely inadequate. To correctly describe deep inelastic scattering, resonance production, and high-energy quasi-elastic scattering, one must employ the Relativistic Fermi Gas (RFG) model.

The RFG model retains that the nucleus is a degenerate gas of independent fermions filling all momentum states up to the Fermi momentum. However, it upgrades the interaction dynamics through the Impulse Approximation. The IA postulates that at high momentum transfers, the incident lepton scatters off a single, independent nucleon inside the nucleus \cite{Benhar2005}. The remaining $A-1$ nucleons are treated as spectators that do not participate in the primary interaction. 

In the RFG formalism, the initial target nucleon possesses a three-momentum $\mathbf{k}$ such that $|\mathbf{k}| \le k_F$. The total energy of this initial bound nucleon, $\epsilon_k$, is governed by the relativistic dispersion relation modified by an average binding energy parameter, $\epsilon_1$ \cite{SmithMoniz1972}:

\begin{equation}
    \epsilon_k = \sqrt{|\mathbf{k}|^2 + m^2} - \epsilon_1.
    \label{eq:rfg_initial_energy}
\end{equation}

\noindent When the neutrino transfers a four-momentum $q = (\omega, \mathbf{q})$ to this specific nucleon, it is ejected from the nuclear core. The total relativistic energy of the final recoiling particle, $\epsilon_{k-q}$, is given by \cite{SmithMoniz1972}:

\begin{equation}
    \epsilon_{k-q} = \sqrt{|\mathbf{k} - \mathbf{q}|^2 + (m')^2} - \epsilon_2,
    \label{eq:rfg_final_energy}
\end{equation}

\noindent where $m'$ is the mass of the final state particle ($m' = m$ for quasi-elastic scattering, or $m' = m_R$ for the production of a baryon resonance of mass $m_R$. For DIS, since it breaks down the nucleus, $m'$ turns into the invariant mass of the hadronic system $W$). The parameter $\epsilon_2$ accounts for any residual potential felt by the outgoing particle. In many standard applications, the interaction energy of the recoil particle is chosen to be zero ($\epsilon_2 = 0$). Relativistic energy conservation requires that the energy of the initial nucleon plus the transferred energy equals the energy of the final state particle:

\begin{equation}
    \epsilon_k + \omega = \epsilon_{k-q}.
    \label{eq:rfg_energy_conservation}
\end{equation}

With the relativistic kinematics established, the neutrino-nucleus cross-section can be constructed. Under the RFG model, the nuclear response is obtained by integrating the neutrino-nucleon interaction over the entire initial momentum distribution, $n_i(\mathbf{k}) = \Theta(k_F - |\mathbf{k}|)$ \cite{SmithMoniz1972}. Crucially, this integration must incorporate a Pauli blocking factor, $(1 - n_f(|\mathbf{k} - \mathbf{q}|))$, ensuring that the recoiling nucleon is not scattered into a momentum state already occupied by the Fermi sea. The subsequent challenge lies in formulating the exact tensor contractions between the weak leptonic current and the relativistic hadronic current, a formalism pioneered by Smith and Moniz (1972) \cite{SmithMoniz1972}, which will be explored later on.

\section{Neutrino-Nucleus Cross-Section in the Relativistic Fermi Gas Model}

The Smith and Moniz formalism uses the Relativistic Fermi Gas (RFG) model to calculate the neutrino-nucleus cross-section by bridging the elementary nucleon interaction with the complex nuclear environment. The mathematical description starts with the differential cross-section:

\begin{equation}
    \frac{d \sigma}{d q^2}=\frac{|\mathcal{M}|^2}{64 \pi M_T^2 E^2_{\nu}} , \nonumber
    \label{eq:diff_cross_section_general}
\end{equation}

where $M_T$ is the target mass, and the matrix element $\mathcal{M}$ contains the dynamics of the interaction, which is expressed as:

\begin{equation}
    |\mathcal{M}|^2 = \frac{G^2_F}{2} \cos^2{\theta_C} N(q^2). \nonumber
\end{equation}

\noindent Here, $N(q^2)$ is the contraction of the leptonic tensor $L^{\mu \nu}$ and the hadronic tensor. To apply this to a nuclear target, one must determine the hadronic tensor for the entire nucleus, denoted by $W_{\mu \nu}$. Smith and Moniz use Lorentz invariance rules to parameterize $W_{\mu \nu}$, since the internal structure of the nucleus is not treated explicitly. This yields the general expression \cite{kuzmin2008quasielastic}:

\begin{equation}
\begin{aligned}
    W_{\mu \nu} = -W_1 g_{\mu \nu} + \frac{W_2}{m_T^2}p_{\mu}p_{\nu} - i\frac{W_3}{2m_T^2}\epsilon_{\mu \nu \sigma \tau}  p^{\sigma}q^{\tau} + \frac{W_4}{m_T^2}q_{\mu}q_{\nu} \\ + \frac{W_5}{2m_T^2}(p_{\mu}q_{\nu} + p_{\nu}q_{\mu}) +\frac{W_6}{2m_T^2}(p_{\mu}q_{\nu} - p_{\nu}q_{\mu}) ,
\end{aligned}
\end{equation}

\noindent where $p$ is the target initial momentum, $q$ is the momentum transfer, $m_T$ is the target nuclear mass, and $\epsilon_{\mu \nu \sigma \tau}$ is the totally antisymmetric Levi-Civita tensor. The tensor is parameterized by five macroscopic nuclear structure functions: $W_1$, $W_2$, $W_3$, $W_4$, $W_5$ and $W_6$. The next step is to calculate the contraction $N(q^2) = L^{\mu \nu} W_{\mu \nu}$. The leptonic tensor is given by Equation \ref{eq:leptonic_tensor}:

\begin{equation}
    L^{\mu \nu} = 8[k^{\mu} k'^{\nu} + k'^{\mu} k^{\nu} - g^{\mu \nu} k \cdot k' \mp i\epsilon^{\mu \nu \rho \lambda} k_{\rho} k'_{\lambda} ]. \nonumber
\end{equation}

\noindent Performing the full tensor contraction $L^{\mu \nu} W_{\mu \nu}$ with the remaining terms yields  \cite{kuzmin2008quasielastic}:

\begin{equation}
\begin{aligned}
    L^{\mu \nu} W_{\mu \nu} = & \left( \frac{E_\ell - P_\ell \cos\theta_\ell}{m_T} \right) (W_1 + 2\kappa^2 W_4) \\
    & \pm \left[ \left( \frac{E_\nu + E_\ell}{m_T} \right) \left( \frac{E_\ell - P_\ell \cos\theta_\ell}{2m_T} \right) - 2\kappa^2 \right] W_3 \\
    & + \frac{E_\ell + P_\ell \cos\theta_\ell}{2m_T} W_2 - 2\kappa^2 W_5,
\end{aligned}
\label{eq:Sm&MoContraction}
\end{equation}

\noindent where $E_\ell$ denote the total energy of the outgoing charged lepton evaluated in the laboratory frame; $P_\ell$ is the magnitude of the outgoing charged lepton's three-momentum ($|\mathbf{p}_\ell|$); $\theta_\ell$ represents the scattering angle of the outgoing lepton with respect to the incident neutrino beam direction and $\kappa$ is a factor given by $\kappa = \frac{m_{\ell}}{2m_T}$. Note that in Equation \ref{eq:Sm&MoContraction}, the sixth structure function, $W_6$, associated with the antisymmetric combination of the target and transfer four-momenta, $(p^\mu q^\nu - p^\nu q^\mu)$, is entirely absent from the final contracted tensors. Mathematically, the leptonic tensor can be decomposed into a symmetric and an antisymmetric part: $L_{\mu\nu} = L_{\mu\nu}^{sym} + L_{\mu\nu}^{asym}$  \cite{kuzmin2008quasielastic}. The contraction of the antisymmetric $W_6$ term with the symmetric part of the leptonic tensor ($L_{\mu\nu}^{sym}$) trivially yields zero. However, the contraction with the antisymmetric part of the leptonic tensor ($L_{\mu\nu}^{asym} \propto \epsilon_{\mu\nu\alpha\beta} k^\alpha k'^\beta$) yields terms proportional to $\epsilon_{\mu\nu\alpha\beta} p^\mu q^\nu k^\alpha k'^\beta$. Because the momentum transfer is defined as $q = k - k'$, the four-vector $q$ lies entirely within the plane spanned by the initial and final lepton momenta, $k$ and $k'$. Consequently, the contraction of the fully antisymmetric Levi-Civita tensor is zero. Therefore, even if $W_6 \neq 0$, it kinematically cannot contribute to the contraction and consequently the cross-section, resulting in the common exclusion of the term in recent nuclear models.

To connect these macroscopic nuclear response functions ($W_i$) to the physics of a single nucleon, the RFG model assumes the nucleus is a collection of non-interacting fermions bound in a potential well. The total nuclear tensor is constructed by integrating the single-nucleon hadronic tensor, $H_{\mu\nu}$, over all possible nucleon momenta within the Fermi sphere \cite{SmithMoniz1972}:

\begin{equation}
    (W_{\mu\nu})_{lab} = \int d^3\mathbf{k} \, f(\mathbf{k}, \mathbf{q}, \omega) H_{\mu\nu}.
\end{equation}

\noindent The function $f(\mathbf{k}, \mathbf{q}, \omega)$ enforces the physical constraints of the nuclear medium:

\begin{equation}
    f(\mathbf{k}, \mathbf{q}, \omega) = \frac{m_T}{(2\pi)^3} \frac{\delta(E_{\mathbf{k}} + \omega - E_{\mathbf{k-q}}) \, n_i(\mathbf{k}) \, [1 - n_f(|\mathbf{k-q}|)]}{E_{\mathbf{k}} E_{\mathbf{k-q}}},
\end{equation}

\noindent this function includes the Dirac delta for energy conservation (accounting for the nucleon binding energy), the initial momentum distribution $n_i(\mathbf{k}) = \Theta(k_F - |\mathbf{k}|)$ representing the filled states up to the Fermi momentum $k_F$, and the Pauli blocking factor $[1 - n_f(|\mathbf{k-q}|)]$, which forbids the interaction if the recoil nucleon attempts to occupy an already filled momentum state. By carrying out this integration, Smith and Moniz demonstrated that the nuclear structure functions $W_i$ are linear combinations of the single-nucleon form factors \cite{kuzmin2008quasielastic}:

\begin{equation}
    W_i(Q^2)=2M^2|V_{ud}|^2 w_i(Q^2) \delta(2(pq) - Q^2),
\end{equation}

\noindent with

\begin{align}
    w_1 &= |F_A|^2 + x' \left( |F_V + F_M|^2 + |F_A|^2 \right), \nonumber  \\
    w_2 &= |F_V|^2 + |F_A|^2 + x' \left( |F_M|^2 + 4|F_T|^2 \right), \nonumber  \\
    w_3 &= -2\text{Re}\left[ F_A^* (F_V + F_M) \right], \nonumber  \\
    w_4 &= \frac{1}{4} \left[ x' \left( |F_M - 2F_S|^2 + 4|F_P + F_T|^2 \right) - |F_M|^2 \right] \nonumber \\
        &\quad + |F_S|^2 + \frac{1}{2}\text{Re}\left[ F_V^* (2F_S - F_M) - 2F_A^* (F_P + F_T) \right], \nonumber  \\
    w_5 &= w_2 + 2\text{Re}\left[ F_S^* (F_V - x'F_M) - F_T^* (F_A - 2x'F_P) \right].
\end{align}

\noindent In these expressions, the terms $w_1$ through $w_6$ represent the single-nucleon structure functions, which parameterize the elementary weak interaction vertex. They are entirely determined by the fundamental weak nucleon form factors: the vector ($F_V$), axial-vector ($F_A$), weak magnetic ($F_M$), pseudoscalar ($F_P$), scalar ($F_S$), and tensor ($F_T$) form factors, each encoding different structural aspects and coupling symmetries of the nucleon. Finally, the parameter $x'$ denotes a dimensionless kinematic variable related to the squared four-momentum transfer.

Substituting these relations back into Eq. (\ref{eq:Sm&MoContraction}) allows to determine the link between the observable neutrino-nucleus differential cross-section and the microscopic weak interactions governed by the elementary nucleon form factors.

\section{The Local Fermi Gas Model}

While the Relativistic Fermi Gas (RFG) model provides a framework for evaluating neutrino-nucleus kinematics, it assumes that the nucleus is a uniform sphere of constant density, representing an oversimplification of nuclear structure. In reality, electron scattering experiments have long established that the distribution of nucleons within a heavy nucleus is not uniform \cite{Hofstadter1956}. The density peaks at the center and smoothly decreases towards the surface. This variation in nuclear density directly affects the momentum of the bound nucleons and the surrounding binding potentials. To address these physical realities and provide a more accurate description of the initial state, modern interaction frameworks employ the Local Fermi Gas (LFG) model. The LFG model retains the basic premise of the Fermi gas: nucleons are treated as non-interacting fermions that fill all available momentum states up to a certain Fermi level according to the Pauli exclusion principle. However, instead of applying a single, global Fermi momentum ($k_F$) to the entire nucleus, the LFG model postulates that the nucleus can be divided into infinitesimal volume elements ($d^3r$). Within each local volume element, the nucleons behave as a standard Fermi gas but with a local Fermi momentum, $k_F(r)$, that depends explicitly on the radial distance $r$ from the center of the nucleus.

To determine the local Fermi momentum, the LFG model utilizes empirically measured nuclear density profiles. For medium and heavy nuclei, the radial density distribution of nucleons is typically parameterized by the Woods-Saxon distribution \cite{Woods1954}:

\begin{equation}
    \rho(r) = \frac{\rho_0}{1 + \exp\left(\frac{r - c}{a}\right)},
    \label{eq:woods_saxon}
\end{equation}

\noindent where $\rho_0$ is the central density of the nucleus, $c$ is the half-density radius, meaning the distance at which the density falls to $\rho_0/2$, and $a$ is the diffuseness parameter, which characterizes the surface thickness of the nucleus. These parameters are precisely constrained by elastic electron-nucleus scattering data. Because protons and neutrons can possess slightly different density distributions (especially in neutron-rich targets), the local density is often evaluated separately for protons ($\rho_p(r)$) and neutrons ($\rho_n(r)$). Following the standard relation between particle number density and the maximum momentum state of a degenerate fermion gas, the local Fermi momentum for a specific nucleon type is given by:

\begin{equation}
    k_{F, p}(r) = (3\pi^2 \rho_p(r))^{1/3}, \quad \quad k_{F, n}(r) = (3\pi^2 \rho_n(r))^{1/3}.
    \label{eq:local_fermi_momentum}
\end{equation}

\noindent This radial dependence alters the initial momentum distribution of the target. In the RFG model, the probability of finding a nucleon with momentum $|\mathbf{k}|$ is given by a rigid step function, $\Theta(k_F - |\mathbf{k}|)$, abruptly cutting off at the global $k_F \approx 250$ MeV/c. In the LFG model, nucleons residing near the nuclear surface ($r > c$) experience a significantly lower local density, restricting them to lower momentum states. Conversely, nucleons deep within the nuclear core ($r < c$) are subjected to higher densities and can reach momenta that exceed the average global $k_F$. Consequently, when integrated over the entire nuclear volume, the LFG produces a smooth, asymmetric momentum distribution, replacing the sharp cutoff of the RFG model.

\subsection{Neutrino-Nucleus Cross-Section in the LFG Framework}

The transition from a global to a local density profile requires a reformulation of how the observable neutrino-nucleus cross-section is calculated. In the LFG approach, the Impulse Approximation is still maintained, meaning the incident neutrino scatters off an individual bound nucleon. The total nuclear response is constructed by evaluating the interaction within each differential volume element and subsequently integrating over the entire spatial extent of the target nucleus. Mathematically, the local hadronic tensor, $W_{\mu\nu}^{local}(r)$, is constructed using the exact same relativistic tensor contractions developed in the Smith and Moniz formalism (Equation \ref{eq:Sm&MoContraction}). However, the integration over the initial nucleon momentum $\mathbf{k}$ is now constrained by the local variables \cite{Nieves2004}:

\begin{equation}
    W_{\mu\nu}^{local}(r) = \int \frac{d^3\mathbf{k}}{(2\pi)^3} \frac{m_N}{E_{\mathbf{k}}} \frac{m_N}{E_{\mathbf{k+q}}} \Theta(k_F(r) - |\mathbf{k}|) \, \Theta(|\mathbf{k} + \mathbf{q}| - k_F(r)) \, \delta(E_{\mathbf{k}} + \omega - E_{\mathbf{k+q}}) H_{\mu\nu},
\end{equation}

\noindent where $H_{\mu\nu}$ is the single-nucleon hadronic tensor. Notice that both the initial state momentum distribution $\Theta(k_F(r) - |\mathbf{k}|)$ and the Pauli blocking factor $\Theta(|\mathbf{k} + \mathbf{q}| - k_F(r))$ are evaluated dynamically based on the local Fermi momentum $k_F(r)$.

The total observable differential cross-section for the nucleus is then obtained by contracting the weak interaction probability with the local nuclear response and integrating over the nuclear volume $V$ \cite{Nieves2004}:

\begin{equation}
    \left( \frac{d\sigma}{d\Omega d\omega} \right)_{LFG} = \int d^3r \, \left[ \frac{d\sigma_{local}(k_F(r))}{d\Omega d\omega} \right].
    \label{eq:lfg_total_cross_section}
\end{equation}

This formulation of the Local Fermi Gas is a theoretical prerequisite for implementing advanced many-body nuclear effects. Models such as the Random Phase Approximation (RPA), which describes how the weak interaction is screened by the polarization of the nuclear medium, depend on the local nuclear density \cite{Gil1997}. By establishing a radially dependent phase space, the LFG framework provides the structural foundation required by modern neutrino event generators to model complex interactions, such as Meson Exchange Currents (MEC) and multi-nucleon (2p2h) excitations, improving the precision of Monte Carlo predictions \cite{Nieves2012, NuSTEC:2018}.

\newpage
\chapter{Neutrino Event Generators}

Modern precision neutrino oscillation experiments employ wide-band neutrino beams peaking in the few-GeV energy range to study flavor transitions over long baselines. Accelerator-based neutrinos are produced with a broad, continuous energy spectrum. Furthermore, to compensate for the low weak interaction cross-sections, these experiments must utilize massive, complex nuclear targets, such as liquid argon ($^{40}\text{Ar}$). Consequently, predicting the observable event rates cannot rely solely on idealized analytical formulas derived for scattering off free nucleons~\cite{Gallagher2011}. While theoretical frameworks like the Relativistic Fermi Gas (RFG) model provide robust calculations for differential cross-sections, as demonstrated in previous sections, they are insufficient to describe fully the final states. When a neutrino interacts within a nucleus, it scatters off a nucleon subjected to nuclear binding potentials. The outgoing hadrons must traverse the dense strongly-interacting medium of the residual nucleus. During this transit, they undergo Final State Interactions, such as pion absorption, charge exchange, and elastic scattering, which drastically alter the kinematics and particle multiplicity emerging from the target~\cite{NuSTEC:2018}. Neutrino event generators serve as the computational bridge between fundamental electroweak theory and experimental measurement. They are sophisticated Monte Carlo (MC) software frameworks designed to simulate the complete neutrino scattering event. Generators fold the analytical cross-section models with the experiment specific incident flux, sample the primary kinematics using probabilistic rejection techniques, and simulate the intra-nuclear hadronic cascades~\cite{Andreopoulos2010}. By providing detailed, particle-by-particle predictions of the final-state topologies, these generators enable the experimental reconstruction of the elusive incident neutrino energy, which is the cornerstone of any precision oscillation analysis. 

Historically, neutrino experiments relied heavily on proprietary, experiment-specific Monte Carlo codes, such as NEUGEN and NUANCE. These early generators were often coupled to their respective detector geometries and flux profiles, making cross-experiment comparisons and global model tuning difficult. As the global neutrino physics program transitioned towards an era of high-precision oscillation measurements, the necessity for a unified, rigorously validated, and universal simulation platform became paramount. To address this demand, the GENIE (\textit{Generates Events for Neutrino Interaction Experiments}) was conceived and developed in the early 2000s. Built from the ground up as a modern, object-oriented C++ framework, GENIE was explicitly designed to be a Monte Carlo platform capable of simulating neutrino scattering, from MeV energy scales relevant to reactor and solar neutrinos, up to PeV scales for astrophysical observations~\cite{Andreopoulos2010}. For long-baseline efforts experiments, such as the DUNE, GENIE serves as the primary standard for modeling the interactions of the neutrino beam with heavy nuclear targets. GENIE compartmentalizes the event generation into independent, interchangeable physics algorithms. This architecture separates the evaluation of primary differential cross-sections, the sampling of initial target nucleon kinematics, such as the implementation of the RFG, and the subsequent intra-nuclear tracking \cite{Andreopoulos2010}. This modular design is advantageous for ongoing phenomenological research. It allows physicists to isolate, modify, or introduce novel cross-section formulations and nuclear dynamics models directly, thereby testing new theoretical hypotheses against complex experimental observables without compromising the integrity of the simulation infrastructure.

\section{Neutrino Event Simulation}

The generation of a neutrino scattering event within a complex nuclear target is a computational and physical challenge. To render this, Monte Carlo generators like GENIE adopt an approach based on the factorization theorem. In this approach, the complete event simulation is subdivided into three sequential and independent stages: incident flux sampling, evaluation of the primary weak interaction, and propagation of the resulting hadrons through the nuclear medium~\cite{katori2018neutrino}.

\subsection{Beam Physics and Flux Modeling}
The first step in the simulation consists of determining the initial conditions of the projectile. Neutrino beams originate from pion and kaon decays so they possess a continuous energy spectrum. The event generator imports the beam information provided by the experimental collaboration, that describe the neutrino flux as a function of energy and flavor~\cite{DUNE_TDR_Vol2}. Utilizing Monte Carlo techniques, the software samples the energy, flavor, and three-momentum vector of the incident neutrino that will interact within the detector volume.

\subsection{Cross-Section Model and Primary Interaction}
Once the incident neutrino is defined, the generator must determine if and how it interacts. The software evaluates the total cross-sections for all possible interaction channels (Quasi-Elastic Scattering, Resonance Production, and Deep Inelastic Scattering) at the sampled energy by incoherently summing the exclusive cross-sections:

\begin{equation}
    \sigma^{tot} = \sigma^{QEL} \oplus \sigma^{RES} \oplus \sigma^{DIS}.
\end{equation}

\noindent The target nucleus is modeled using approximations such as the RFG and other models discussed in the previous section. 

The kinematics of the primary interaction are then resolved using the leptonic and hadronic tensors discussed previously. If the interaction is kinematically allowed, respecting, for instance, Pauli blocking for the ejected nucleon, the final-state particles of the primary vertex are generated. It is at this stage that the nuclear structure functions and form factors dictate the momentum transfer and the initial particle multiplicity.

\subsection{Final State Interactions}
The simulation does not conclude at the primary interaction. The recently created hadronic particles (protons, neutrons, pions) are born within the potential of the residual nucleus. As they travel, these particles are subjected to strong nuclear forces, undergoing what are known as Final State Interactions. 

In GENIE, this process is handled by intranuclear cascade models, such as the INTRANUKE package. FSI drastically alter the event topology through processes, such as exemplified in Figure \ref{fig:FSI}, like elastic scattering, charge exchange, where a $\pi^+$ may scatter into a $\pi^0$, thereby escaping track-based detection, and absorption, where pions are completely annihilated by the nucleus~\cite{Dytman2011}. Consequently, the kinematics measured experimentally in the detector can differ significantly from the original kinematics. Making accurate FSI modeling is one of the most significant challenges for precision neutrino oscillation measurements.

\begin{figure}
    \centering
    \includegraphics[width=0.65\linewidth]{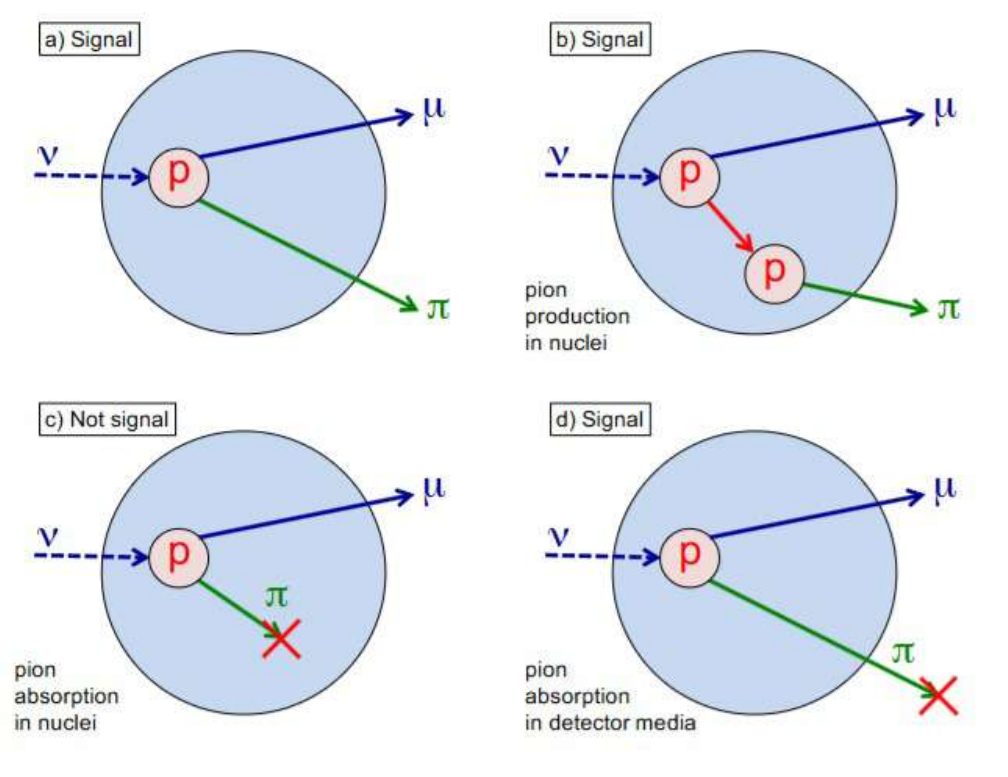}
    \caption{Schematic representation of a neutrino-nucleus interaction within a heavy target, such as Argon. The diagram illustrates the phenomenological complexities of the nuclear environment. The incident neutrino interacts with a bound nucleon subjected to initial-state dynamics. During this intra-nuclear propagation, they are subject to Final State Interactions, such as elastic scattering, charge exchange, and pion absorption, which significantly modify the outgoing particle kinematics and the final observable event topology. \cite{katori2018neutrino}.}
    \label{fig:FSI}
\end{figure}

\section{GENIE Neutrino Monte Carlo Event Generator}

The GENIE is a universal, object-oriented C++ Monte Carlo framework designed to simulate neutrino-nucleus scattering across a expansive kinematic regime. It works as the standard simulation platform for long-baseline neutrino efforts and serves as the computational bridge between fundamental electroweak scattering theory and the particle detectors. The software is built upon a modular architecture that compartmentalizes the event generation pipeline into distinct, sequential stages: the initial sampling of the incident neutrino flux, the evaluation of the primary weak interaction cross-sections, and the propagation of the resulting hadronic final states through the nuclear medium. This structural division not only provides a unified computational environment for global model tuning, but also allows physicists to isolate, test, and implement new theoretical cross-section models without disrupting the simulation infrastructure. The following subsections detail the core mechanisms that drive this generation pipeline and the specific models responsible for simulating the intra-nuclear transport of the final-state particles.

\subsection{Monte Carlo Techniques}

The GENIE framework relies on Monte Carlo integration and sampling techniques to evaluate the phase space of the final-state particles~\cite{Andreopoulos2010}. The core computational engine for kinematic generation within GENIE is based on the Acceptance-Rejection Method. In experimental particle physics, generating unweighted events, where each simulated event corresponds physically to one expected interaction in the detector, is preferred over weighted generation, as it allows for direct statistical comparison with raw detector data. To sample a specific kinematic variable, such as the outgoing lepton's energy or scattering angle, the generator must draw random values distributed according to the differential cross-section, denoted here as $f(x) \propto d\sigma/dx$. Because $f(x)$ for weak interactions is often complex, the acceptance-rejection technique introduces a bounding function, $C \cdot g(x)$, which must satisfy the condition $C \cdot g(x) \geq f(x)$ for all possible values of $x$ within the allowed kinematic limits~\cite{James1980}. The Monte Carlo sampling proceeds through the following algorithmic steps:

\begin{enumerate}
    \item A random kinematic configuration $x'$ is sampled from the simpler distribution $g(x)$.
    \item A uniform random number $u$ is drawn from the interval $[0, 1]$.
    \item The algorithm evaluates the physical cross-section at $x'$ and tests the condition:
    \begin{equation}
        u \leq \frac{f(x')}{C \cdot g(x')}
    \end{equation}
    \item If the condition is met, the kinematic configuration $x'$ is \textit{accepted} and the event is generated. If not, the configuration is \textit{rejected}, and the algorithm restarts by drawing a new $x'$ and $u$.
\end{enumerate}

\noindent To maximize computational efficiency and minimize the rejection rate, GENIE pre-computes the absolute maximum differential cross-section for a given interaction channel across a grid of neutrino energies during its initialization phase. These pre-calculated maxima form the constant envelope $C$, allowing $g(x)$ to be a uniform distribution across the available phase space~\cite{Andreopoulos2010}. The efficiency of this method drops significantly if the bounding envelope is poorly optimized or if the cross-section contains sharp peaks, such as narrow resonances. To mitigate this, GENIE restricts the phase-space generation limits based on exact kinematics, ensuring that computational time is not wasted evaluating the hadronic tensor in strictly forbidden kinematic regimes.

\subsection{Final State Interactions Models in GENIE}

Within the GENIE framework, the intra-nuclear propagation of hadrons is modeled using the INTRANUKE simulation package, which is fundamental for predicting the correct final state topologies. INTRANUKE steps the generated hadrons through the nuclear density profile, evaluating the probability of secondary strong interactions based on parameterized hadron-nucleus cross sections.

The literature dedicated to modeling and validating generator models focuses predominantly on the production and propagation of pions~\cite{Le2015, Eberly2015, McGivern2016}. This emphasis is not coincidental. First, in the energy range relevant to long-baseline experiments like DUNE, the resonance production channel, specifically the excitation of the target nucleon to the $\Delta(1232)$ state and its subsequent decay into a pion and a nucleon, constitutes a dominant fraction of the total cross section, as discussed previously in the neutrino nucleon interaction section. Second, due to the $\Delta$ resonance itself, pions exhibit an large cross section with nuclear matter interactions~\cite{McGivern2016}. Consequently, a pion has a high probability of undergoing an interaction that alters its state before exiting the nucleus.

The primary FSI processes that affect pions and distort experimental observables include:
\begin{itemize}
    \item Pion Absorption: The pion generated is entirely absorbed by the target nucleus, typically transferring its rest mass and kinetic energy to the ejection of multiple correlated nucleons~\cite{Eberly2015}. Experimentally, this effect can be misidentified as a CCQE-like interaction, contaminating signal samples~\cite{McGivern2016}.
    
    \item Charge Exchange (CEX): The pion interacts with a nucleon and alters its isospin state (e.g., $\pi^+ n \rightarrow \pi^0 p$ or $\pi^- p \rightarrow \pi^0 n$)~\cite{Le2015}. This process constitutes a major source of uncertainty for oscillation analyses. Neutral pions ($\pi^0$) decay almost instantaneously into photon pairs ($\gamma\gamma$). In LArTPCs, the electromagnetic showers induced by these photons can mimic the signature of an electron, creating a background for $\nu_e$ appearance searches~\cite{Le2015}.
    
    \item Elastic and Inelastic Scattering: The pion survives the nuclear traversal but transfers kinetic energy to the nucleus and has its trajectory deflected~\cite{Eberly2015}.
\end{itemize}

Although pions dominate the literature due to their direct impact on backgrounds, FSI are not limited to them. The propagation of nucleons ejected from the target nucleus is important, especially for the energy reconstruction techniques employed in Argon targets~\cite{McGivern2016}. Protons originating from quasi-elastic interactions or resonance decays undergo intense nucleon-nucleon scattering within the nuclear potential. These FSI cause severe degradation of the proton's kinetic energy and frequently induce nuclear evaporation, a process in which the excited residual nucleus emits multiple low-energy neutrons or protons~\cite{McGivern2016}. Because neutral particles like neutrons and escape detection, low-energy protons yield tracks too short to be accurately reconstructed, this ``missing" energy biases the measurement of the total incident neutrino energy. The INTRANUKE package includes dedicated transport algorithms to simulate this nucleon cascade and ensure overall energy conservation~\cite{McGivern2016}.

Studies conducted by MINERvA for charged pions ($\pi^\pm$)~\cite{Eberly2015, McGivern2016} and neutral pions ($\pi^0$)~\cite{Le2015} demonstrate that while the GENIE FSI model successfully predicts the qualitative behavior and overall cross-section scale, significant tensions remain regarding the exact shapes of the differential spectra. The generator overpredicts pion production in certain kinematic regions while underpredicting in others~\cite{McGivern2016}.

\newpage
\chapter[Axial Mass in $\nu$-A Scattering]{Impact of the Axial Mass in Neutrino-Nucleus Scattering}

As discussed in previews sections, in neutrino interactions, the precise determination of the nucleon structure is fundamental for predicting interaction rates and event topologies. The hadronic weak current is parameterized by a set of vector and axial-vector form factors. While the vector form factors ($F_V^1, F_V^2$) are well-constrained by electron scattering experiments through the Conserved Vector Current (CVC) hypothesis, the axial-vector form factor ($F_A$) and its associated parameter, the axial mass ($M_A$), remain one of the most critical and debated sources of systematic uncertainty in modern neutrino physics. This section provides a theoretical discussion on the analytical role of the axial mass, the historical ``Axial Mass Anomaly," and its phenomenological impact on the models evaluated within the GENIE framework.

\section{The Axial Form Factor}

As demonstrated mathematically, the differential cross sections $d\sigma/dQ^2$ is dependent on the axial form factor, $F_A(Q^2)$. At zero momentum transfer ($Q^2 = 0$), the value of the axial form factor is determined by neutron beta decay measurements, yielding $F_A(0) = g_A \approx -1.272$ \cite{PDG2022}. However, its dependency as a function of the squared four-momentum transfer, $Q^2$, cannot be derived from first principles using perturbative Quantum Chromodynamics (QCD) due to the non-perturbative nature of the strong interaction at the energy scales relevant to neutrino oscillation experiments (few-GeV) \cite{NuSTEC:2018}.

 $F_A(Q^2)$ is modeled using a phenomenological dipole parameterization, mirroring the functional form that describes the electromagnetic form factors of the nucleon. The dipole axial form factor is expressed as \cite{LlewellynSmith:1972zz}:

\begin{equation}
    F_A(Q^2) = \frac{F_A(0)}{\left( 1 + \frac{Q^2}{M_A^2} \right)^2},
    \label{eq:dipole_axial}
\end{equation}

\noindent where $M_A$ is the axial mass. Physically, the axial mass is inversely proportional to the axial root-mean-square radius of the nucleon, $\langle r_A^2 \rangle$, which characterizes the spatial distribution of the weak axial charge within the proton or neutron \cite{NuSTEC:2018}:

\begin{equation}
    \langle r_A^2 \rangle = -\frac{6}{F_A(0)} \left. \frac{dF_A(Q^2)}{dQ^2} \right|_{Q^2=0} = \frac{12}{M_A^2}.
\end{equation}

According to Equation \ref{eq:dipole_axial}, the value of $M_A$ controls the $Q^2$ evolution of the interaction. A larger value of $M_A$ flattens the $Q^2$ distribution, suppressing the high-$Q^2$ decay of the form factor. Consequently, increasing $M_A$ enhances the probability of interactions with large momentum transfers, leading to a higher total integrated cross section.

\begin{figure}
    \centering
    \includegraphics[width=0.65\linewidth]{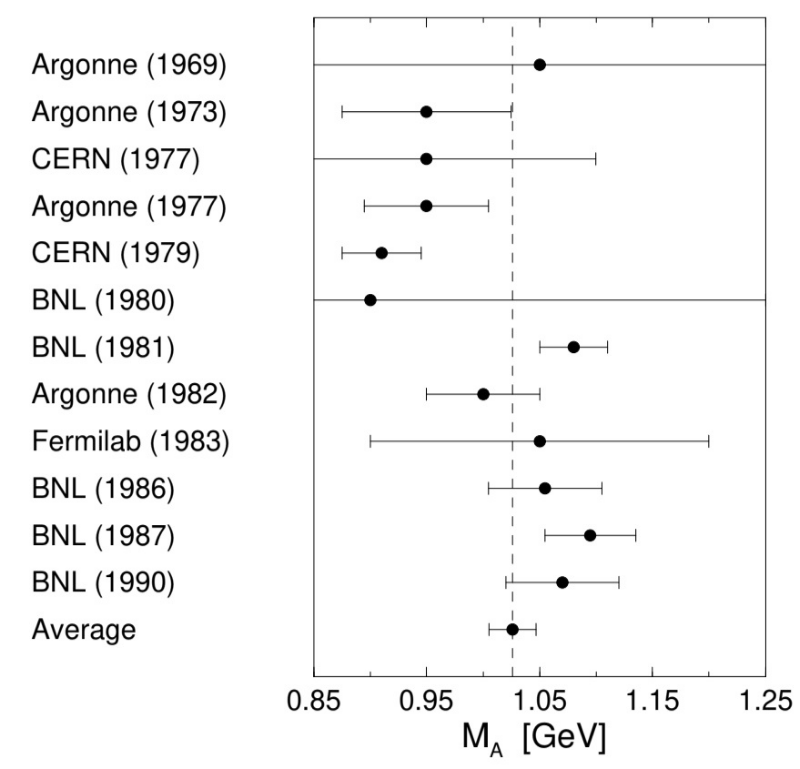}
    \caption{Historical experimental measurements of the axial mass ($M_A$) obtained from bubble chamber experiments using neutrino interactions on hydrogen and deuterium targets \cite{bernard2002axial}.}
    \label{fig:FSI}
\end{figure}

Figure \ref{fig:FSI} summarizes the primary experimental measurements that directly determined the axial mass observable in quasi-elastic neutrino scattering. Because these early experiments utilized bubble chambers filled with light target nuclei, the interactions were free from complex nuclear medium effects, providing a clean environment to extract the $M_A$ observable.

\section{The Axial Mass Puzzle in Heavy Nuclei}

The free nucleon value of the axial mass was established in the 1970s and 1980s through neutrino-deuterium and neutrino-hydrogen bubble chamber experiments, such as those conducted at Argonne National Laboratory (ANL) and Brookhaven National Laboratory (BNL) \cite{Radecky1982, Kitagaki1983}. Because deuterium nuclear medium effects are minimal, allowing for a nearly direct measurement of the fundamental nucleon property. A global fit to these data sets yielded an average value of $M_A = 1.03 \pm 0.06 \text{ GeV/c}^2$ \cite{abe2023updated}. However, modern neutrino experiments, such as MiniBooNE, K2K, and MINER$\nu$A, which utilize heavy nuclear targets like Carbon ($^{12}$C), Oxygen ($^{16}$O), and Argon ($^{40}$Ar) to maximize interaction rates, observed a significant discrepancy. When applying the standard Relativistic Fermi Gas (RFG) model to fit their high-statistics CCQE-like data, these experiments found a severe underprediction of the total cross-section and a distortion in the $Q^2$ shape. To reconcile the RFG theoretical predictions with the experimental data, it was necessary to artificially inflate the axial mass parameter to an effective value of $M_A = 1.20 \pm 0.09 \text{ GeV/c}^2$ \cite{gran2006measurement}. This discrepancy between the $M_A \approx 1.0 \text{ GeV/c}^2$ and the $M_A \approx 1.2 \text{ GeV/c}^2$ value became known as the ``Axial Mass Anomaly." This problem highlighted that an inflated $M_A$ is not a reflection of a change in the fundamental structure of the nucleon itself, but rather an artifact masking incomplete nuclear modeling. 

\section{Impact of $M_A$ Variations}

\begin{figure}
    \centering
    \begin{subfigure}[b]{0.48\textwidth}
        \centering
        \includegraphics[width=1\linewidth]{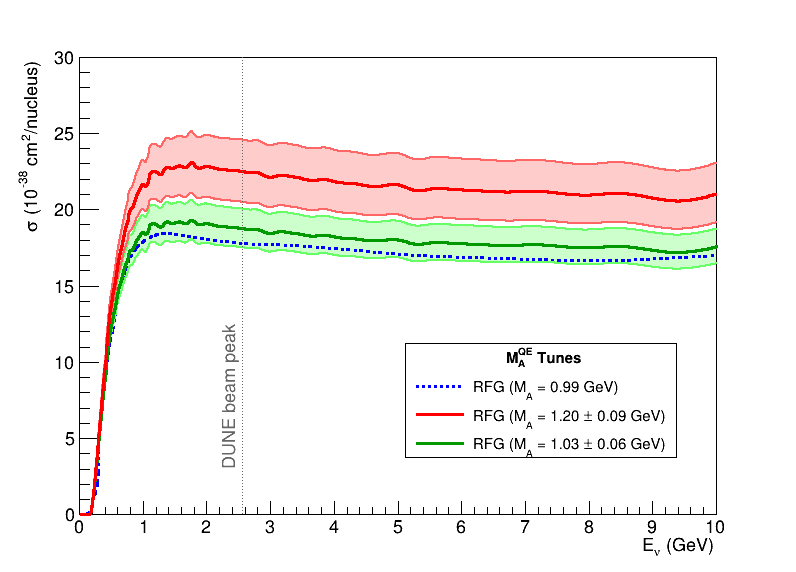}
        \caption{}
        \label{fig:axial_mass_comp}
    \end{subfigure}
    \hfill
    \begin{subfigure}[b]{0.48\textwidth}
        \centering
        \includegraphics[width=1\linewidth]{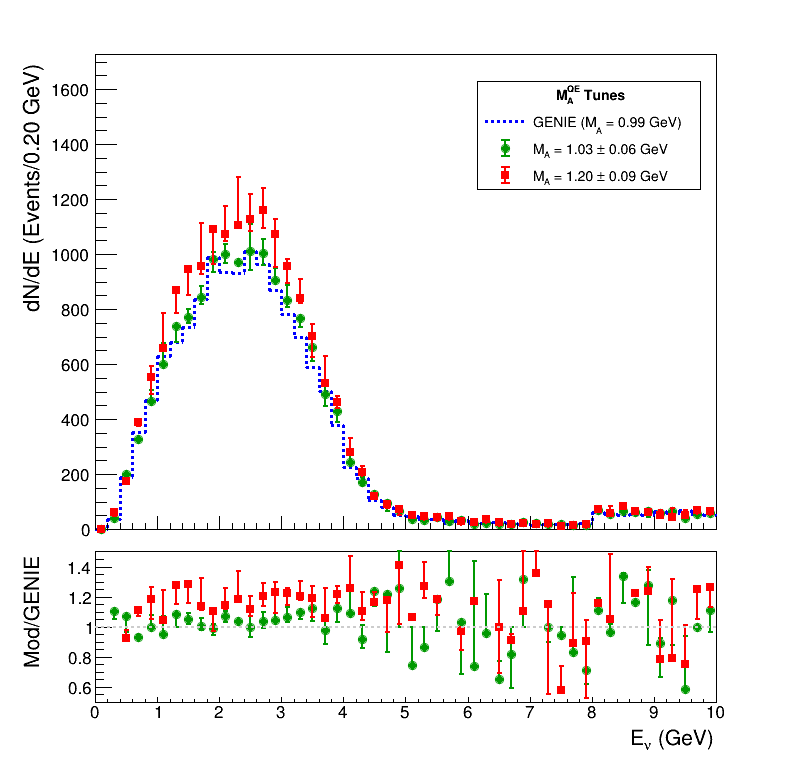}
        \caption{}
        \label{fig:event_rate_MA}
    \end{subfigure}
    
    \caption{\small{(a) Sensitivity of the $\nu_\mu$ CCQE cross-section to axial mass ($M_A$) variations simulated with the GENIE event generator. The blue dashed line represents the Relativistic Fermi Gas (RFG) prediction using a standard axial mass ($M_A = 0.99$ GeV). The solid red and green curves illustrate the enhanced cross-sections utilizing axial masses derived from heavy-target experiments ($M_A = 1.20 \pm 0.09$ GeV \cite{gran2006measurement}) and recently constrained fits ($M_A = 1.03 \pm 0.06$ GeV \cite{abe2023updated}), respectively. The vertical dashed line indicates the 2.5 GeV peak energy of the DUNE neutrino flux. (b) Simulated event rate distribution for $\nu_\mu$ interactions $Ar^{40}$ as a function of neutrino energy ($E_\nu$), utilizing the DUNE flux profile. The top panel displays the absolute number of generated events per 0.20 GeV bin. The blue dashed histogram represents the GENIE prediction, while the green and red markers represent the simulated events using different axial masses. The bottom panel displays the ratio of the modified models to the GENIE prediction (Mod/GENIE). (Author, 2026).}}
    \label{fig:ccqe_analysis}
\end{figure}

Using the GENIE event generator, the sensitivity of the total CCQE and RES cross-sections to variations in the axial mass was simulated for muon neutrino ($\nu_\mu$) using the values presented previously. The results are then presented in Figure \ref{fig:ccqe_analysis}. The simulation (Figure \ref{fig:axial_mass_comp}) establishes the GENIE curve utilizing an axial mass of $M_A = 0.99$ GeV and the \texttt{G18\_02a\_00\_000} tune. Against this, the cross-section is re-evaluated using two distinct effective parameterizations. The first modification ($M_A = 1.20 \pm 0.09$ GeV) represents the high values from experiments K2K and MiniBooNE. The second modification ($M_A = 1.03 \pm 0.06$ GeV) illustrates a value from recent constrained models. The simulated data explicitly shows the cross-section enhancement. Focusing on the DUNE peak flux energy (indicated by the vertical dashed line at approximately 2.5 GeV), the GENIE cross-section rests at approximately $18 \times 10^{-38} \text{ cm}^2$. By inflating the axial mass to $1.20$ GeV, the cross-section at this same energy is drastically boosted to approximately $24.5 \times 10^{-38} \text{ cm}^2$.

To evaluate how these propagate to actual experimental observables, a full event generation was simulated by using the modified cross-sections with the DUNE neutrino flux and the same tune configuration. The resulting event rate distributions are presented in Figure \ref{fig:event_rate_MA}. The top panel represents the absolute number of generated events over the incident neutrino energy. As expected, the event yield peaks in the $2.0 - 3.0$ GeV region, mirroring the profile of the LBNF wide-band beam. The effective mass $M_A = 1.20$ GeV produces a higher event yield across the entire active spectrum compared to the GENIE curve. To visually quantify this discrepancy, the bottom panel plots the ratio of the modified models to the GENIE prediction. In the region of the DUNE flux peak ($1.5 \le E_\nu \le 3.5$ GeV), the $M_A = 1.20$ GeV tune induces an excess of approximately $20\%$ to $40\%$. Conversely, the $M_A = 1.03$ GeV tune results in a shift of roughly $10\%$. These results demonstrate the sensitivity of the models with respect to the axial mass parameter. While utilizing the effective axial mass within the RFG model recovers the total event rate observed in heavy-target experiments, it achieves this by inflating the parameter, leading to a bias. 

\begin{figure}
    \centering
    \begin{subfigure}[b]{0.48\textwidth}
        \centering
        \includegraphics[width=1\linewidth]{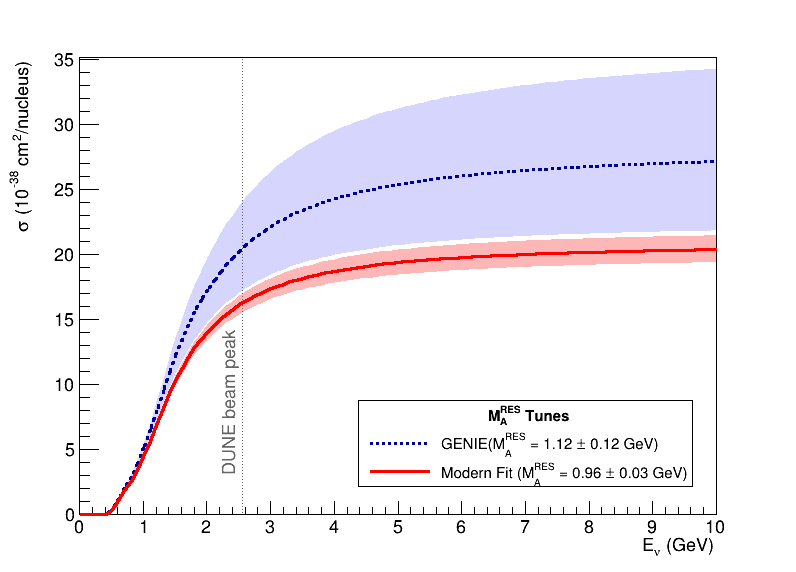}
        \caption{}
        \label{fig:ccres_xsec}
    \end{subfigure}
    \hfill
    \begin{subfigure}[b]{0.48\textwidth}
        \centering
        \includegraphics[width=1\linewidth]{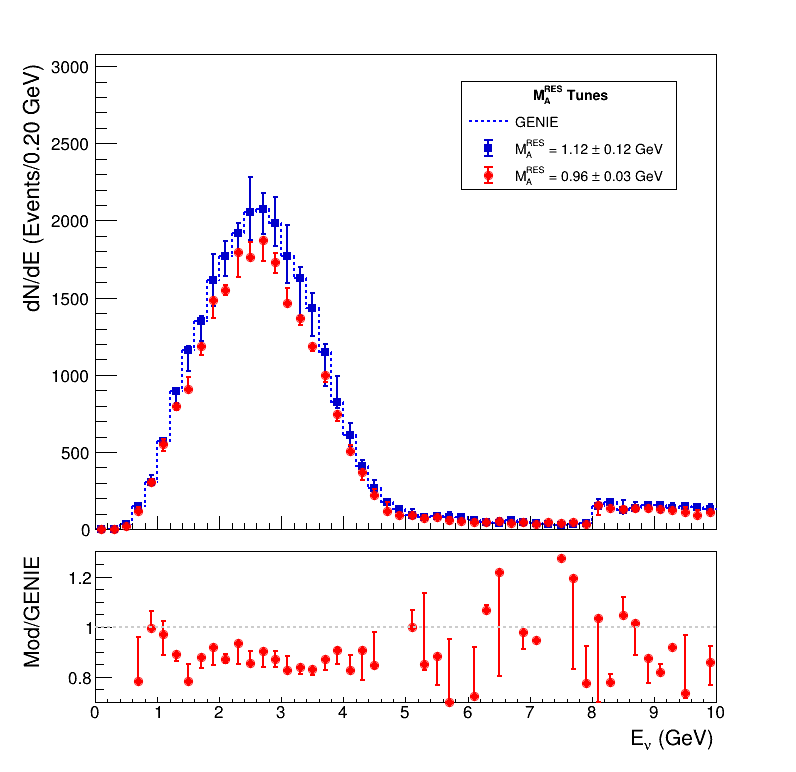}
        \caption{}
        \label{fig:ccres_events}
    \end{subfigure}

    \caption{\small{(a) The integrated cross-section per nucleon as a function of energy. The blue dotted curve represents the GENIE default value ($M_A^{RES} = 1.12 \pm 0.12$ GeV \cite{tena2021neutrino}), which yields a larger cross-section with broad parameter uncertainties. The red solid curve illustrates the cross-section using a modern phenomenological fit ($M_A^{RES} = 0.96 \pm 0.03$ GeV \cite{rodrigues2016constraining}). (b) The simulated CCRES event rate distribution utilizing the DUNE flux. The bottom ratio panel (Mod/GENIE) demonstrates that adopting the modern $M_A^{RES}$ fit induces a systematic deficit, reducing the predicted resonant event yield by approximately $10\%$ to $20\%$ in the flux peak region ($1.5 - 3.5$ GeV). (Author, 2026).}}
    \label{fig:ccres_analysis}
\end{figure}

Beyond the CCQE channel, CCRES constitutes another dominant fraction of the total cross-section at DUNE energies. The default GENIE configuration utilizes $M_A^{RES} = 1.12 \pm 0.12$ GeV, a historical value derived from early ANL and BNL deuterium bubble chamber experiments \cite{tena2021neutrino}. However, modern phenomenological re-analyses of these datasets suggest a significantly lower and more constrained value of $M_A^{RES} = 0.96 \pm 0.03$ GeV \cite{rodrigues2016constraining}. To evaluate the phenomenological impact of this difference in the CCRES channel, the cross-section and the respective events were simulated using the \texttt{G18\_02a\_00\_000} tune and DUNE Flux, as shown in Figure \ref{fig:ccres_analysis}. The implementation of the re-analyzed parameter reduces the axial mass and, consequently, the cross-section, as shown in Figure \ref{fig:ccres_xsec}. When generating events distribution (Figure \ref{fig:ccres_events}) reveals a systematic deficit. The ratio panel indicates that the implementation of the modern fit reduces the predicted resonant events by $10\%$ two $20\%$ across the statistically significant region of the energy spectrum. These evaluations emphasizes the necessity for accurate parameterization to control systematic backgrounds in event analyses.

Furthermore, the enhanced $Q^2$ distribution, as according to Eq. \ref{eq:dipole_axial}, alters the predicted outgoing lepton momentum and scattering angle. For experiments like DUNE, to achieve a more precise measurement for Charge Parity (CP) violation and other oscillation parameters, it requires an low systematic uncertainty margin, targeting approximately $2\%$ for the effective uncorrelated normalization uncertainty on the $\nu_e$ appearance channel and $5\%$ for the $\nu_\mu$ sample \cite{NuSTEC:2018}. As demonstrated by the simulations, relying on phenomenological adjustments like an inflated $M_A$ introduces variations in the total event rate ranging from $20\%$ to $40\%$ at the flux peak, which can impact the uncertainties of the measurements. Utilizing a higher value of $M_A$, boosting the CCQE cross-section can create a severe bias in neutrino energy reconstruction. Relying on an inflated axial mass is an compensation for fitting experimental data. Because the Relativistic Fermi Gas (RFG) omits multi-nucleon correlations, such as two-particle-two-hole ($2\text{p}2\text{h}$) and Meson Exchange Currents, changing $M_A$ forces the cross-section to match the total observed event rate, introducing significant systematic biases in both energy reconstruction and topology classification. When a 2p2h event is falsely modeled and reconstructed as a standard 1p1h CCQE interaction, the ``invisible energy", the energy expended in breaking nucleon correlations or carried away by multiple low-energy ejected nucleons below detection thresholds, is entirely ignored \cite{NuSTEC:2018}. Therefore, robust nuclear models is not merely a theoretical refinement, but a fundamental requirement to mitigate background uncertainties, classify event topologies correctly, and ensure the success of DUNE's precision oscillation physics program.

\section{Comparison with MicroBooNE 2022 Experimental Data}

\begin{figure}
\centering\includegraphics[width=0.75\linewidth]{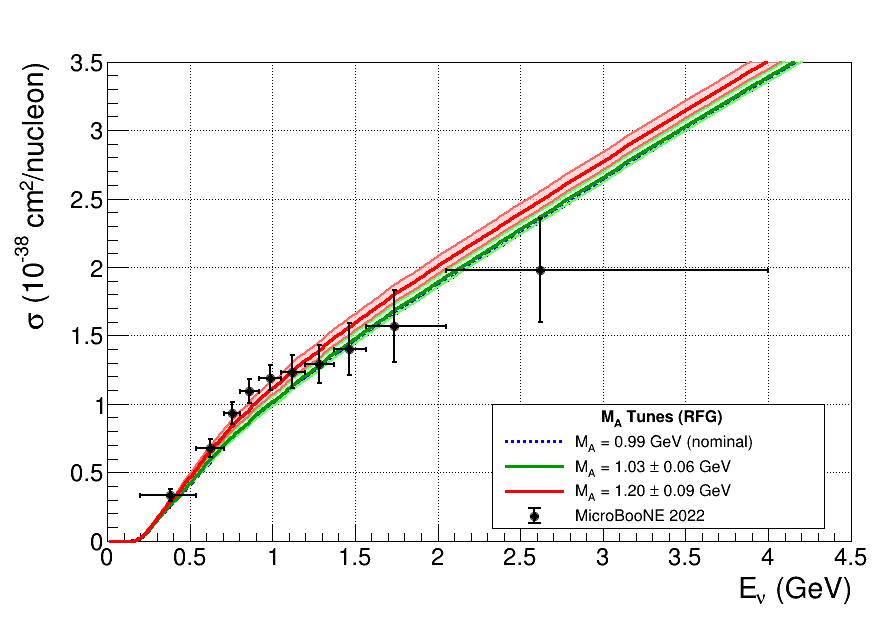}
\caption{\small{Comparison of the simulated $\nu_\mu$ inclusive cross-section per nucleon on Argon against experimental data from MicroBooNE (2022) \cite{MicroBooNE2022}. The curves represent GENIE RFG predictions evaluated at different axial mass values: $M_A = 0.99$ GeV (blue dashed), $M_A = 1.03 \pm 0.06$ GeV (green solid), and $M_A = 1.20 \pm 0.09$ GeV (red solid). (Author, 2026).}}
\label{fig:mboone_comparison}
\end{figure}

To contextualize the impact of the axial mass parameter within Liquid Argon (LAr) targets, the GENIE Relativistic Fermi Gas (RFG) predictions are directly compared against the measured $\nu_\mu$ cross-section from the MicroBooNE experiment. Figure \ref{fig:mboone_comparison} presents the integrated cross-section per nucleon alongside the MicroBooNE 2022 data points \cite{MicroBooNE2022}. The simulated curves evaluate the RFG model across the three distinct parameterizations of the axial mass presented in this section. As illustrated in the distribution, the experimental data exhibits agreement for the $M_A = 1.20 \pm 0.09$ GeV parameterization (red curve), particularly in the low-energy spectrum ($E_\nu \le 1.5$ GeV). In this specific kinematic region, both the nominal GENIE tune and the $1.03$ GeV modification systematically underpredict the observed interaction rate. The MicroBooNE data points align closely with the upper trajectory, demonstrating that the $1.20$ GeV setting phenomenologically recovers the cross-section magnitude measured by the LArTPC detector. While the $M_A = 1.20$ GeV curve provides a much closer description of the experimental measurements, this requirement exposes underlying limitations within the default GENIE RFG framework. The necessity to rely on a specific parameter adjustment to reconcile theoretical predictions with heavy-target Argon data indicates that the baseline model may not capture the complete dynamics of the nuclear medium. For next-generation, high-precision experiments such as DUNE, depending on the phenomenological tuning of a single parameter rather than employing explicit physical models could introduce systematic biases in energy reconstruction. This behavior emphasizes that the current default modeling approach in GENIE may not be sufficient for high-precision oscillation physics, reinforcing the urgent need to transition towards more sophisticated theoretical frameworks.

\newpage
\chapter[Nuclear Models]{Evaluation of Advanced Nuclear Models in the CCQE Channel}

As demonstrated in the previous sections, to resolve the discrepancies between theoretical predictions and heavy-target experimental data requires abandoning the oversimplified assumptions of the Relativistic Fermi Gas (RFG) model and implementing advanced theoretical frameworks capable of explicitly describing the complex many-body dynamics within the nuclear medium. 

This section presents a computational evaluation of the Charged Current Quasi-Elastic (CCQE) channel using the GENIE event generator. The performance of the RFG model (Llewellyn-Smith) is directly compared against two state-of-the-art nuclear formulations: the Local Fermi Gas (LFG) supplemented with Random Phase Approximation (RPA) correlations (Nieves model), and the SuperScaling Approximation (SuSAv2). By analyzing the initial state dynamics, integrated cross-sections, and differential kinematic observables, this comparative study quantifies the phenomenological impact of these models on DUNE experiment.

\section{Initial State Dynamics: Nucleon Momentum Distribution}

The difference between nuclear models begins at the description of the initial state of the target nucleus. As discussed previously, in the RFG model, the nucleus is treated as a sphere of constant density, where non-interacting nucleons fill all available momentum states up to the Fermi momentum. Conversely, the Local Fermi Gas (LFG) model considers that the nuclear density, $\rho(r)$, is not constant but rather follows a Woods-Saxon distribution.

\begin{figure}
    \centering
    \includegraphics[width=0.75\linewidth]{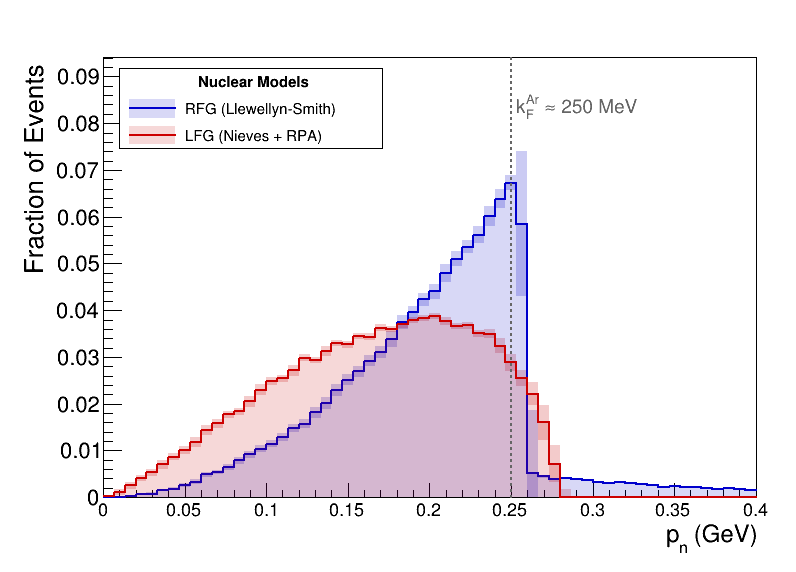}
    \caption{Simulated initial momentum distribution ($p_n$) of the target nucleons in $^{40}\text{Ar}$ for CCQE interactions. The blue histogram represents the Relativistic Fermi Gas (RFG) model, exhibiting a characteristic physical cutoff at the global Fermi momentum $k_F^{Ar} \approx 250$ MeV. The red histogram illustrates the Local Fermi Gas (LFG) model, which incorporates radial density variations, resulting in a broader, smoother momentum distribution that extends into higher momentum regimes. (Author, 2026).}
    \label{fig:nucleon_momentum}
\end{figure}

Figure \ref{fig:nucleon_momentum} presents the simulated initial momentum distribution ($p_n$) of the bound nucleons within the $^{40}\text{Ar}$ target. The RFG model (blue curve) is characterized by a cutoff near the standard Argon Fermi momentum limit, highlighting the boundaries of the potential well. In contrast, the LFG model (red curve) produces a broader and more physically realistic distribution. Because the LFG uses the local density, nucleons residing near the nuclear surface experience a lower effective Fermi momentum, shifting the distribution peak to lower values.

\section{Integrated Cross-Sections and Event Rate}

\begin{figure}
    \centering
    \includegraphics[width=0.75\linewidth]{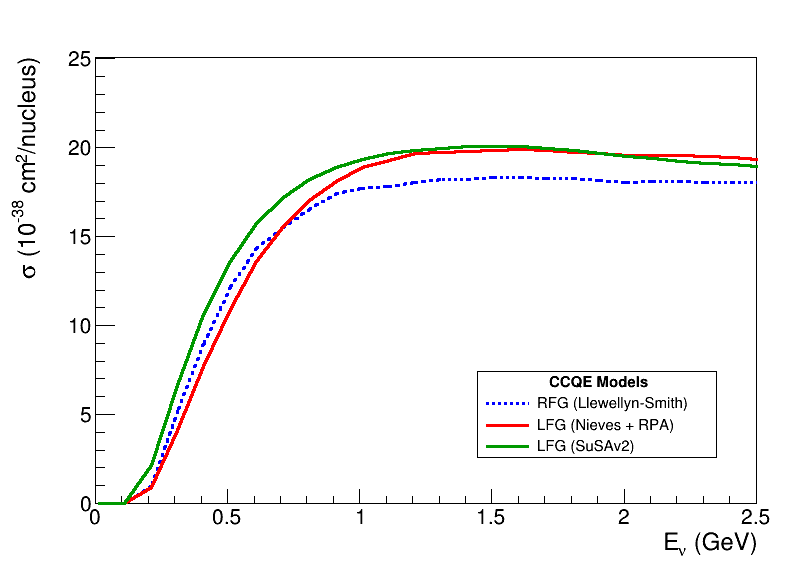}
    \caption{Integrated $\nu_\mu$ CCQE cross-section per  $^{40}\text{Ar}$ nucleus as a function of neutrino energy ($E_\nu$). The plot compares the standard RFG model (blue dotted line) against the advanced LFG with Nieves + RPA corrections (red solid line) and the SuSAv2 model (green solid line). Both advanced models demonstrate significant cross-section enhancements at higher energies relative to the RFG baseline, while exhibiting varying degrees of threshold suppression. (Author, 2026).}
    \label{fig:ccqe_model_comp}
\end{figure}

The modifications in the initial state dynamics propagate directly into the calculation of the interaction probabilities. Figure \ref{fig:ccqe_model_comp} presents the integrated cross-section per nucleon as a function of the incident neutrino energy for the three evaluated nuclear models. The simulation reveals that while the RFG model (blue dotted line) quickly plateaus at approximately $18 \times 10^{-38} \text{ cm}^2 / \text{nucleus}$, both the LFG (Nieves + RPA) and SuSAv2 models predict an enhanced cross-section across the energy spectrum. The SuSAv2 model (green line), which is built upon the phenomenological ``superscaling" behavior observed in electron-nucleus scattering data, predicts the highest overall cross-section in the peak flux region. The Nieves LFG model (red line) exhibits a pronounced suppression at the low-energy threshold compared to the other models.

\begin{figure}[htbp]
    \centering
    \begin{subfigure}[b]{0.48\textwidth}
        \centering
        \includegraphics[width=1\linewidth]{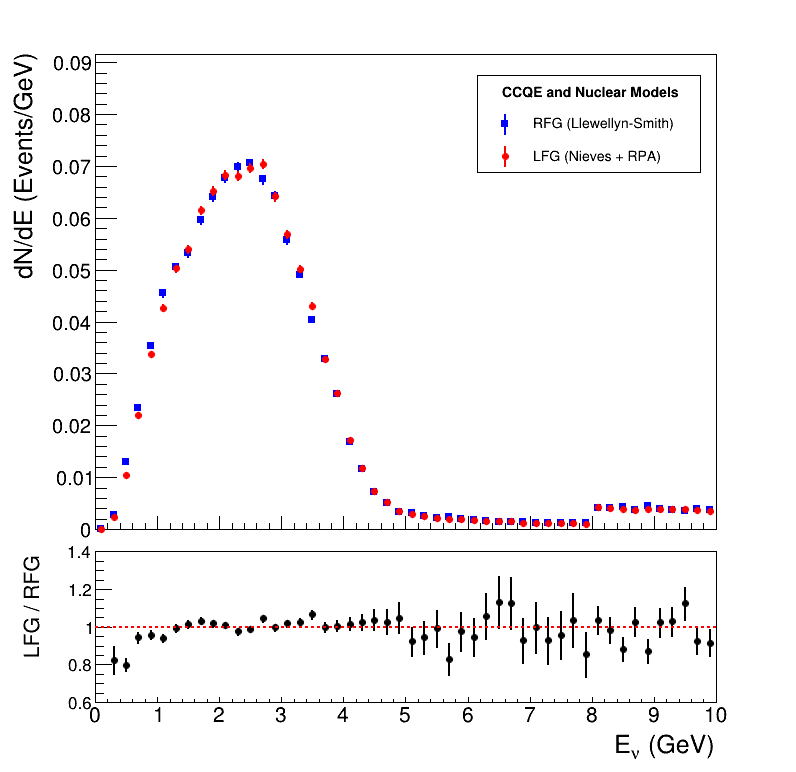}
        \caption{RFG vs. LFG (Nieves + RPA)}
        \label{fig:ev_ratio_lfg}
    \end{subfigure}
    \hfill
    \begin{subfigure}[b]{0.48\textwidth}
        \centering
        \includegraphics[width=1\linewidth]{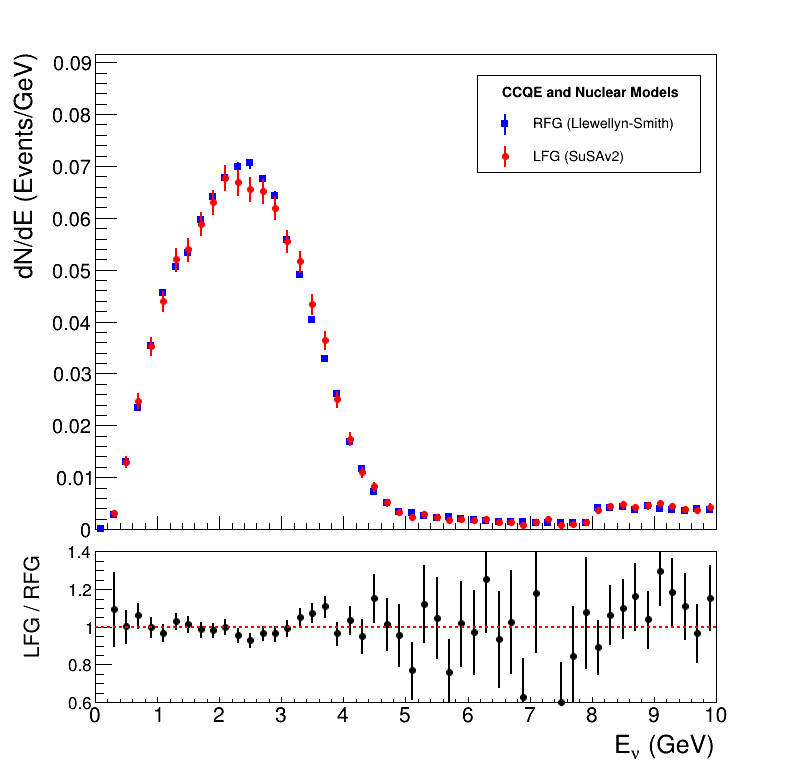}
        \caption{RFG vs. LFG (SuSAv2)}
        \label{fig:ev_ratio_susav2}
    \end{subfigure}
    \caption{Simulated event rate distributions for $\nu_\mu$ CCQE interactions utilizing the DUNE flux. (a) Comparison between the baseline RFG model and the LFG model with Nieves RPA corrections. (b) Comparison between the RFG and the SuSAv2 model. The bottom panels display the ratio of the advanced models to the RFG baseline. (Author, 2026).}
    \label{fig:event_rates_advanced}
\end{figure}

To assess the practical impact of these models on experimental measurements, the cross-sections were convoluted with the DUNE flux, generating the absolute event rate distributions shown in Figure \ref{fig:event_rates_advanced}. According to the ratio panels in Figure \ref{fig:event_rates_advanced}, neither the LFG nor the SuSAv2 formulation induces a systematic enhancement in the event rate.

\section{Kinematic Observables and the $Q^2$ Spectrum}

\begin{figure}[htbp]
    \centering
    \begin{subfigure}[b]{0.48\textwidth}
        \centering
        \includegraphics[width=1\linewidth]{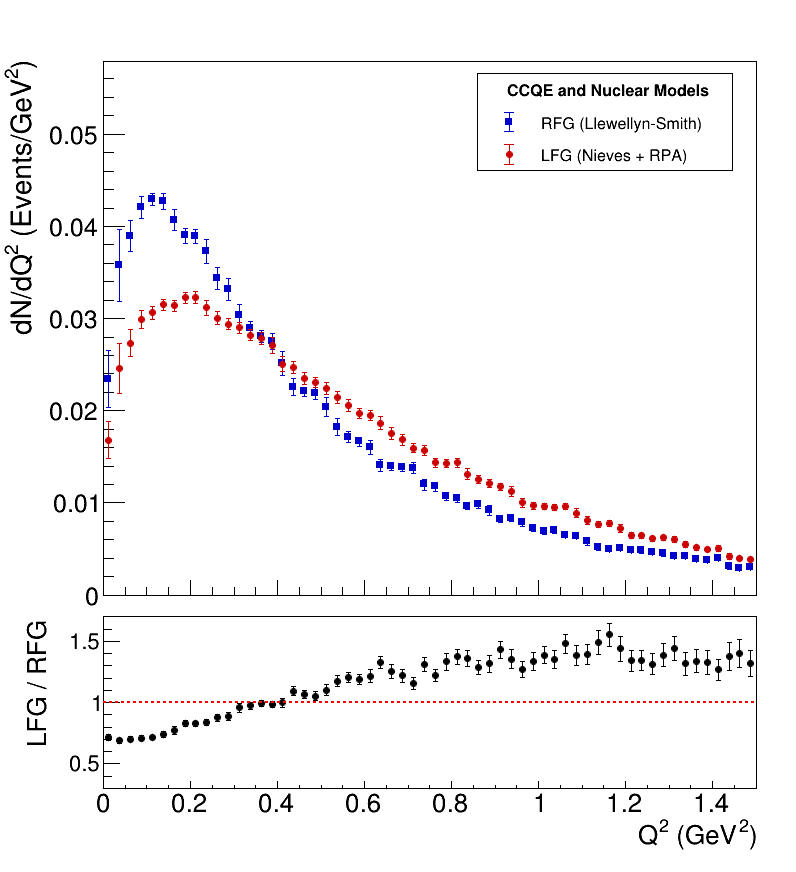}
        \caption{RFG vs. LFG (Nieves + RPA)}
        \label{fig:q2_lfg}
    \end{subfigure}
    \hfill
    \begin{subfigure}[b]{0.48\textwidth}
        \centering
        \includegraphics[width=1\linewidth]{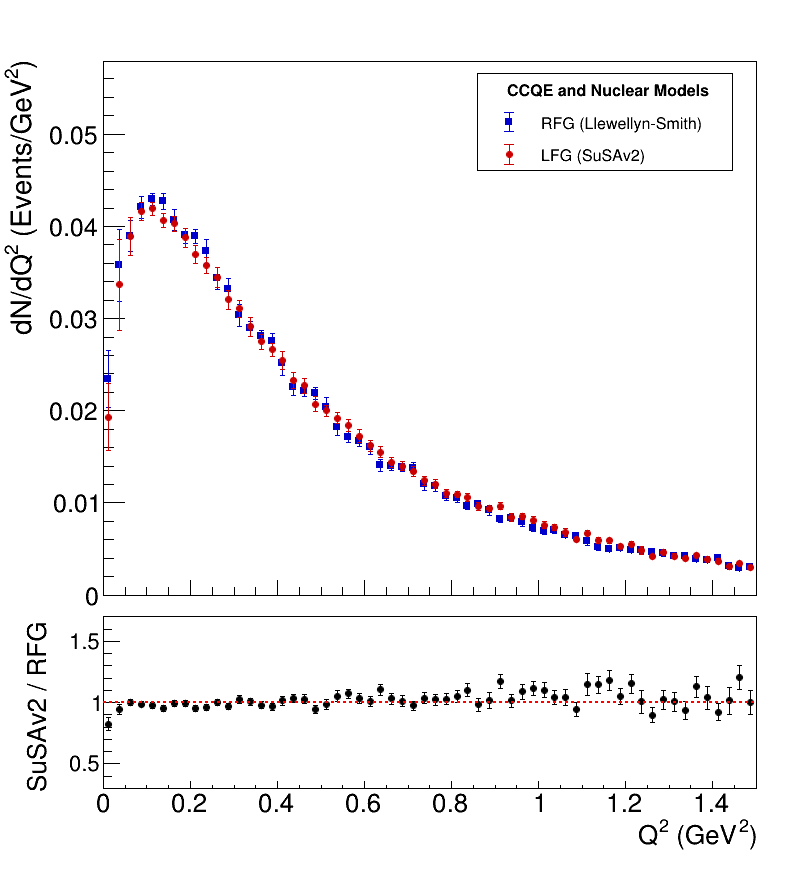}
        \caption{RFG vs. LFG (SuSAv2)}
        \label{fig:q2_susav2}
    \end{subfigure}
    \caption{Simulated $Q^2$ distributions for $\nu_\mu$ CCQE interactions. The bottom panels highlight the ratio between the advanced nuclear models and the RFG baseline. (a) The Nieves LFG model demonstrates a severe, characteristic RPA suppression at $Q^2 < 0.3 \text{ GeV}$, followed by a systematic enhancement at higher momentum transfers. (b) The SuSAv2 model exhibits a relatively flatter ratio, maintaining closer structural agreement with the RFG shape while globally normalizing the yield. (Author, 2026).}
    \label{fig:q2_ratios}
\end{figure}

To underscore the differences between the models and to discuss the importance for oscillation experiments, a simulation of the squared four-momentum transfer ($Q^2$) distribution is presented in \ref{fig:q2_ratios}. In the low-$Q^2$ regime, long-range nucleon-nucleon correlations become highly significant. The Random Phase Approximation (RPA) models these correlations. The comparison between the RFG and the Nieves LFG model (Figure \ref{fig:q2_lfg}) illustrates this effect. In the low momentum transfer region ($Q^2 < 0.3 \text{ GeV}^2$), the LFG model exhibits a suppression relative to the RFG, dropping to nearly $40\%$ of the uncorrected yield. As $Q^2$ increases beyond $0.4 \text{ GeV}$, the screening effect dissipates, and the LFG model induces an enhancement due to the altered available phase space derived from the local Fermi momentum distribution. Conversely, the SuSAv2 approach (Figure \ref{fig:q2_susav2}) provides a kinematic mapping that is indistinguishable from the RFG (Llewellyn-Smith) model.

\section{Differential Cross-Sections and Lepton Angular Distributions}

\begin{figure}[htbp]
    \centering
    \begin{subfigure}[b]{0.48\textwidth}
        \centering
        \includegraphics[width=1\linewidth]{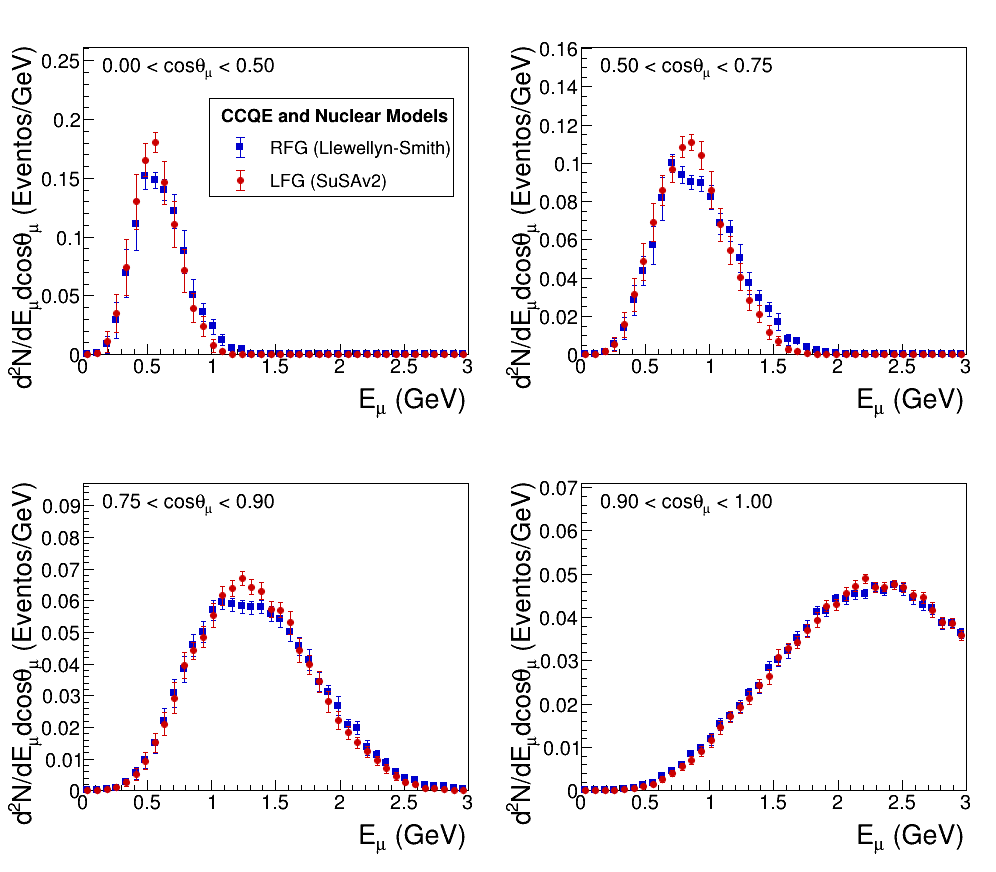}
        \caption{Angular slices: RFG vs. LFG (Nieves + RPA)}
        \label{fig:d2sigma_lfg}
    \end{subfigure}
    \hfill
    \begin{subfigure}[b]{0.48\textwidth}
        \centering
        \includegraphics[width=1\linewidth]{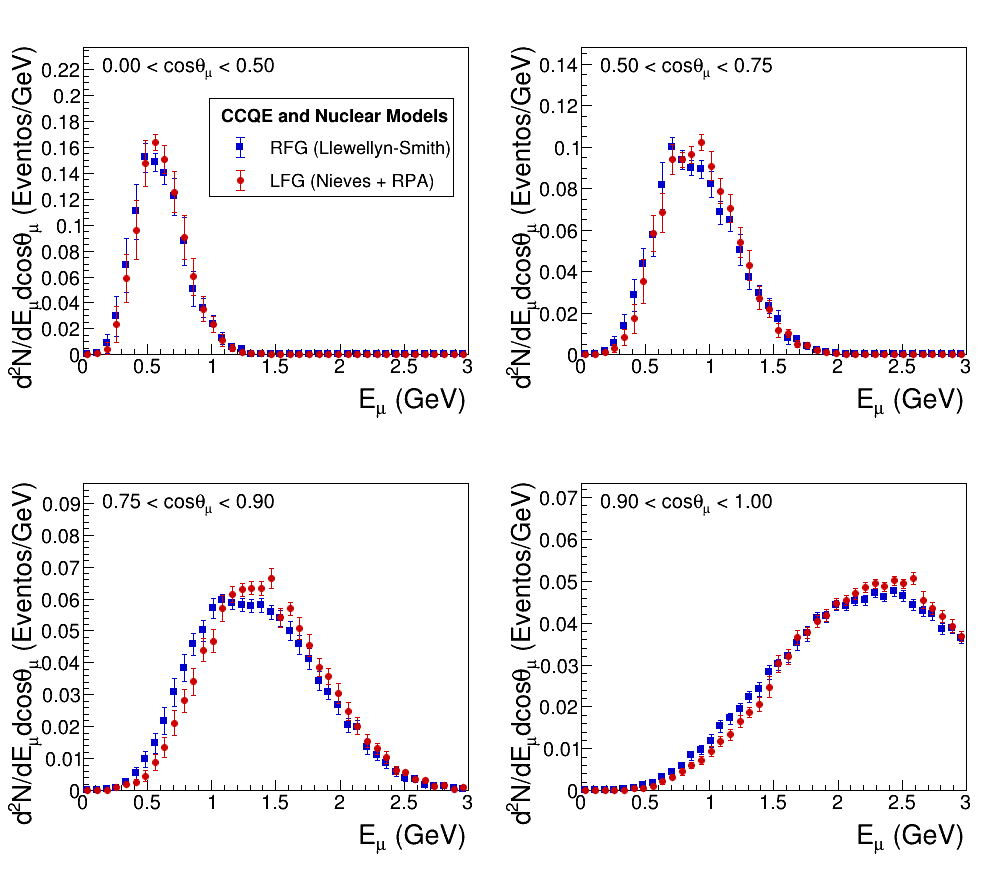}
        \caption{Angular slices: RFG vs. LFG (SuSAv2)}
        \label{fig:d2sigma_susav2}
    \end{subfigure}
    \caption{Simulated outgoing muon energy ($E_\mu$) spectra divided into four distinct scattering angle ($\cos\theta_\mu$) bins. The highly forward scattering bin ($0.90 < \cos\theta_\mu < 1.00$) corresponds to the low-$Q^2$ regime, where the models exhibit their most fundamental differences due to nuclear screening effects. (Author, 2026).}
    \label{fig:d2sigma_models}
\end{figure}

To further disentangle the kinematic implications of the nuclear models, it is necessary to examine the double-differential cross-section, specifically evaluating the outgoing lepton energy ($E_\mu$) mapped across different scattering angle ($\cos\theta_\mu$). Figure \ref{fig:d2sigma_models} presents the kinematic slices, comparing the RFG with the advanced formulations. The angular dependency confirms that the most forward scattering bin ($0.90 < \cos\theta_\mu < 1.00$) geometrically corresponds to events with the lowest momentum transfers. In this specific panel (Figure \ref{fig:d2sigma_lfg}, bottom-right), the red curve (LFG + RPA) exhibits an enhancement at the peak of the muon energy distribution relative to the blue RFG curve. As the scattering angle increases (moving towards transverse or backward scattering, corresponding to $\cos\theta_\mu < 0.90$), the momentum transfer $Q^2$ increases, and the RPA screening effect vanishes. Consequently, in the other angular bins, the LFG and SuSAv2 models overtake the RFG baseline, yielding higher interaction probabilities.

The simulation of these models in GENIE demonstrates that the transition from the Relativistic Fermi Gas to local nuclear frameworks is not merely an exercise in overall normalization, it redistributes the final-state kinematics of the observable particles. In particular, the notable variations observed in the $Q^2$ spectra and outgoing muon angular distributions highlight the dependence of predicted observables on the chosen nuclear model.

\newpage
\chapter[DIS Update]{Bodek-Yang Model Update}

The modeling of the DIS/SIS within the GENIE event generator relies on the Bodek-Yang (BY) formalism, as seen in previous sections. This section details an evaluation where the GENIE configuration is compared against the updated 2021 Bodek-Yang parameters \cite{bodek2021inelastic}, quantifying the impact of refined corrections on the expected DUNE cross-sections.

In the standard high-$Q^2$ DIS regime, interaction cross-sections are calculated using the Quark-Parton Model (QPM) in combination with Parton Distribution Functions (PDFs) governed by perturbative Quantum Chromodynamics (pQCD). However, a significant portion of the DUNE neutrino flux resides in the few-GeV transition region, often referred to as Shallow Inelastic Scattering (SIS). In this low-$Q^2$ and low-$W$ domain, standard pQCD calculations fail due to large non-perturbative effects. To bridge the gap between the resonance region and the pure DIS regime, GENIE employs the Bodek-Yang model, as already discussed. See section 4. This approach extends the validity of standard high-energy PDFs down to the limit ($Q^2 \to 0$). The model achieves this through two primary modifications to the standard formalism:

\begin{enumerate}
    \item \textbf{The Effective Scaling Variable ($\xi_w$):} The standard Bjorken scaling variable ($x = Q^2 / 2M\nu$) is replaced by a new kinematic variable, $\xi_w$, which accounts for the initial state target mass and transverse momentum of the quarks. It is defined as \cite{bodek2002modeling}:
    \begin{equation}
        \xi_w = \frac{Q^2 + M_f^2 + B}{M\nu \left[1 + \sqrt{1 + Q^2/\nu^2}\right] + A},
        \label{eq:xi_w}
    \end{equation}
    where $M_f$ is the mass of the final state quark, $\nu$ is the energy transfer, and $M$ is the nucleon mass. The empirical parameters $A$ and $B$ account for higher-order QCD terms.
    
    \item \textbf{Multiplicative $K$-factors:} At very low $Q^2$, the vector structure function $\mathcal{F}_2$ must vanish to preserve the conservation of the electromagnetic current. To enforce this physical constraint, the PDFs are multiplied by empirical $K$-factors. For instance, the vector $K$-factor for valence quarks is parameterized using the nucleon dipole form factor $G_D(Q^2)$ \cite{bodek2002modeling}:
    \begin{equation}
        K_{valence}^{vector}(Q^2) = [1 - G_D^2(Q^2)] \left( \frac{Q^2 + C_{v2}}{Q^2 + C_{v1}} \right),
        \label{eq:k_factor_val}
    \end{equation}
    where $C_{v1}$ and $C_{v2}$ are tunable parameters derived from charged-lepton scattering data.
\end{enumerate}

By evaluating the leading-order PDFs at $\xi_w$ and applying these $K$-factors, the Bodek-Yang model provides a continuous description of the inelastic cross-section across the entire DUNE kinematic phase space.

\section{The 2021 Parameter Update}

The default GENIE configuration relies on an earlier iteration of the Bodek-Yang model, which utilizes the GRV98 parton distribution functions. Specifically, the scaling variable parameters are set to $A=0.538 \text{ GeV}^2$ and $B=0.305 \text{ GeV}^2$, with the valence $K$-factors characterized by $C_{v1u}=0.291 \text{ GeV}^2$ and $C_{v2u}=0.189 \text{ GeV}^2$. However, recent re-evaluations, incorporating modern high-precision electron, muon, and photo-production scattering datasets, have led to a significant refinement of the model. In this updated iteration, the enhanced target mass terms are determined to be $A = 0.621 \text{ GeV}^2$ and $B = 0.380 \text{ GeV}^2$. The increase in these parameters shifts the effective momentum fraction ($\xi_w$), thereby altering the cross-section primarily in the low-$Q^2$ and low-$W$ boundaries of the inelastic continuum. Furthermore, the $K$-factor parameters for the up and down quarks have been decoupled and updated ($C_{v1d}=0.341$, $C_{v1u}=0.417$, $C_{v2d}=0.323$, $C_{v2u}=0.264 \text{ GeV}^2$ \cite{bodek2021inelastic}), modifying the threshold suppression of the structure functions.

\section{Computational Analysis of the DIS Cross-Section}

\begin{figure}[htbp]
    \centering
    \includegraphics[width=0.75\linewidth]{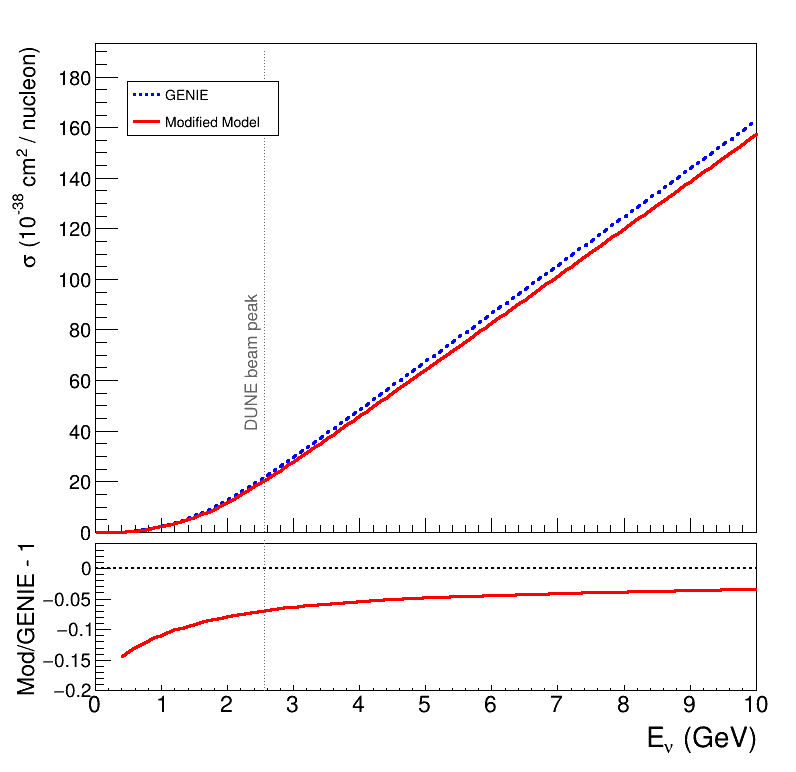}
    \caption{Integrated $\nu_\mu$ Deep Inelastic Scattering (DIS) cross-section per nucleon as a function of incident neutrino energy ($E_\nu$). The top panel compares the baseline GENIE prediction utilizing the legacy Bodek-Yang parameters (blue dotted line) against the modified model implementing the 2021 parameter update (red solid line). The bottom panel displays the relative fractional difference between the models. A systematic, energy-dependent suppression is observed in the modified model, reflecting the impact of the updated higher-twist coefficients and enhanced target mass parameters. The vertical line designates the peak flux energy of the DUNE experiment at approximately 2.5 GeV. (Author, 2026).}
    \label{fig:dis_cross_section_ratio}
\end{figure}

Therefore, to quantitatively evaluate the impact of this refinement on neutrino scattering, the default GENIE Bodek-Yang implementation was compared against a modified simulation incorporating the 2021 updated parameters. The integrated cross-section per nucleon was calculated for the muon neutrino DIS channel across the energy spectrum relevant to DUNE ($0 - 10$ GeV). The computational results are presented in Figure \ref{fig:dis_cross_section_ratio}.

The top panel of Figure \ref{fig:dis_cross_section_ratio} illustrates the characteristic linear rise of the DIS cross-section as a function of the incident neutrino energy. While both models follow the expected fundamental behavior derived from the Quark-Parton Model, a clear divergence in the absolute normalization is visible. The modified model incorporating the 2021 parameters (red solid line) predicts a systematically lower cross-section across the entire evaluated energy spectrum compared to the GENIE baseline (blue dotted line). The magnitude of this theoretical shift is explicitly quantified in the bottom panel, which plots the relative fractional difference between the two configurations. The suppression is highly energy-dependent, being most pronounced near the kinematic threshold of the inelastic continuum. At $E_\nu = 1.0$ GeV, the updated parameters induce a deficit of approximately $12\%$ relative to the default GENIE tune. As the neutrino energy increases and the influence of the non-perturbative corrections (higher-twist and target mass effects) diminishes, the models begin to converge. However, even in the deep continuum, a residual difference persists. Crucially, at the peak energy of the DUNE wide-band flux ($E_\nu \approx 2.5$ GeV), the update suppresses the DIS cross-section by approximately $7\%$.

These findings highlight the profound sensitivity of the simulation framework to the phenomenological parameters governing the SIS region. Utilizing the outdated default parameters may lead to an overestimation of the inelastic background in the oscillation sample. Implementing the modernized 2021 Bodek-Yang parameters provides a more rigorously constrained foundation, ensuring that the simulated hadronic showers and energy deposition profiles reflect the most current global scattering data.

\newpage
\chapter[Scientific Outreach]{University Extension and Scientific Outreach}

The Masters Program in Physics at the Federal University of Alfenas (UNIFAL-MG), located in Poços de Caldas, incorporates university extension activities as a requirement to fill the gap between academic research and the local community. Engaging in scientific outreach is crucial in the field of high-energy physics, where the abstract nature of subatomic phenomena and the scale of experimental facilities can often seem inaccessible to the general public. The main component of this extension work involved the design, production, and presentation of educational materials for the \textit{Mais Ciência Por Favor} event. This initiative brought secondary school students from various local schools in Poços de Caldas to the university laboratories, providing them a interactive experience of the science produced within the institution \cite{unifal_mais_ciencia}. To translate the complex methodology of particle physics into an accessible format, two complementary macroscopic models were developed: a 3D-printed Linear Accelerator (LINAC) and a macroscopic Scattering Experiment model. 

\section{Connecting Acceleration and Detection in the Classroom}

In experimental particle physics, the investigative pipeline relies on accelerating a particle beam to high energies and subsequently colliding it with a target to probe the fundamental structure of matter. The outreach activities mirrored this exact pipeline using 3D-printed macroscopic analogs.

\subsection{The 3D-Printed LINAC Model}
\begin{figure}
    \centering
    \includegraphics[width=0.65\linewidth]{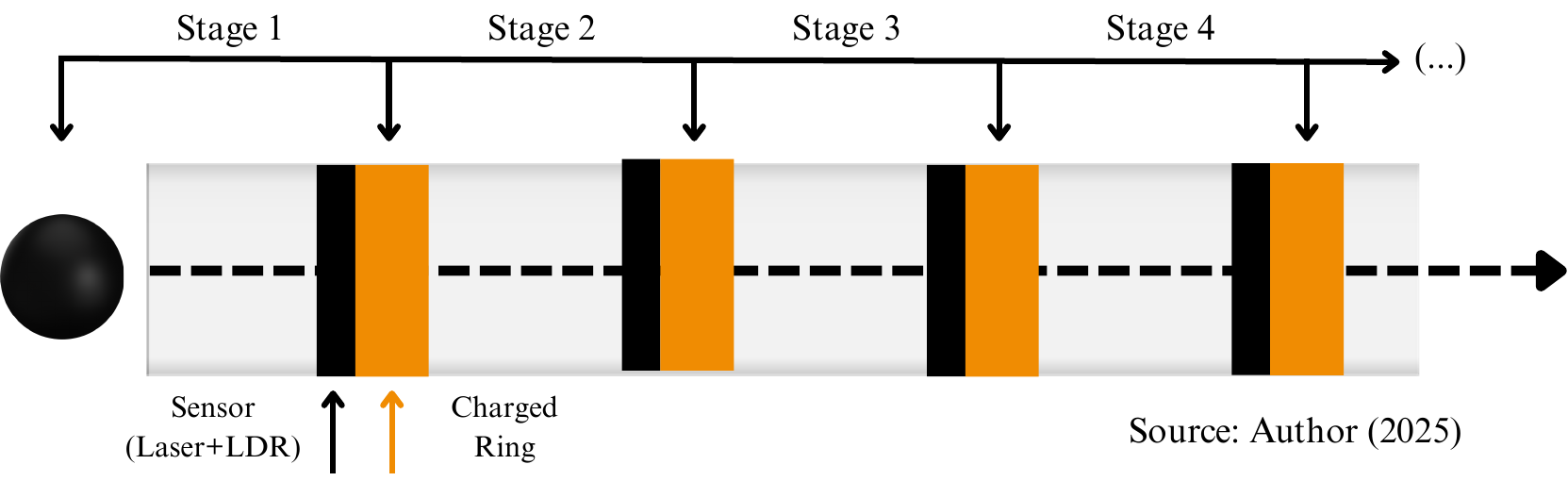}
    \caption{Graphical representation of a 3D-printed model linear accelerator. It consist in a plexiglass tube, a graphite coated ping pong ball, a sensor composed by a laser and a LDR represented by the black range, and a charged ring represented by the orange range. The arrow pointing to the right in the figure represents the direction that the ball will accelerate. The electric field between the ball and the ring and the stage 1 gives the ball an amount of kinetic energy, as it pass through the sensor the polarity of the field inverts, consequently consequently taking the ball to the second stage and so on  (Author).}
    \label{fig:LINACScheme}
\end{figure}

\begin{figure}
    \centering
    \includegraphics[width=0.65\linewidth]{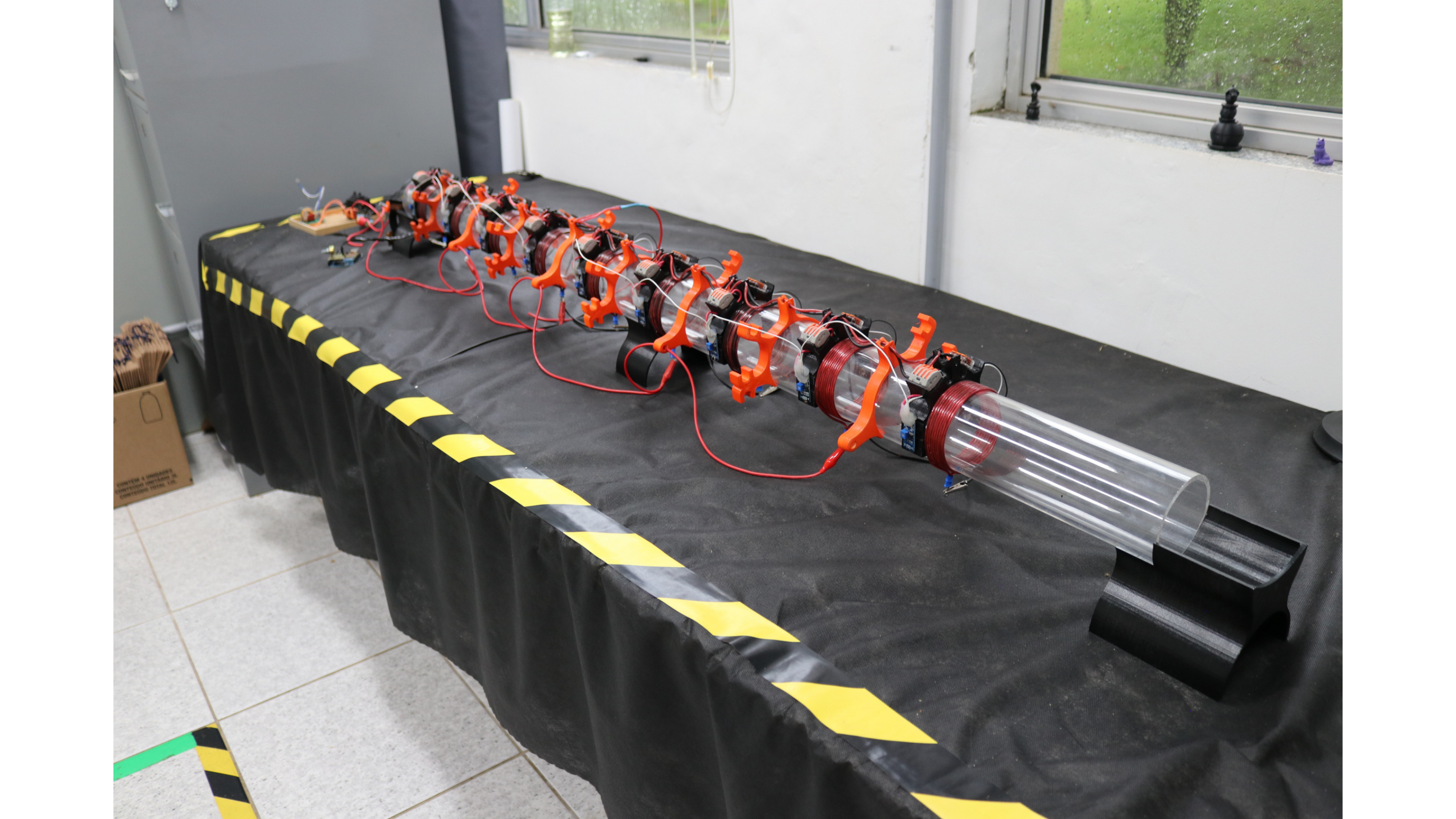}
    \caption{Real picture of the 3D-printed linear accelerator constructed for the ``Mais Ciencia Por Favor" event, where the Federal University of Alfenas (UNIFAL-MG) brought students of Poços de Caldas-MG to the university to encourage the scientific outreach (Author).}
    \label{fig:LINACPic}
\end{figure}

The first step involves particle acceleration. To demonstrate this, an interactive, 3D-printed LINAC model, originally designed by CERN's S'Cool LAB \cite{ScoolLabLINAC}, was assembled and utilized. The model uses a high-voltage potential difference across conductive copper tape segments within a plexiglass tube. The Figure \ref{fig:LINACScheme} demonstrates this construction. As a ping-pong ball coated with conductive material passes through the tube, it triggers light-dependent resistors that swhich the electric fields, accelerating the ball through the stages. This macroscopic setup provides a direct visual representation of how linear accelerators utilize synchronized electromagnetic fields to increase kinetic energy to charged particles. Figure \ref{fig:LINACPic} is a picture of the completed constructed LINAC.

\subsection{The Scattering Experiment}
\begin{figure}[htpb]
    \centering
    
    \begin{subfigure}[b]{0.45\textwidth}
        \centering
        \includegraphics[width=\textwidth]{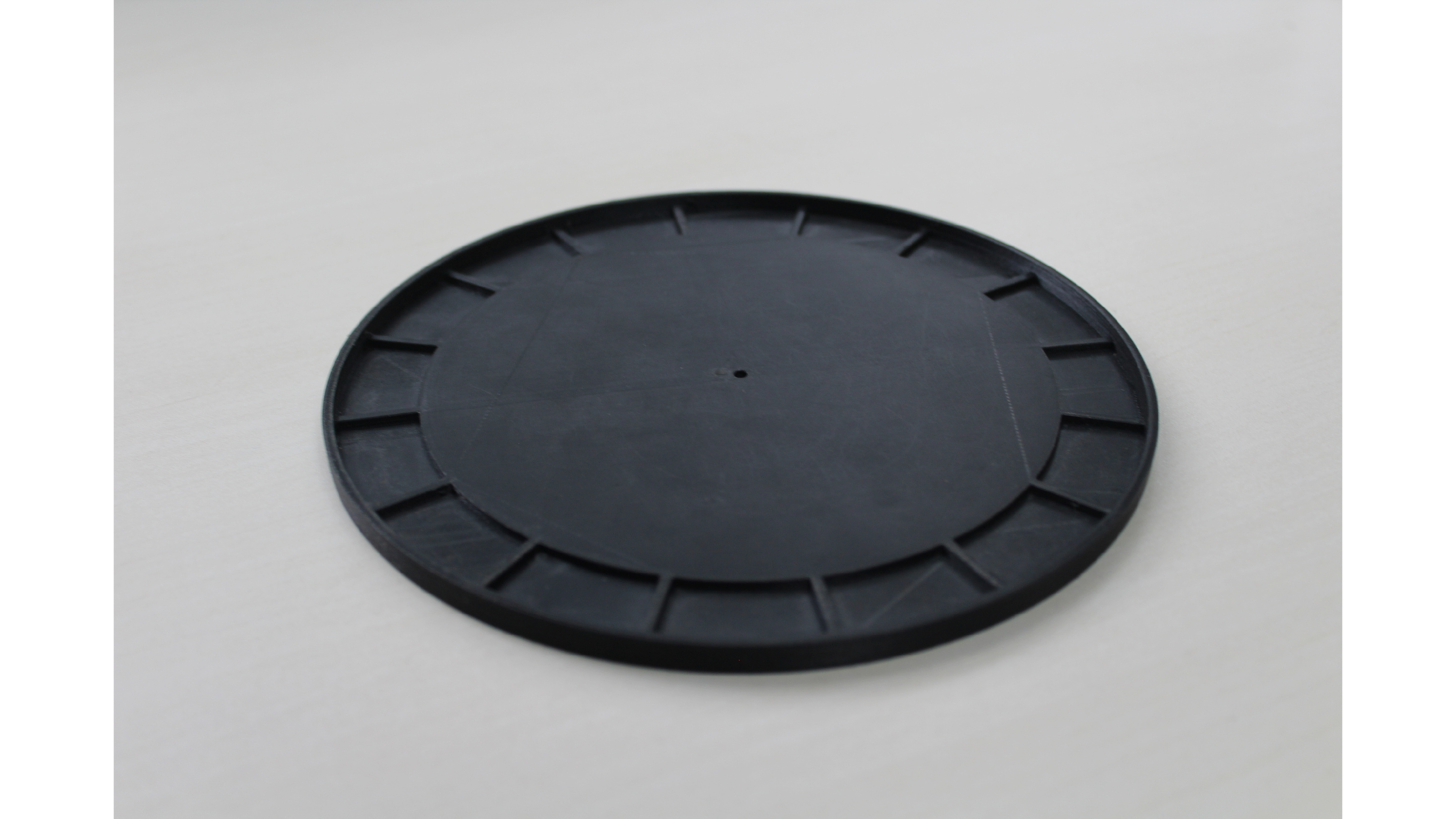}
        \caption{Base}
        \label{fig:sub_a}
    \end{subfigure}
    \hfill
    \begin{subfigure}[b]{0.45\textwidth}
        \centering
        \includegraphics[width=\textwidth]{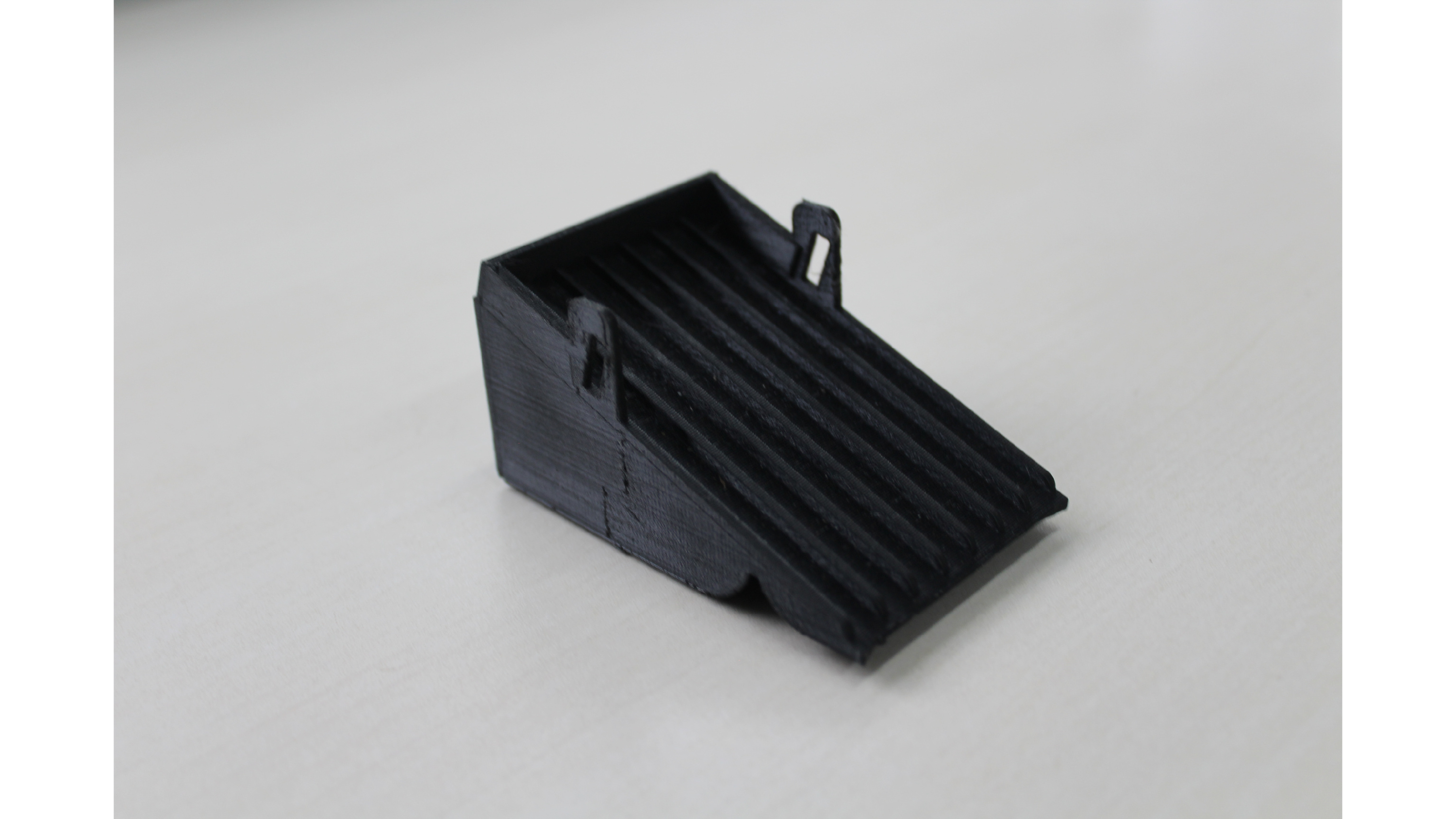}
        \caption{Ramp}
        \label{fig:sub_b}
    \end{subfigure}
    
    \vspace{0.5cm} 
    
    \begin{subfigure}[b]{0.45\textwidth}
        \centering
        \includegraphics[width=\textwidth]{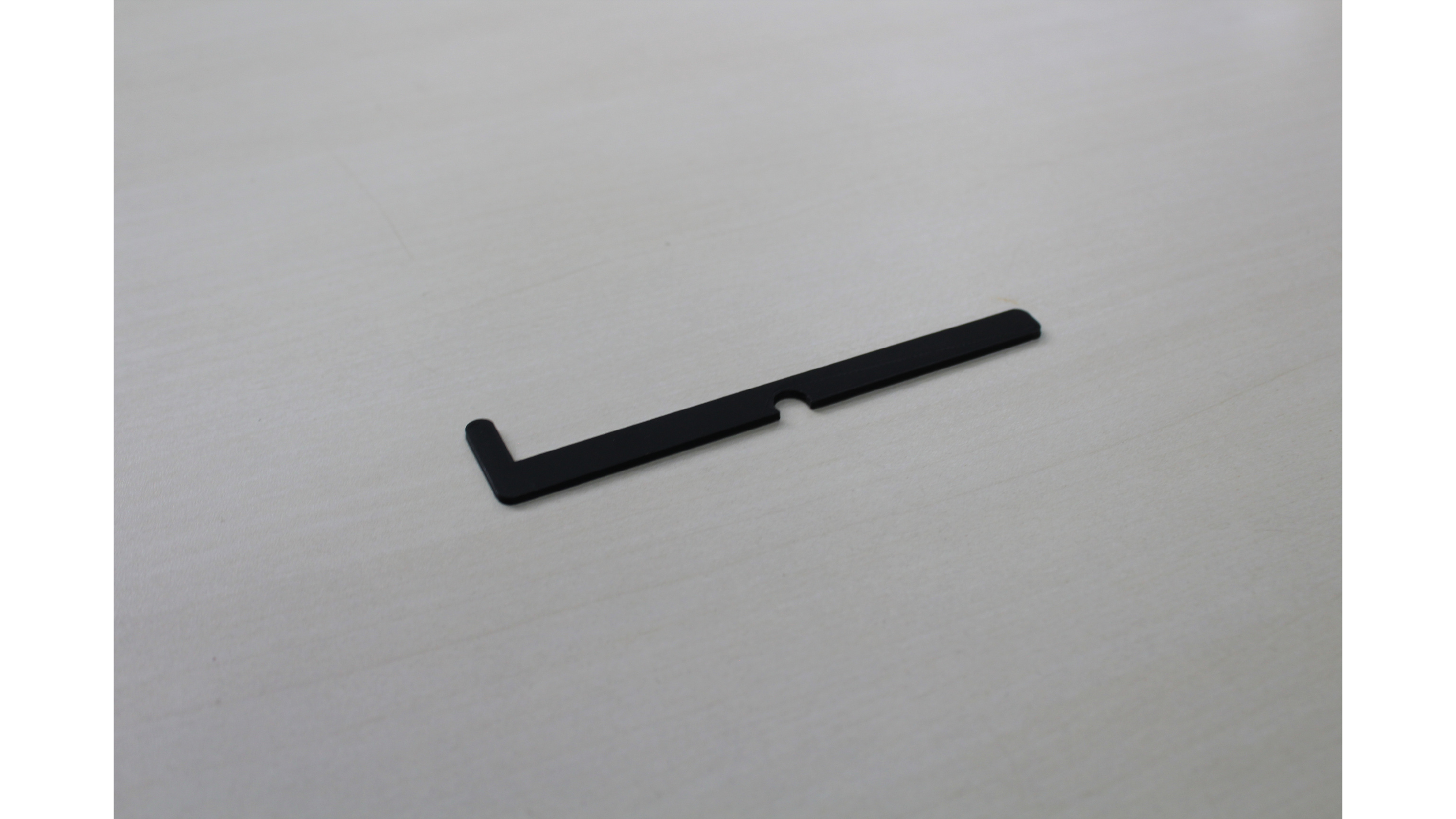}
        \caption{Slider}
        \label{fig:sub_c}
    \end{subfigure}
    \hfill
    \begin{subfigure}[b]{0.45\textwidth}
        \centering
        \includegraphics[width=\textwidth]{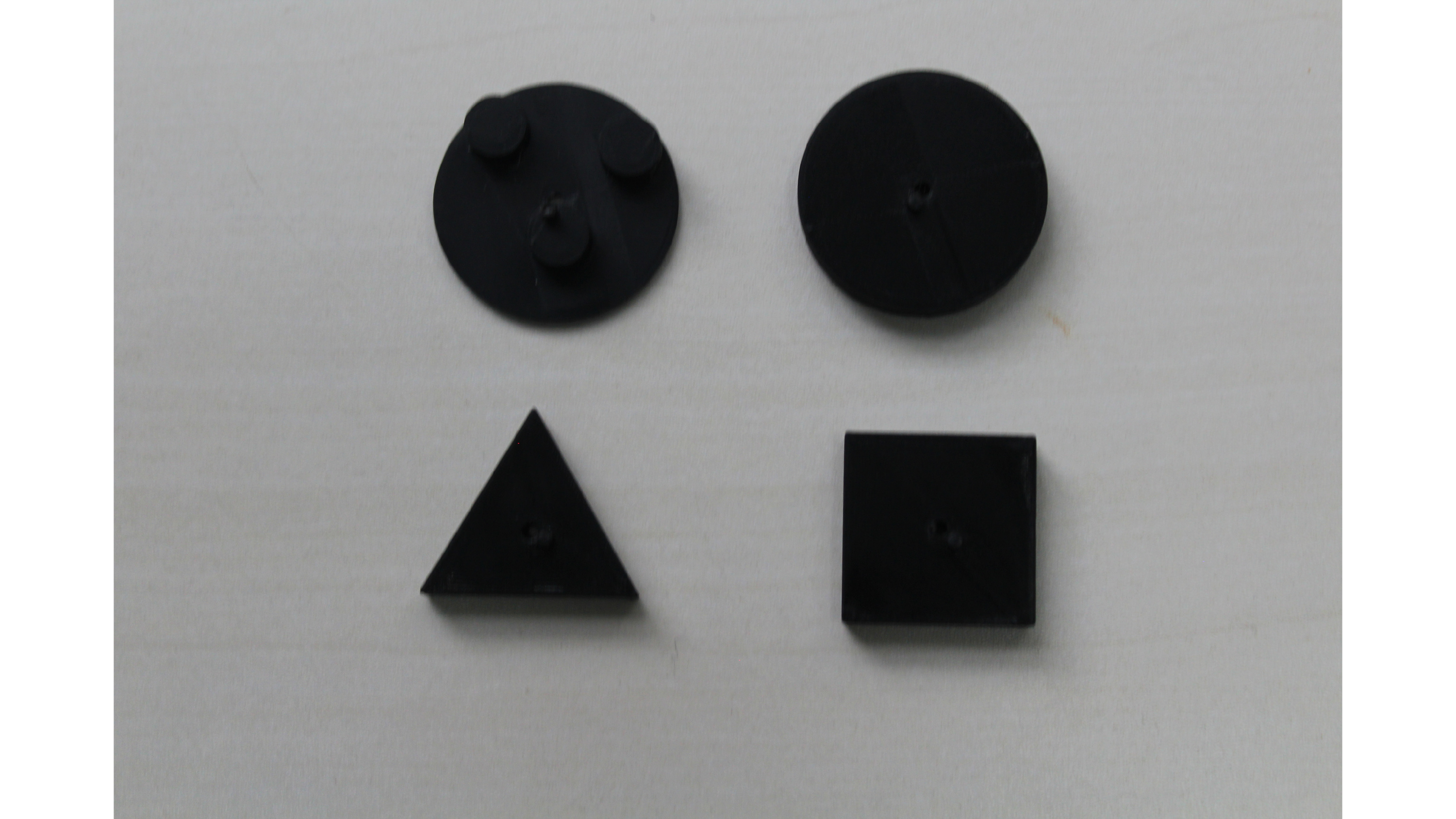}
        \caption{Objects}
        \label{fig:sub_d}
    \end{subfigure}
    
    \vspace{0.5cm} 
    
    \begin{subfigure}[b]{0.45\textwidth}
        \centering
        \includegraphics[width=\textwidth]{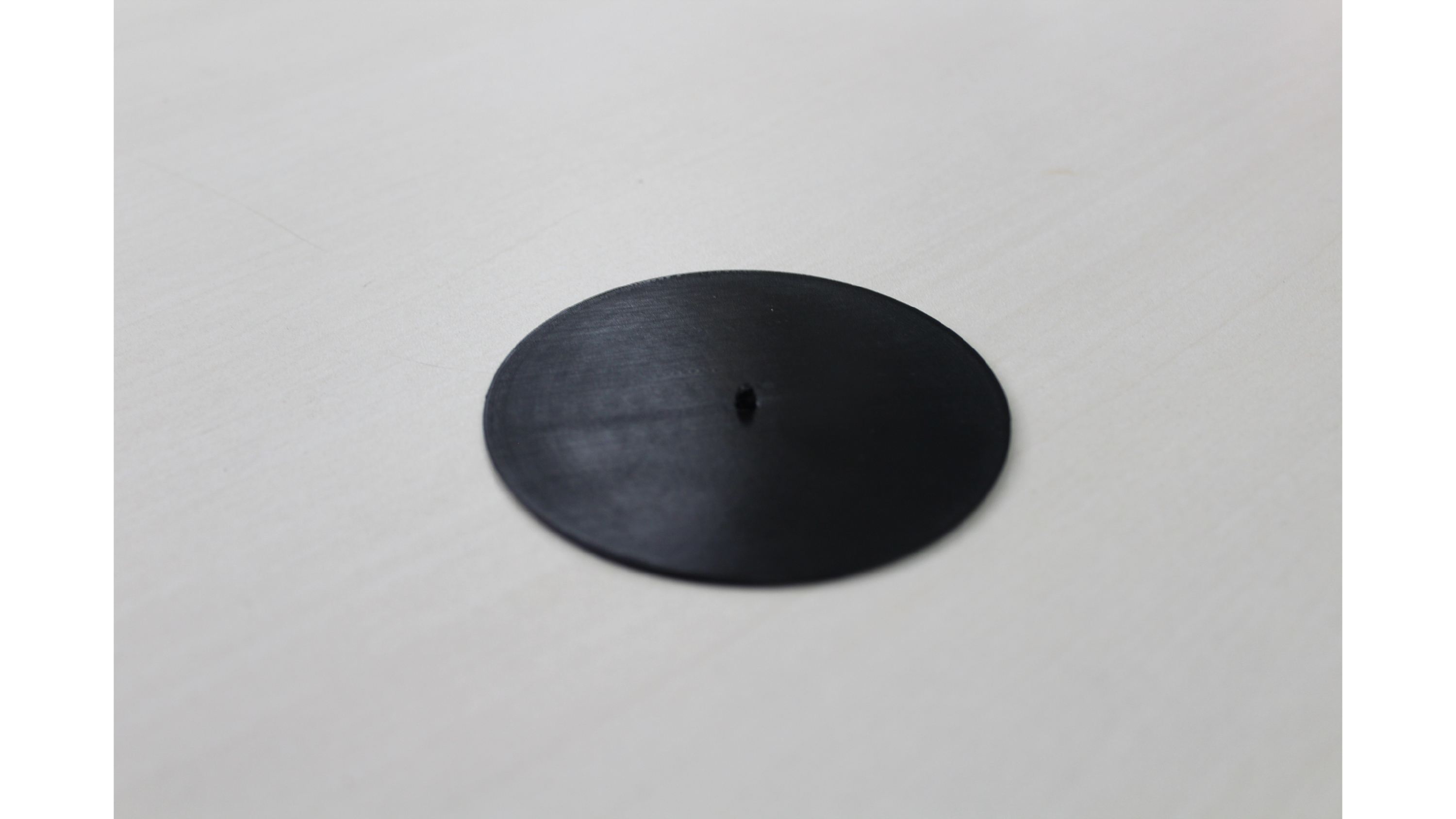}
        \caption{Lid}
        \label{fig:sub_e}
    \end{subfigure}
    \hfill
    \begin{subfigure}[b]{0.45\textwidth}
        \centering
        \includegraphics[width=\textwidth]{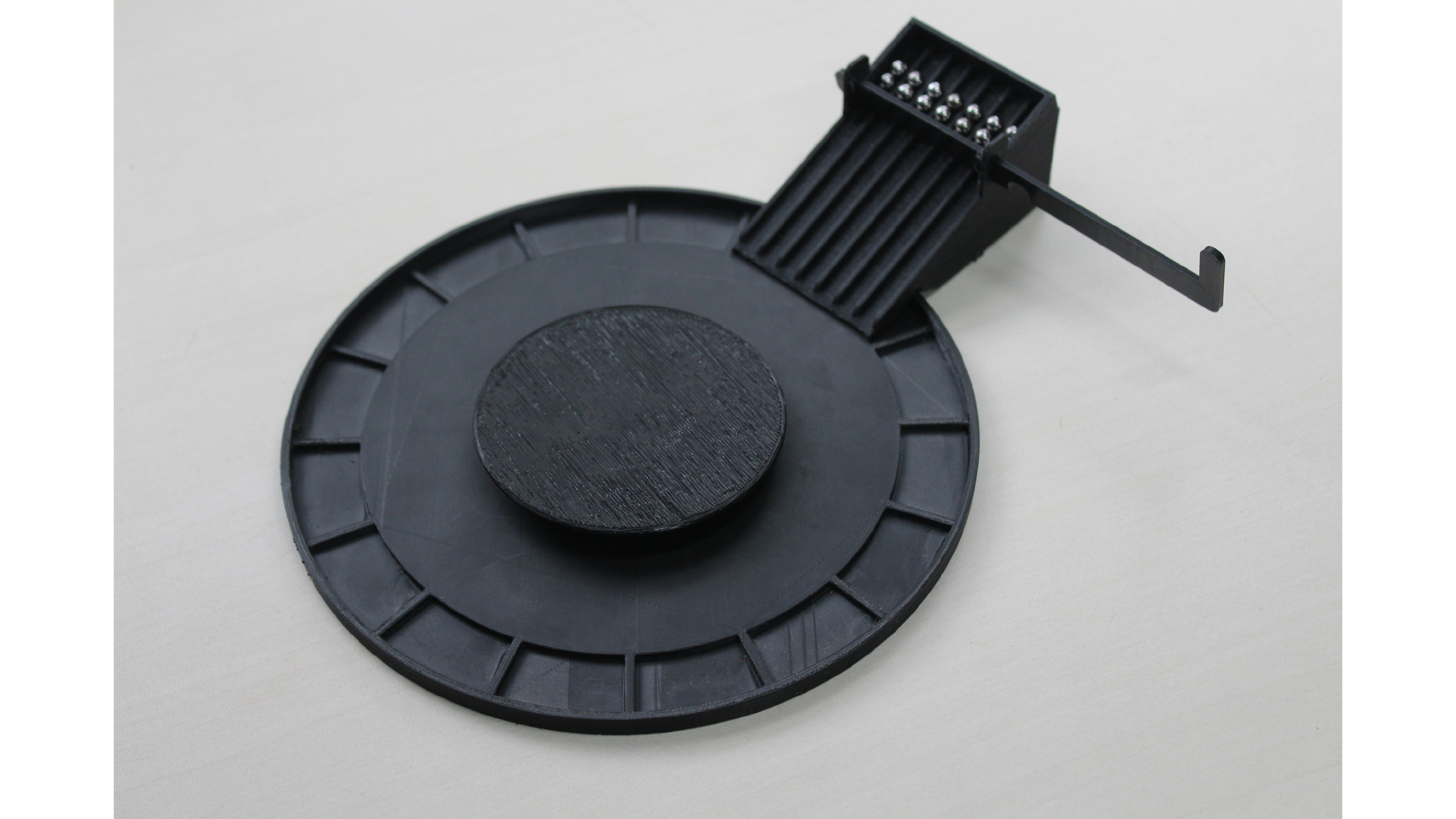}
        \caption{Assembled Experiment}
        \label{fig:sub_f}
    \end{subfigure}
    
    \caption{Pictures of the 3D-printed Scattering Experiment components. (a) serves as the base for the experiment, containing a hole in the center to fit the scattering target and cavities on the edges to collect the scattered balls. (b) is the ramp, utilized for placing the 5mm steel balls, have cavities on both sides to fit the slider. (c) is the slider, an component that, when fitted into the ramp, allows to systematically control the start of the run. (d) are the scattering objects, which are the main component in the experiment, placed in the middle of the base for the balls to scatter. The objects are separated in four objects being a circular cylinder, a cuboid, a triangular prism and a substructre. (e) is the Lid, an component that offers a great educational value, since it allows the conductor of the experiment to hide the scattering objects by placing the lid on top of it. (f) is the Assembled experiment. (Author).}
    \label{fig:ScatteringExperiment}
\end{figure}

Once particles are accelerated, they are used as probes. To illustrate the principles of particle detection and target analysis, a scattering experiment was constructed. This analog model, also developed by CERN's S'Cool LAB, consists in 3D-printed components, such as a base, a ramp, a slider, four scattering objects and a lid. All represented respectivelly in Figure \ref{fig:ScatteringExperiment}. The run is given by the launch of steel balls through the ramp, that collides with hidden object placed under a covering lid, that scatter off the balls that are collected in the cavities of the base. By observing the deflection angles and the scattering patterns of the steel balls as they exit the target area, students can infer the shape and size of the hidden object without directly seeing it. This hands-on activity introduces students to the scientific methodology of scattering and cross-sections, allowing them to perform both qualitative and semi-quantitative analyses.

\begin{figure}
    \centering
    \includegraphics[width=0.75\linewidth]{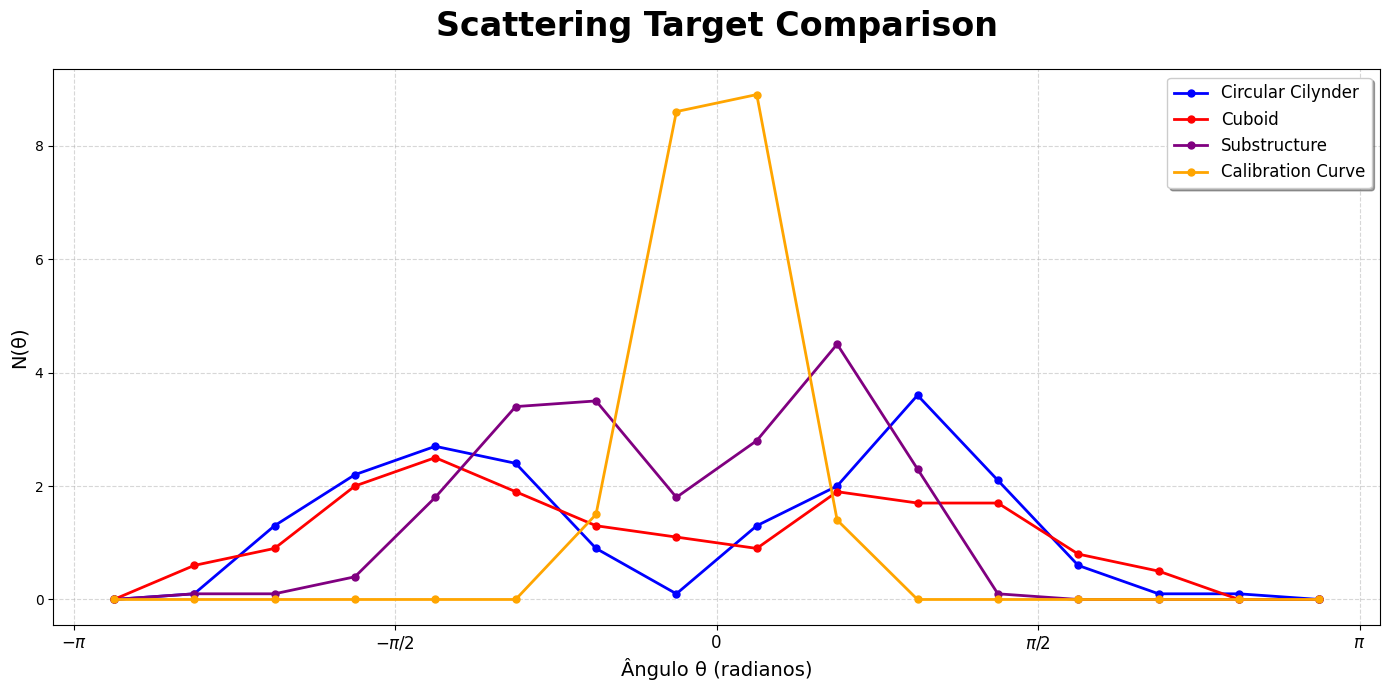}
    \caption{Scattering comparison in the 3D-printed Scattering experiment between the targets: circular cylinder (blue), cuboid (red), substructre (purple) and the calibration curve (yellow) that consists in the run of the experiment without target. The experiment run consists in 5mm steel balls accelerated with a ramp, towards the center of a base that contains the target. The collision between the balls and the target create the scattering pattern (Author).}
    \label{fig:ScatteringComparison}
\end{figure}

 To encompass students of all levels of education, the activity was divided into three distinct levels of complexity. The first tier consists of a qualitative analysis aimed at students with no prior knowledge of particle physics, seeking to introduce fundamental scattering concepts intuitively. In this stage, the emphasis is placed on the observation of scattering patterns and the inference of the target's properties without relying on complex mathematical formulations. During the activity, an educator conceals a specific geometric target under a lid. As a limited number of steel spheres are launched down the ramp, students must observe the deflection angles and formulate hypotheses regarding the hidden object's shape. This exercise emphases the distinction between direct observation and scientific inference, simulating how physicists investigate invisible subatomic structures. The second pedagogical level introduces a semi-quantitative approach, designed for students with basic experience in data collection and graphical analysis. This stage deepens on the comprehension of the scattering phenomena by requiring a more rigorous experimental procedure without demanding detailed theoretical cross-section calculations. Students perform at least ten launches for each scattering object to calculate the average number of spheres scattered at specific angles. The collected data is systematically organized into tables and subsequently represented using histograms, plotting the number of scattered spheres as a function of the scattering angle. By comparing these graphical representations, for instance, as demonstrated in Figure \ref{fig:ScatteringComparison}, students can conduct a discussion on how the geometry of the target directly influences the angular distribution of the projectiles.

\noindent The final level consists of a quantitative analysis intended for advanced students who possess greater mathematical rigor and prior familiarity with the concepts of scattering and cross-sections. The focus of this stage is the application of a theoretical framework to allow a direct comparison between theoretical predictions and the acquired experimental data. Students utilize classical mechanics to determine the impact parameter and the differential cross-section. This theoretical formalism enables the derivation of the expected number of scattered particles as a function of the angle, effectively bridging the gap between theoretical calculations and the experimental histograms generated in the previous stage.

\section{``Mais Ciência Por Favor" Festival}
The 3D-printed LINAC and scattering experiment models were implemented during the \textit{Mais Ciência Por Favor} festival at UNIFAL-MG, engaging secondary school students in hands-on activities. This initiative demonstrated significant pedagogical value by combining these macroscopic analogs with fundamental physics concepts, effectively catching the attention of the local community and demystifying the principles of particle physics.

\begin{figure}
    \centering
    \includegraphics[width=0.75\linewidth]{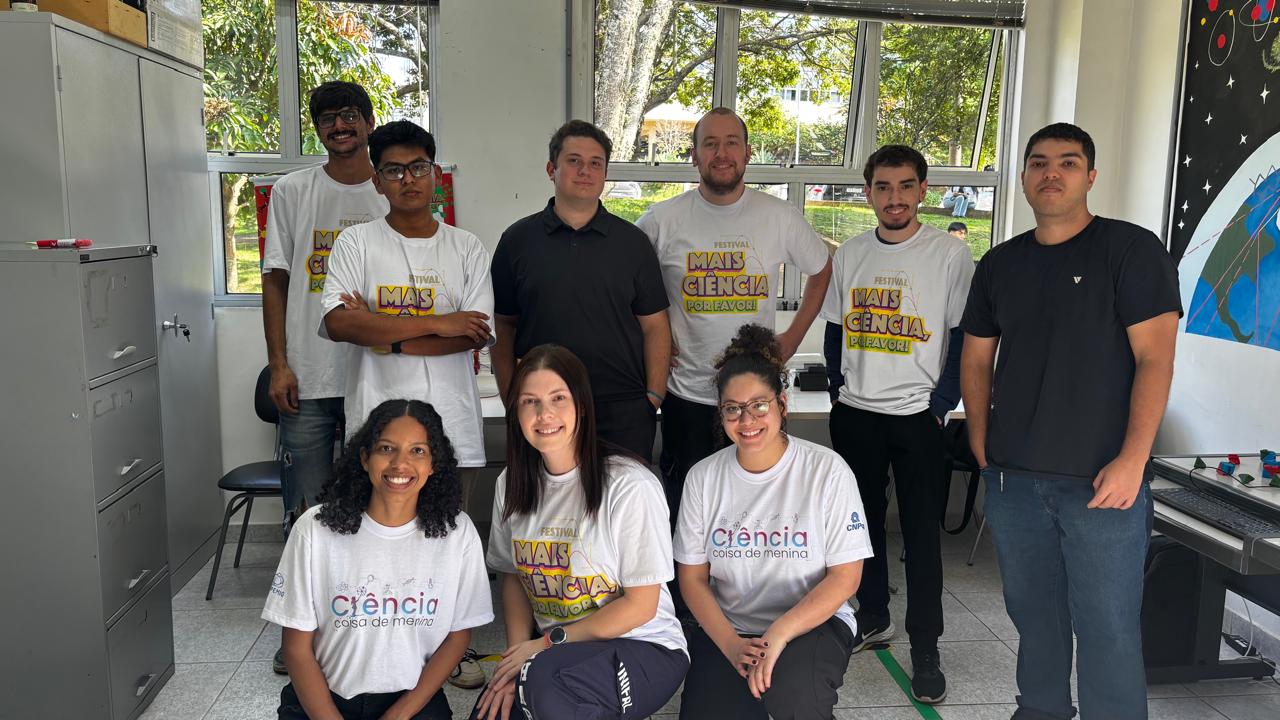}
    \caption{Picture of the collaboration presenting the projects of the Nulab and QUANTUM Lab at the ``Mais Ciência Por Favor" Festival (Author).}
    \label{fig:MCPF}
\end{figure}

\section{Presentation at IPPOG Brazil Conference (2025)}
The methodology, educational framework, and results of this outreach initiative were subsequently presented at the International Particle Physics Outreach Group (IPPOG) Brazil conference in 2025. This presentation highlighted the role of accessible 3D-printed models in promoting scientific literacy, inspiring future scientists, and fulfilling the university's mandate for public social engagement.

\section{Presentation at IPPOG Brazil Conference (2025)}

The implementation of these interconnected models during the \textit{Mais Ciência Por Favor} event demonstrated their high pedagogical value in education. The integration of 3D modeling technologies with fundamental physics concepts effectively demystified the operation of particle accelerators and detectors. The methodology and framework of this outreach initiative were subsequently presented at the International Particle Physics Outreach Group (IPPOG) Brazil conference (2025), reinforcing the critical role of accessible analog models in inspiring the next generation of scientists and fulfilling the university's mandate for social engagement.

\newpage
\chapter{Conclusion}

The transition to a high-precision era in neutrino physics relies on the ability to accurately reconstruct incident neutrino energies from complex final-state topologies. As established in the introduction of this work, the necessary utilization of heavy nuclear targets, specifically Liquid Argon ($^{40}\text{Ar}$), introduces profound theoretical and experimental challenges. Within the dense nuclear environment, initial-state nucleon dynamics, multi-nucleon correlations (such as Meson Exchange Currents), and Final State Interactions systematically distort experimental observables and threaten to bias the extraction of fundamental parameters like Charge-Parity violation. To confront these challenges, this dissertation systematically evaluated the theoretical frameworks governing neutrino-nucleus interactions within the GENIE Monte Carlo event generator. By conducting a rigorous comparative study between established baseline models and modern theoretical implementations, this research quantified the phenomenological impact of advanced nuclear modeling on DUNE's expected cross-sections and event rates.

The investigation into the initial-state dynamics revealed the severe limitations of the standard Relativistic Fermi Gas (RFG) model. The reliance on a nuclear parameter, called axial mass, to reconcile heavy-target data anomalies was shown to be an artifact masking incomplete nuclear dynamics. The evaluated transition to the Local Fermi Gas model, supplemented with Random Phase Approximation and the Nieves modelling, proved to be a vital advancement. These frameworks properly account for the radial density of the nucleus and explicitly simulate long-range nucleon correlations, naturally suppressing the cross-section at low momentum transfers and redistributing the kinematic phase space without relying on unphysical parameter inflation.

Furthermore, this work directly addressed the interaction channels responsible for major background topologies, specifically resonant pion production and Deep Inelastic Scattering. Implementing modernized phenomenological constraints, such as the updated $M_A^{RES} = 0.96 \text{ GeV/c}^2$ and the 2021 Bodek-Yang parameters, demonstrated systematic suppressions in the predicted event rates. As emphasized initially, mismodeling the pion production vertices compromises the uncertainties introduced by FSI. Therefore, adopting these rigorously constrained models is an absolute prerequisite to prevent the misclassification of event topologies and to mitigate the severe biases in calorimetric energy reconstruction that threaten DUNE's oscillation sensitivity.

Beyond the computational and theoretical physics landscape, this master's program successfully fulfilled its university extension mandate. The abstract methodologies of particle physics and scattering kinematics were translated into tangible educational tools through the design, production, and presentation of 3D-printed macroscopic models. The interactive Linear Accelerator analog and the scattering experiment successfully engaged secondary school students, effectively demystifying the scientific method and fulfilling the university's commitment to societal scientific literacy.

\section*{Next Steps and Future Work}

While this dissertation has established a robust evaluation of advanced nuclear frameworks, such as the LFG model + RPA , and modernized DIS parameters, mitigating theoretical uncertainties in neutrino-nucleus interactions remains an ongoing task. Building upon the computational foundation laid in this study, future research will focus on the direct implementation and fine-tuning of comprehensive multi-nucleon interaction models, specifically two-particle-two-hole and Meson Exchange Currents, directly within the GENIE source code, optimizing them for Liquid Argon targets. Furthermore, subsequent efforts must prioritize the validation of these simulated cross-sections against emerging high-precision scattering data from LArTPC detectors, such as those from DUNE's Near Detector prototypes. This empirical validation is critical to quantitatively assess the impact of Final State Interactions and missing energy on DUNE's specific calorimetric energy reconstruction algorithms. 

Finally, the university extension and scientific outreach initiatives developed herein will be expanded. To maximize the pedagogical impact of the macroscopic 3D-printed LINAC and scattering experiment models, future activities will include the production and publication of digital educational content on platforms such as YouTube and Instagram. Translating these physical analogs into accessible multimedia resources will broaden the reach of particle physics education, continuing to bridge the gap between advanced academic research and public scientific literacy.

\label{sec:conclusion}

\clearpage
\bibliography{Bibliography}

@article{neutrinoHistory,
  title={Neutrino. History of a unique particle},
  author={Bilenky, SM},
  journal={The European Physical Journal H},
  volume={38},
  number={3},
  pages={345--404},
  year={2013},
  publisher={Springer}
}

@article{MestradoGustavo,
author = {Valdiviesso, Gustavo and Guzzo, M.},
year = {2005},
month = {12},
pages = {495-506},
title = {Compreendendo a Oscilação de Neutrinos},
volume = {27},
journal = {Revista Brasileira de Ensino de Física},
doi = {10.1590/S1806-11172005000400002}
}

@misc{nu-fit-data,
  author       = {{The Nu-Fit Collaboration}},
  title        = {{Nu-FIT}},
  howpublished = {\url{http://www.nu-fit.org}},
  year         = {2025},
  note         = {Accessed: August 1, 2025}
}

@misc{cern_sm,
  author = {{CERN}},
  title = {The Standard Model},
  year = {2025},
  url = {https://home.cern/science/physics/standard-model},
  urlaccessdate = {23 nov. 2025},
  note = {European Organization for Nuclear Research}
}

@misc{fnal_dune_layout,
    author = "{Fermi National Accelerator Laboratory (FNAL)}",
    title = "{Deep Underground Neutrino Experiment - Layout and Oscillation Probability}",
    year = "2025",
    url = "https://lbnf-dune.fnal.gov/",
    note = "[Acessado em 2 de março de 2026]"
}

@misc{sacerdoti2021vdtpc,
    author = "Sacerdoti, Sabrina and {DUNE Collaboration}",
    title = "{A LArTPC with Vertical Drift for the DUNE Far Detector}",
    howpublished = "Apresentação no \textit{NuFact 2021: The 22nd International Workshop on Neutrinos from Accelerators}",
    year = "2021",
    month = "Sep",
    publisher = "CERN Indico",
    url = "https://indico.cern.ch/event/855372/contributions/4450651/attachments/2304163/3919805/vdtpc_nufact.pdf",
    note = "[Acessado em 2 de março de 2026]"
}

@book{griffiths2020introduction,
  title={Introduction to elementary particles},
  author={Griffiths, David},
  year={2020},
  publisher={John Wiley \& Sons}
}

@article{NaturalUnits,
  title={Understanding the natural units and their hidden role in the laws of physics},
  author={Humpherys, David},
  journal={European Journal of Physics},
  volume={45},
  number={5},
  pages={055802},
  year={2024},
  publisher={IOP Publishing}
}

@book{crossection,
  title={Particles and nuclei: an introduction to the physical concepts},
  author={Povh, Bogdan and Rith, Klaus and Scholz, Christoph and Zetsche, Frank},
  year={2015},
  publisher={Springer}
}

@article{electroweak,
  title={The Standard model of electroweak interactions},
  author={Pich, Antonio},
  journal={arXiv preprint arXiv:1201.0537},
  year={2012}
}

@book{peskin2018introduction,
  title={An Introduction to quantum field theory},
  author={Peskin, Michael E},
  year={2018},
  publisher={CRC press}
}

@article{mandelstam1958determination,
  title={Determination of the pion-nucleon scattering amplitude from dispersion relations and unitarity. General theory},
  author={Mandelstam, Stanley},
  journal={Physical Review},
  volume={112},
  number={4},
  pages={1344},
  year={1958},
  publisher={APS}
}

@article{FermiGoldenRule,
  title={From Fermi’s golden rule to cross sections},
  author={Santos, Joao Miguel Costa},
  year={2025}
}

@article{DUNE_TDR_Vol2,
    author = "Abi, B. and others",
    collaboration = "DUNE",
    title = "{Deep Underground Neutrino Experiment (DUNE), Far Detector Technical Design Report, Volume II: DUNE Physics}",
    eprint = "2002.03005",
    archivePrefix = "arXiv",
    primaryClass = "hep-ex",
    doi = "10.1088/1748-0221/15/08/T08010",
    journal = "JINST",
    volume = "15",
    number = "08",
    pages = "T08010",
    year = "2020"
}

@article{LBNF_CDR_Vol3,
    author = "Acciarri, R. and others",
    collaboration = "DUNE",
    title = "{Long-Baseline Neutrino Facility (LBNF) and Deep Underground Neutrino Experiment (DUNE) Conceptual Design Report, Volume 3: Long-Baseline Neutrino Facility}",
    eprint = "1601.05471",
    archivePrefix = "arXiv",
    primaryClass = "physics.ins-det",
    reportNumber = "FERMILAB-DESIGN-2016-03",
    year = "2016"
}

@article{DUNE_ND_CDR,
    author = "Abed Abud, A. and others",
    collaboration = "DUNE",
    title = "{Deep Underground Neutrino Experiment (DUNE) Near Detector Conceptual Design Report}",
    eprint = "2103.13910",
    archivePrefix = "arXiv",
    primaryClass = "physics.ins-det",
    doi = "10.3390/instruments5040031",
    journal = "Instruments",
    volume = "5",
    number = "4",
    pages = "31",
    year = "2021"
}

@book{Halzen:1984mc,
    author = "Halzen, F. and Martin, A. D.",
    title = "{Quarks and Leptons: An Introductory Course in Modern Particle Physics}",
    publisher = "John Wiley \& Sons",
    year = "1984",
    isbn = "978-0-471-88741-6"
}

@article{LlewellynSmith:1972zz,
    author = "Llewellyn Smith, C. H.",
    title = "{Neutrino Reactions at Accelerator Energies}",
    reportNumber = "SLAC-PUB-0958",
    doi = "10.1016/0370-1573(72)90010-5",
    journal = "Phys. Rept.",
    volume = "3",
    number = "5",
    pages = "261--379",
    year = "1972"
}

@book{Athar:2020kqz,
    author = "Athar, M. Sajjad and Singh, S. K.",
    title = "{The Physics of Neutrino Interactions}",
    publisher = "Cambridge University Press",
    year = "2020",
    address = "Cambridge",
    isbn = "978-1-108-48906-5",
    doi = "10.1017/9781108963480"
}

@article{rein1981neutrino,
  title={Neutrino-excitation of baryon resonances and single pion production},
  author={Rein, Dieter and Sehgal, Lalit M},
  journal={Annals of Physics},
  volume={133},
  number={1},
  pages={79--153},
  year={1981},
  publisher={Elsevier}
}

@article{CTEQ2002,
    author = "Pumplin, J. and Stump, D. R. and Huston, J. and Lai, H. L. and Nadolsky, P. and Tung, W. K.",
    title = "{New generation of parton distributions with uncertainties from global QCD analysis}",
    journal = "JHEP",
    volume = "07",
    pages = "012",
    year = "2002",
    eprint = "hep-ph/0201195",
    doi = "10.1088/1126-6708/2002/07/012"
}

@article{bodek2002modeling,
  title={Modeling deep inelastic cross sections in the few GeV region},
  author={Bodek, Arie and Yang, UK},
  journal={Nuclear Physics B-Proceedings Supplements},
  volume={112},
  number={1-3},
  pages={70--76},
  year={2002},
  publisher={Elsevier}
}

@article{Bodek2021,
    author = "Bodek, A. and Christy, M. E. and Coopersmith, B.",
    title = "{Inelastic Axial and Vector Structure Functions for Lepton-Nucleon Scattering 2021 Update}",
    eprint = "2108.09240",
    archivePrefix = "arXiv",
    primaryClass = "hep-ph",
    doi = "10.1140/epjc/s10052-021-09787-z",
    journal = "Eur. Phys. J. C",
    volume = "81",
    year = "2021"
}

@article{NuSTEC:2018,
    author = "Alvarez-Ruso, L. and others",
    collaboration = "NuSTEC",
    title = "{NuSTEC White Paper: Status and challenges of neutrino–nucleus scattering}",
    journal = "Prog. Part. Nucl. Phys.",
    volume = "100",
    pages = "1--68",
    year = "2018",
    doi = "10.1016/j.ppnp.2018.01.006"
}

@article{Hofstadter1956,
    author = "Hofstadter, Robert",
    title = "{Electron Scattering and Nuclear Structure}",
    journal = "Rev. Mod. Phys.",
    volume = "28",
    number = "3",
    pages = "214--254",
    year = "1956",
    doi = "10.1103/RevModPhys.28.214"
}

@article{Moniz1971,
    author = "Moniz, E. J. and Sick, I. and Whitney, R. R. and Ficenec, J. R. and Kephart, R. D. and Trower, W. P.",
    title = "{Quasielastic Electron Scattering in Complex Nuclei}",
    journal = "Phys. Rev. Lett.",
    volume = "26",
    issue = "8",
    pages = "445--448",
    year = "1971",
    doi = "10.1103/PhysRevLett.26.445"
}

@article{SmithMoniz1972,
    author = "Smith, R. A. and Moniz, E. J.",
    title = "{Neutrino reactions on nuclear targets}",
    journal = "Nucl. Phys. B",
    volume = "43",
    pages = "605--622",
    year = "1972"
}

@article{Benhar2005,
    author = "Benhar, Omar and Farina, N. and Nakamura, H. and Sakuda, M. and Seki, R.",
    title = "{Electron- and neutrino-nucleus scattering in the impulse approximation regime}",
    journal = "Phys. Rev. D",
    volume = "72",
    pages = "053005",
    year = "2005",
    doi = "10.1103/PhysRevD.72.053005"
}

@article{kuzmin2008quasielastic,
  title={Quasielastic axial-vector mass from experiments on neutrino--nucleus scattering},
  author={Kuzmin, Konstantin S and Lyubushkin, Vladimir V and Naumov, Vadim A},
  journal={The European Physical Journal C},
  volume={54},
  number={4},
  pages={517--538},
  year={2008},
  publisher={Springer}
}

@article{Gallagher2011,
    author = "Gallagher, Hugh and Garvey, G. and Zeller, G. P.",
    title = "{Neutrino-Nucleus Interactions}",
    journal = "Ann. Rev. Nucl. Part. Sci.",
    volume = "61",
    pages = "355--378",
    year = "2011",
    doi = "10.1146/annurev-nucl-102010-130434"
}

@article{Andreopoulos2010,
    author = "Andreopoulos, C. and others",
    title = "{The GENIE Neutrino Monte Carlo Generator}",
    journal = "Nucl. Instrum. Meth. A",
    volume = "614",
    pages = "87--104",
    year = "2010",
    doi = "10.1016/j.nima.2009.12.009"
}

@article{katori2018neutrino,
  title={Neutrino--nucleus cross sections for oscillation experiments},
  author={Katori, Teppei and Martini, Marco},
  journal={Journal of Physics G: Nuclear and Particle Physics},
  volume={45},
  number={1},
  pages={013001},
  year={2018},
  publisher={IOP Publishing}
}

@inproceedings{Dytman2011,
    author = "Dytman, Steve",
    title = "{Final state interactions in neutrino-nucleus experiments}",
    booktitle = "{12th International Workshop on Neutrino Factories, Superbeams and Beta Beams (NuFact 10)}",
    year = "2011",
    volume = "1382",
    pages = "156--158",
    publisher = "AIP Conference Proceedings",
    doi = "10.1063/1.3644290"
}

@article{Le2015,
      author         = "Le, T. and others",
      title          = "{Single neutral pion production by charged-current $\bar{\nu}_\mu$ interactions on hydrocarbon at $\langle E_\nu \rangle = 3.6$ GeV}",
      collaboration  = "MINERvA",
      journal        = "Phys. Lett. B",
      volume         = "749",
      pages          = "130--136",
      year           = "2015",
      doi            = "10.1016/j.physletb.2015.07.039"
}

@article{Eberly2015,
      author         = "Eberly, B. and others",
      title          = "{Charged pion production in $\nu_\mu$ interactions on hydrocarbon at $\langle E_\nu \rangle= 4.0$ GeV}",
      collaboration  = "MINERvA",
      journal        = "Phys. Rev. D",
      volume         = "92",
      pages          = "092008",
      year           = "2015",
      doi            = "10.1103/PhysRevD.92.092008"
}

@article{McGivern2016,
      author         = "McGivern, C. L. and others",
      title          = "{Cross sections for $\nu_\mu$ and $\bar{\nu}_\mu$ induced pion production on hydrocarbon in the few-GeV region using MINERvA}",
      collaboration  = "MINERvA",
      journal        = "Phys. Rev. D",
      volume         = "94",
      pages          = "052005",
      year           = "2016",
      doi            = "10.1103/PhysRevD.94.052005"
}

@article{James1980,
    author = "James, F.",
    title = "{Monte Carlo theory and practice}",
    journal = "Reports on Progress in Physics",
    volume = "43",
    number = "9",
    pages = "1145--1189",
    year = "1980",
    doi = "10.1088/0034-4885/43/9/002"
}

@misc{unifal_mais_ciencia,
    author = "{Universidade Federal de Alfenas (UNIFAL-MG)}",
    title = "{Festival Mais Ciência Por Favor}",
    year = "2026",
    url = "https://www.unifal-mg.edu.br/festivalmaiscienciapfv/",
    note = "[Acessado em 6 de março de 2026]"
}

@manual{ScoolLabLINAC,
    author = "Bernstein, Fabian and Keller, Oliver",
    title = "{3D printed LINAC: Assembly Instructions (Version 1.1)}",
    organization = "CERN S'Cool LAB",
    year = "2020",
    note = "Open Hardware licensed under CERN-OHL-P v2"
}

@inproceedings{P5,
  author = "Carneiro, Mateus Fernandes",
  title = "{Neutrino interaction physics and the DUNE near detector}",
  doi = "10.22323/1.390.0119",
  booktitle = "Proceedings of 40th International Conference on High Energy physics {\textemdash} PoS(ICHEP2020)",
  year = 2021,
  volume = "390",
  pages = "119"
}

@techreport{U11,
  title={Expanding the physics reach of DUNE in the near and far detectors},
  author={Mart{\'\i}nez L{\'o}pez, Francisco},
  year={2025},
  institution={Queen Mary, U. of London}
}

@article{P4,
    author = "Bathe-Peters, Lars",
    title = "{2p-2h Cross-Section Systematics in DUNE}",
    year = "2025",
    eprint = "2501.08725",
    archivePrefix = "arXiv"
}

@inproceedings{P12,
  title={Neutrino-Argon Interaction with GENIE Event Generator},
  author={Chesneanu, Daniela},
  booktitle={AIP Conference Proceedings},
  volume={1304},
  number={1},
  pages={489--493},
  year={2010},
  organization={American Institute of Physics}
}

@article{P2,
    author = "GENIE Collaboration",
    title = "{Implementation of the Martini-Ericson-Chanfray-Marteau RPA-based (anti)neutrino cross-section model in the GENIE neutrino event generator}",
    year = "2025",
    eprint = "2508.13939",
    archivePrefix = "arXiv"
}

@article{U12,
    author = "Lalnuntluanga, R. and Pradhan, R. K. and Giri, A.",
    title = "{Probing neutrino-nucleus interaction in DUNE and MicroBooNE}",
    journal = "Nucl. Phys. B",
    year = "2024",
    url = "https://scoap3-prod-backend.s3.cern.ch/media/files/88768/10.1016/j.nuclphysb.2024"
}

@article{P7,
    author = "Sharma, H. R. and Nagu, S. and Singh, J. and Singh, R. B.",
    title = "{Study of Pion Production in $\nu_\mu$ Interactions on $^{40}$Ar in DUNE Using GENIE and NuWro Event Generators}",
    year = "2022",
    url = "https://link.springer.com/article/10.1134/S1547477122060267"
}

@article{P8,
    author = "Sharma, H. R. and others",
    title = "{Quantifying Effects Of Final State Interactions On Pion Production In DUNE Using Monte Carlo Event Generators}",
    journal = "J. Phys. Conf. Ser.",
    volume = "2426",
    year = "2023"
}

@article{P14,
    author = "Andreopoulos, C. and others",
    title = "{The GENIE Neutrino Monte Carlo Generator}",
    year = "2015",
    eprint = "1510.05494",
    archivePrefix = "arXiv"
}

@misc{U7,
    author = "{GENIE Collaboration}",
    title = "{GENIE-MC/Generator - GitHub}",
    year = "2026",
    url = "https://github.com/GENIE-MC/Generator"
}

@article{P1,
    author = "GENIE Collaboration",
    title = "{Test of the GENIE neutrino event generator against $^{16}$O(e,e'p) data: Reduced cross section}",
    year = "2025",
    eprint = "2511.03274",
    archivePrefix = "arXiv"
}

@article{P3,
    author = "Butkevich, A. V. and Luchuk, S. V.",
    title = "{Assessing the accuracy of the GENIE event generator with electron scattering data: Reduced cross section and nuclear transparency}",
    year = "2025",
    eprint = "2506.19355",
    archivePrefix = "arXiv"
}

@misc{U17,
    author = "{DUNE Collaboration}",
    title = "{Neutrino Interaction Physics \& the DUNE Near Detector}",
    year = "2026",
    note = "Indico presentation",
    url = "https://indico.cern.ch/event/868940/contributions/3816980/attachments/2082771/34"
}

@article{Sjostrand:2006za,
    author = "Sjostrand, Torbjorn and Mrenna, Stephen and Skands, Peter Z.",
    title = "{PYTHIA 6.4 Physics and Manual}",
    journal = "JHEP",
    volume = "05",
    pages = "026",
    year = "2006",
    doi = "10.1088/1126-6708/2006/05/026"
}

@book{Galassi:2009,
    author = "Galassi, Mark and Davies, Jim and Theiler, James and Gough, Brian and Jungman, Gerard and Alken, Patrick and Booth, Michael and Rossi, Fabrice",
    title = "{GNU Scientific Library Reference Manual}",
    publisher = "Network Theory Ltd.",
    year = "2009",
    isbn = "0954612078"
}

@misc{log4cpp,
    author = "Bastiaan Bakker",
    year = "2026",
    title = "{Log4cpp: A library of C++ classes for flexible logging}",
    howpublished = "\url{http://log4cpp.sourceforge.net/}",
    note = "Accessed: 01 October 2024"
}

@article{Buckley:2014ana,
    author = "Buckley, Andy and Ferrando, James and Lloyd, Stephen and Nordstrom, Karl and Page, Ben and Rufenacht, Martin and Smith, Marek and Watt, Graeme",
    title = "{LHAPDF6: parton density access in the LHC precision era}",
    journal = "Eur. Phys. J. C",
    volume = "75",
    number = "3",
    pages = "132",
    year = "2015",
    doi = "10.1140/epjc/s10052-015-3318-8"
}

@article{Andreopoulos:2009rq,
    author = "Andreopoulos, C. and others",
    collaboration = "GENIE",
    title = "{The GENIE Neutrino Monte Carlo Generator}",
    journal = "Nucl. Instrum. Meth. A",
    volume = "614",
    pages = "87--104",
    year = "2010",
    doi = "10.1016/j.nima.2009.12.009"
}

@article{Woods1954,
  title = {Diffuse Surface Optical Model for Nucleon-Nuclei Scattering},
  author = {Woods, R. D. and Saxon, David S.},
  journal = {Phys. Rev.},
  volume = {95},
  issue = {2},
  pages = {577--578},
  year = {1954},
  month = {Jul},
  publisher = {American Physical Society},
  doi = {10.1103/PhysRev.95.577}
}

@article{Nieves2004,
  title = {Inclusive quasielastic neutrino reactions},
  author = {Nieves, J. and Amaro, J. E. and Valverde, M.},
  journal = {Phys. Rev. C},
  volume = {70},
  issue = {5},
  pages = {055503},
  year = {2004},
  month = {Nov},
  publisher = {American Physical Society},
  doi = {10.1103/PhysRevC.70.055503}
}

@article{Gil1997,
  title = {Many body effects on inclusive muon capture in 12C},
  author = {Gil, A. and Nieves, J. and Oset, E.},
  journal = {Nuclear Physics A},
  volume = {627},
  number = {3},
  pages = {543--619},
  year = {1997},
  issn = {0375-9474},
  doi = {10.1016/S0375-9474(97)00405-2}
}

@article{Nieves2012,
  title = {Inclusive nucleon emission induced by neutrino interactions},
  author = {Nieves, J. and S\'anchez, F. and Ruiz Simo, I. and Vicente Vacas, M. J.},
  journal = {Phys. Rev. D},
  volume = {85},
  issue = {11},
  pages = {113008},
  year = {2012},
  month = {Jun},
  publisher = {American Physical Society},
  doi = {10.1103/PhysRevD.85.113008}
}

@article{Brun:1997pa,
    author = "Brun, R. and Rademakers, F.",
    title = "{ROOT: An object oriented data analysis framework}",
    journal = "Nucl. Instrum. Meth. A",
    volume = "389",
    pages = "81--86",
    year = "1997",
    doi = "10.1016/S0168-9002(97)00048-X"
}

@article{PDG2022,
  author = {Workman, R. L. and others},
  collaboration = {Particle Data Group},
  title = {Review of Particle Physics},
  journal = {Prog. Theor. Exp. Phys.},
  volume = {2022},
  pages = {083C01},
  year = {2022},
  doi = {10.1093/ptep/ptac097}
}

@article{Radecky1982,
  title = {Study of single-pion production by weak charged currents in low-energy $\nu d$ interactions},
  author = {Radecky, G. M. and others},
  journal = {Phys. Rev. D},
  volume = {25},
  issue = {5},
  pages = {1161--1173},
  year = {1982},
  month = {Mar},
  publisher = {American Physical Society},
  doi = {10.1103/PhysRevD.25.1161}
}

@article{Kitagaki1983,
  title = {High-energy quasielastic $\nu_{\mu} n \rightarrow \mu^- p$ scattering in deuterium},
  author = {Kitagaki, T. and others},
  journal = {Phys. Rev. D},
  volume = {28},
  issue = {3},
  pages = {436--442},
  year = {1983},
  month = {Aug},
  publisher = {American Physical Society},
  doi = {10.1103/PhysRevD.28.436}
}

@article{abe2023updated,
  title={Updated T2K measurements of muon neutrino and antineutrino disappearance using 3.6$\times$ 10 21 protons on target},
  author={Abe, Ko and Akhlaq, N and Akutsu, Ryosuke and Ali, A and Alonso Monsalve, S and Alt, C and Andreopoulos, C and Antonova, M and Aoki, S and Arihara, T and others},
  journal={Physical Review D},
  volume={108},
  number={7},
  pages={072011},
  year={2023},
  publisher={APS}
}

@article{gran2006measurement,
  title={Measurement of the quasielastic axial vector mass in neutrino interactions on oxygen},
  author={Gran, R and Jeon, EJ and Aliu, E and Andringa, S and Aoki, S and Argyriades, J and Asakura, K and Ashie, R and Berghaus, F and Berns, H and others},
  journal={Physical Review D—Particles, Fields, Gravitation, and Cosmology},
  volume={74},
  number={5},
  pages={052002},
  year={2006},
  publisher={APS}
}

@article{rodrigues2016constraining,
  title={Constraining the GENIE model of neutrino-induced single pion production using reanalyzed bubble chamber data},
  author={Rodrigues, Philip and Wilkinson, Callum and McFarland, Kevin},
  journal={The European Physical Journal C},
  volume={76},
  number={8},
  pages={474},
  year={2016},
  publisher={Springer}
}

@article{tena2021neutrino,
  title={Neutrino-nucleon cross-section model tuning in GENIE v3},
  author={Tena-Vidal, J{\'u}lia and Andreopoulos, Costas and Ashkenazi, Adi and Barry, Christopher and Dennis, Steve and Dytman, Steve and Gallagher, Hugh and Gardiner, Steven and Giele, Walter and Hatcher, Robert and others},
  journal={Physical Review D},
  volume={104},
  number={7},
  pages={072009},
  year={2021},
  publisher={APS}
}

@article{bodek2021inelastic,
  title={Inelastic axial and vector structure functions for lepton-nucleon scattering 2021 update},
  author={Bodek, Arie and Yang, Un Ki and Xu, Yang},
  journal={arXiv preprint arXiv:2108.09240},
  year={2021}
}

@article{MicroBooNE2022,
  author = {Abratenko, P. and others},
  collaboration = {MicroBooNE},
  title = {First Measurement of Energy-Dependent Inclusive Muon Neutrino Charged-Current Cross Sections on Argon with the MicroBooNE Detector},
  journal = {Phys. Rev. Lett.},
  volume = {128},
  issue = {15},
  pages = {151801},
  year = {2022},
  month = {Apr},
  publisher = {American Physical Society},
  doi = {10.1103/PhysRevLett.128.151801},
  url = {https://link.aps.org/doi/10.1103/PhysRevLett.128.151801},
  eprint = {2110.14023},
  archivePrefix = {arXiv},
  primaryClass = {hep-ex}
}

@book{Cowan1998,
  title = {Statistical Data Analysis},
  author = {Cowan, Glen},
  publisher = {Oxford University Press},
  year = {1998},
  address = {Oxford, UK},
  isbn = {978-0198501558}
}

@misc{Rodrigues2026_GENIE_Apptainer,
  author       = {Rodrigues, Luiz Felipe Brandão Magalhães},
  title        = {GENIE MC Neutrino Generator (Version: 3.06.02) - Apptainer},
  year         = {2026},
  publisher    = {Zenodo},
  doi          = {10.5281/zenodo.22310542},
  url          = {https://doi.org/10.5281/zenodo.22310542},
  note         = {Software container}
}

@article{bernard2002axial,
  title={Axial structure of the nucleon},
  author={Bernard, Veronique and Elouadrhiri, Latifa and Mei{\ss}ner, Ulf-G},
  journal={Journal of Physics G: Nuclear and Particle Physics},
  volume={28},
  number={1},
  pages={R1--R35},
  year={2002}
}

@misc{sdumont,
    author       = {{Laboratório Nacional de Computação Científica (LNCC/MCTI, Brazil)}},
    title        = {Santos Dumont Supercomputer},
    year         = {2026},
    howpublished = {\url{http://sdumont.lncc.br}},
    note         = {HPC resources provided by LNCC}
}

\chapter[Appendix A: GENIE Environment]{Appendix A: GENIE Environment Setup and Architecture}

To ensure a stable and reproducible simulation pipeline, the installation of the GENIE event generator requires a environment with several high-energy physics (HEP) and mathematical software packages. Since these are interconnected tools, the compilation must be performed using compatible versions. 

\section{Software Dependencies}

The core execution and analysis capabilities of GENIE depend on the following primary software packages.

\begin{description}
    \item[CERN ROOT - Build with GSL and Mathmore (Version: 6.30.02):] An object-oriented framework developed by CERN, optimized for big data processing, statistical analysis, data visualization, and hierarchical storage in high-energy physics \cite{Brun:1997pa}. GENIE relies on ROOT for its primary input/output operations and graphical analysis frameworks.
    
    \item[Pythia (Version: 6.428):] A standard Monte Carlo event generator utilized for simulating high-energy particle collisions, parton showers, and hadronization processes. Within the GENIE pipeline, Pythia is invoked to handle particle fragmentation and final-state decay processes \cite{Sjostrand:2006za}.
    
    \item[GNU Scientific Library - GSL (Version: 2.7):] A numerical library providing a wide range of mathematical routines in C++ \cite{Galassi:2009}. GENIE utilizes GSL for critical numerical routines, including multi-dimensional integration algorithms, random number generation, and differential equation solvers required by various nuclear models.
    
    \item[Log4cpp (Version: 1.1.3):] A library of C++ classes that enables flexible logging to files, consoles, and other destinations \cite{log4cpp}. It manages GENIE's complex logging and verbosity levels.

    \item[LHAPDF (Version: 6.5.0):] A standard C++ library designed for evaluating Parton Distribution Functions (PDFs) \cite{Buckley:2014ana}. In the GENIE framework, LHAPDF is essential for calculating Deep Inelastic Scattering (DIS) cross-sections, as it provides the necessary parameterizations of the quark and gluon momentum distributions within the nucleon.

    \item[GENIE - Generator (Version: 3.06.02):] The core universal Monte Carlo event generator framework specifically designed to simulate neutrino-nucleus interactions \cite{Andreopoulos:2009rq}. It acts as the central engine that integrates the mentioned libraries to compute cross-sections, sample initial kinematics, model complex intranuclear hadronic cascades, and output the final simulated event records.
\end{description}

Once the dependencies are correctly installed, the system architecture requires the configuration of specific environment variables. These variables direct the compiler and linkers to the appropriate headers and shared libraries.

The configuration should be defined in the environment script or user profile (e.g., \texttt{.bashrc}) as follows:

\begin{scriptsize}
\estiloJava
\begin{lstlisting} [caption={Required environment variables configuration for GENIE execution. Source: Author.}]
    # ROOT Environment Setup
    export ROOTSYS=/path/to/root
    export PATH=$ROOTSYS/bin:$PATH
    export LD_LIBRARY_PATH=$ROOTSYS/lib:$LD_LIBRARY_PATH
    
    # Pythia Configuration
    export PYTHIA6=/path/to/pythia6 # If using Pythia6 legacy libraries
    
    # GENIE Core Environment Setup
    export GENIE=/path/to/genie/source_code
    export PATH=$GENIE/bin:$PATH
    export LD_LIBRARY_PATH=$GENIE/lib:$LD_LIBRARY_PATH
\end{lstlisting}
\end{scriptsize}

Where \texttt{GENIE} Points directly to the top-level directory of the GENIE installation, \texttt{ROOTSYS} Defines the root directory of the CERN and ROOT installation, \texttt{PATH} Appends the execution binaries from both GENIE (\texttt{\$GENIE/bin}) and ROOT (\texttt{\$ROOTSYS/bin}) to the system's command line interface, \texttt{LD\_LIBRARY\_PATH} Explicitly informs the dynamic linker where to locate the shared objects for ROOT, Pythia, and GENIE at runtime.

\section{Installing GENIE}
The installation of the GENIE event generator can be approached in different ways depending on the user's requirements. In this manual, we outline two primary methods: a traditional source compilation and a containerized approach.

\subsection*{Option A: Building from Source}

The standard approach involves manually downloading the source code and compiling it within the required dependencies discussed previously. This method provides the highest degree of customization, allowing the user to enable or disable specific physics models, neutrino fluxes, and external cross-section libraries during the configuration stage. However, it requires strict adherence to version compatibility and environment configuration. 

Detailed, step-by-step instructions to follow this method are thoroughly documented in the official GENIE Physics and User Manual \cite{Andreopoulos2010}. Users opting for this route, especially on local machines, are highly encouraged to consult the official documentation to ensure a stable and successful build.

\subsection*{Option B: Apptainer Utilization}

Although this work initially involved a build-from-source installation, a containerized solution was developed to bypass the complexities of dependency resolution and manual compilation, ensuring the easy reproducibility of the software. This method utilizes Apptainer (formerly Singularity) to create a custom image of the complete GENIE environment. This image is pre-configured with the latest version of the GENIE software and all its required libraries, natively compiled for a Linux environment.

Using the Apptainer image guarantees that the simulation environment remains perfectly consistent across different computing clusters or personal machines, without risking conflicts with the host operating system. To utilize this image, the user simply needs to download the container file available in the reference \cite{Rodrigues2026_GENIE_Apptainer}.

The brief installation guide of the container and utilization is shown below:

\begin{scriptsize}
\estiloJava
\begin{lstlisting}[caption={GENIE installation guide using Apptainer software. Author, 2026}]
    # Pull the container image
    sudo apt update
    sudo apt install -y apptainer
    apptainer pull genie_v3_06_02.sif docker://your-registry-url/genie:latest
    
    # Execute a GENIE command inside the containerized environment
    apptainer exec genie_v3_06_02.sif bash -c "source setup_genie.sh && gevgen -n 1000 -e 1.0 -p 14 -t 1000180400"
\end{lstlisting}
\end{scriptsize}

This containerized approach is highly recommended for new users who wish to focus immediately on physics analysis, cross-section evaluation, and event generation rather than software troubleshooting.

\section{Configuration and XML File Structure}

The GENIE framework is modular, relying on an extensive Extensible Markup Language (XML) architecture to manage its physics models, interaction cross-sections, and algorithmic configurations. Rather than hard-coding physical constants and model parameters into the compiled C++ binaries, GENIE externalizes these variables into XML files located within the \texttt{\$GENIE/config} directory.

This architecture allows a flexible configuration of the simulation environment. Some of the most critical XML files include:
\begin{itemize}
    \item \texttt{CommonParam.xml} and \texttt{UserPhysicsOptions.xml}: Contains the tunable free parameters for various physics models. This includes values such as the axial mass ($M_A$) for Quasi-Elastic, and the Fermi momentum ($k_F$) for different target nuclei.
    \item \texttt{EventGeneratorList.xml}: Controls which specific interaction channels (e.g., CCQE, CCRES, CCDIS) are active and assembled during the event generation phase.
    \item \texttt{Messenger.xml}: Manages the verbosity and message thresholds of the \texttt{Log4cpp} framework, aiding in debugging and pipeline monitoring.
\end{itemize}

\begin{figure}
    \centering
    \includegraphics[width=0.7\linewidth]{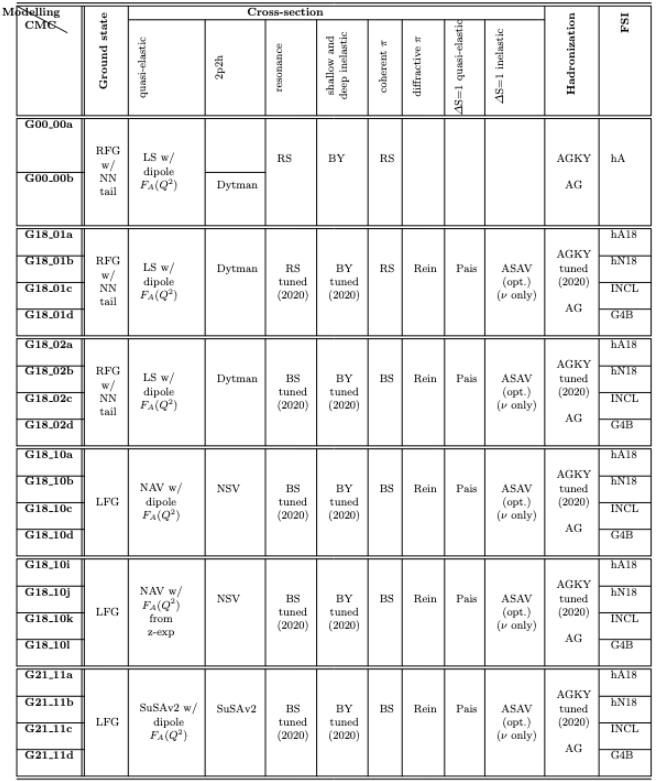}
    \caption{Summary of the Tunes, available within the GENIE event generator framework. The table details the specific theoretical models and phenomenological parameterizations assigned to each tune across the entire event generation pipeline \cite{Andreopoulos2010}.}
    \label{fig:tunes}
\end{figure}

GENIE employs standardized configurations known as ``Tunes'' (e.g., \texttt{G18\_10a\_02\_11b}, where the differences are shown in Figure \ref{fig:tunes}). A Tune defines a comprehensive set of baseline models and parameters within these XML files. 

Users can override these default behaviors without recompiling the source code by placing a modified XML file with the same name in their local execution directory, or by defining the \texttt{\$GXMLPATH} environment variable to point to a custom configuration folder.

\section{Splines and Cross-Section Configuration}

In neutrino event generators like GENIE, evaluating cross-sections numerically during event generation is computationally prohibitive due to the complex multi-dimensional integrals required by advanced nuclear models. To overcome this, GENIE utilizes a pre-computation strategy known as cross-section splines. Cross-sections are evaluated across a predetermined grid of neutrino energy points (knots) for specific neutrino-target combinations and saved as interpolation tables within XML files. During the event generation phase, the software performs rapid spline interpolation to determine the interaction probabilities instantaneously.

\subsection{Generating Cross-Section Splines with \texttt{gmkspl}}

The generation of cross-section splines is handled by the \texttt{gmkspl} executable. This command must be executed prior to running any event generation pipeline unless a pre-calculated spline file matching the desired physics tune and target composition is already available. A standard execution command for generating splines for a muon neutrino ($\nu_\mu$) interacting with an Argon-40 ($^{40}\text{Ar}$) target is:

\begin{scriptsize}
\estiloJava
\begin{lstlisting}[caption={Generating cross-section splines using the \texttt{gmkspl} utility. Author.}]
    gmkspl -n 100 -e 10 -p 14 -t 1000180400 --tune G18_10a_02_11b -o nu_mu_Ar40_splines.xml
\end{lstlisting}
\end{scriptsize}

Where \texttt{-n 100}: Specifies the number of knots (evaluation points) distributed along the energy grid. A higher number of knots increases the precision of the subsequent interpolation at the expense of longer pre-computation times. \texttt{-e 10}: Defines the maximum neutrino energy limit for the spline grid in units of GeV. In this example, the grid spans from the interaction threshold up to 10 GeV. \texttt{-p 14}: Identifies the incoming projectile using its standard Particle Data Group (PDG) code. The code \texttt{12} corresponds to the electron neutrino ($\nu_e$), the \texttt{14} and \texttt{16} stands for muon ($\nu_\mu$) and tau ($\nu_\tau$) neutrino, respectively. For antineutrino configurations, \texttt{-12,-14,-16} ($\bar{\nu}_l$) would be specified. \texttt{-t 1000180400}: Identifies the target nucleus using the 10-digit PDG ion numbering convention scheme ($10LZZZAAAI$). For Argon-40 ($^{40}\text{Ar}$), $Z=18$ and $A=40$, yielding \texttt{1000180400}. \texttt{-o nu\_mu\_Ar40\_splines.xml}: Dictates the output filename where the calculated cross-section matrices will be stored in an Extensible Markup Language (XML) format. \texttt{--tune G18\_10a\_02\_11b}: Selects the physics configuration tune (see Table \ref{fig:tunes}). This ensures that the cross-sections are computed using a consistent set of underlying models (e.g., the specific treatment of CCQE, resonant pion production, or deep inelastic scattering).

While the XML format produced is optimized for the engine during event generation, it is not useful for direct data visualization. To facilitate the visual inspection and phenomenological analysis of the pre-computed cross-sections, GENIE provides the \texttt{gspl2root} utility. This command parses the XML spline files and converts them into standard ROOT format (\texttt{.root}), extracting the energy-dependent cross-section data into plottable \texttt{TGraph} objects. For instance, executing the command:

\begin{scriptsize}
\estiloJava
\begin{lstlisting}[caption={Converting cross-section splines using the \texttt{gspl2root} command. Author.}]
    gspl2root -p 14 -t 1000180400 -f nu\_mu\_Ar40\_splines.xml -o splines.root
\end{lstlisting}
\end{scriptsize} 

It generates a structured ROOT file containing the total cross-sections as a function of the incident neutrino energy. This conversion is an essential data analysis.

\subsection{Cross-Section Configuration via XML Files}

The behavior of the cross-section calculation and the management of the generated splines are governed by Model XML files (e.g., \texttt{LlewellynSmithCCQE.xml}, \texttt{ReinSehgalRES.xml}). These files contain the fine-grained physical parameters utilized during the spline evaluation. For instance, the axial vector form factor or resonance decay parameters are loaded from these files to compute the absolute cross-section values at each knot. By modifying these values and running \texttt{gmkspl}, researchers can generate custom spline sets to study how variations in fundamental physics parameters propagate into the total and differential cross-sections.

\section{Event Generation Workflow}

Once the cross-section splines have been successfully evaluated and stored, the simulation pipeline advances to the generation of the primary neutrino interactions and the subsequent intranuclear hadronic cascades. This phase utilizes the Monte Carlo sampling techniques discussed in previous chapters to determine the exact kinematics of each event based on the supplied nuclear models.

\subsection{The \texttt{gevgen} Command}

The executable responsible for standard event generation in GENIE is \texttt{gevgen}. This application reads the pre-calculated cross-section splines, samples the interaction probabilities, and outputs the simulated events. Depending on the experimental setup being modeled, \texttt{gevgen} can be configured to simulate either a monochromatic neutrino beam or a realistic continuous energy spectrum. For isolated theoretical studies, such as evaluating the cross-section of a specific channel at a fixed energy, a monochromatic beam is utilized. By passing a single numerical value to the energy flag (\texttt{-e}), the user forces all incident neutrinos to possess the exact same initial kinetic energy: 

\begin{scriptsize}
\estiloJava
\begin{lstlisting}[caption={Generating 10,000 monochromatic $\nu_\mu$ events at 2.5 GeV (DUNE peak energy) on an Argon-40 target.}]
gevgen -n 10000 -p 14 -t 1000180400 -e 2.5 --cross-sections nu_mu_Ar40_splines.xml
\end{lstlisting}
\end{scriptsize}

Realistic experimental environments exhibit continuous energy distributions. To accurately reproduce the event rates expected in the detector, GENIE must sample the incoming neutrino energies from a flux histogram. This is achieved by providing an energy range to the \texttt{-e} flag and specifying the ROOT file containing the flux distribution via the \texttt{-f} flag:

\begin{scriptsize}
\estiloJava
\begin{lstlisting}[caption={Generating events using a continuous energy flux described by a ROOT histogram.}]
gevgen -n 10000 -p 14 -t 1000180400 -e 10 -f /path/to/flux.root,histogram_name --cross-sections nu_mu_Ar40_splines.xml
\end{lstlisting}
\end{scriptsize}

\subsection{The GHEP Format}

The native output produced by the \texttt{gevgen} application is a ROOT file containing a \texttt{TTree} named \texttt{gtree}. Each entry in this tree is an \texttt{genie::EventRecord} object, commonly referred to as the GENIE Hadron Event Record (GHEP). The GHEP format encapsulates the complete quantum mechanical history of the interaction. It records the full particle lineage, from the initial pre-interaction state to the final observable particles emerging from the nucleus. The particles within a GHEP record are organized in an array and are assigned specific status codes that define their physical role in the event:

\begin{description}
    \item[Status 0 (Initial State):] Represents the incoming particles before the weak interaction occurs. This includes the incident neutrino and the target nucleus (e.g., $^{40}\text{Ar}$), or the specific bound nucleon targeted by the Impulse Approximation.
    \item[Status 11 (Intermediate State):] Represents transient particles that exist only for an infinitesimal time or within the nuclear medium. Examples include the struck nucleon prior to intranuclear transport, the exchanged $W^\pm$ or $Z^0$ bosons, and intermediate baryonic resonances like the $\Delta(1232)$.
    \item[Status 1 (Final State):] Denotes the stable particles that successfully escape the nuclear potential and emerge into the detector volume. These are the particles (e.g., outgoing muons, emitted protons, and surviving pions) that will subsequently be passed to Geant4 for detector geometry tracking and calorimetric simulation.
\end{description}

\subsection{Flattening the Event Record with \texttt{gntpc}}

The complexity of the GHEP format can be cumbersome for subsequent phenomenological analysis. Furthermore, reading and parsing a GHEP may be inefficient or impractical on data analysis. To simplify these process, GENIE provides the \texttt{gntpc} (GENIE N-Tuple Converter) utility. This application parses the complex GHEP objects, extracts the most relevant observable data, such as the initial state kinematics, the invariant scattering variables ($Q^2, W, x, y$), and the four-momenta of the final-state particles (Status 1), and flattens them into a simplified, standard ROOT \texttt{TTree}.

The conversion is executed using the following command structure:

\begin{scriptsize}
\estiloJava
\begin{lstlisting}[caption={Converting a complex GHEP event record into a simplified GENIE Summary Tree (GST) format.}]
gntpc -i gntp.ghep.root -f gst -o events.root
\end{lstlisting}
\end{scriptsize}

A detailed breakdown of the conversion arguments is provided below:
\begin{description}
    \item[\textbf{-i gntp.ghep.root}:] Specifies the input ROOT file containing the original hierarchical GHEP event records generated by \texttt{gevgen}.
    \item[\textbf{-f gst}:] Defines the target output format. The \texttt{gst} (GENIE Summary Tree) format produces a optimized, flat ROOT tree composed of standard data types and arrays, making it fast to read.
    \item[\textbf{-o events.root}:] Dictates the filename for the simplified output file.
\end{description}

By converting the data into the GST format, the file size is drastically reduced. More importantly, researchers can perform rapid, high-statistics data visualization and kinematic plotting using standard analysis, entirely independent of the GENIE software environment.

\section{Uncertainty Evaluation in Monte Carlo Simulations}
\label{sec:uncertainty_evaluation}

Quantifying the uncertainties of Monte Carlo simulations is as important as determining their central predictive values. The evaluation of interaction cross-sections, final-state particle kinematics, and detector acceptances relies on highly parameterized theoretical models. The propagation of the model uncertainties to the final observable event spectra is a fundamental requirement for minimizing the systematic error in next-generation experiments like DUNE \cite{NuSTEC:2018}. Within the context of MC event generators, these errors are broadly categorized into statistical fluctuations and systematic parameter uncertainties.

The algorithm of Monte Carlo event generation evaluates the phase space using pseudo-random number sequences to accept or reject kinematic configurations based on the differential cross-section probability density function \cite{Cowan1998}. Consequently, the number of simulated events populating a specific kinematic bin (e.g., an outgoing muon energy bin) is a discrete random variable that follows a Poisson distribution \cite{Cowan1998}. For a given histogram bin containing $N$ generated events, the standard deviation representing the statistical uncertainty of the Monte Carlo sample is given by the Poisson error:
\begin{equation}
    \sigma_{stat} = \sqrt{N}.
\end{equation}
When converting the event counts into differential cross-sections or flux-normalized event rates, this statistical error must be properly propagated. To ensure that the statistical fluctuations of the simulation do not artificially inflate the overall experimental error, it is a standard practice in the field to generate MC samples with a statistical magnitude significantly larger than the expected experimental data \cite{PDG2022}.

\section{Computational Resources}

Due to the extensive simulation times and the computational resources required by the Monte Carlo event generation, all simulations presented in this work were performed using the high-performance computing infrastructure of the Santos Dumont supercomputer \cite{sdumont}. The complexity of the models evaluated in the GENIE framework, particularly those describing advanced nuclear effects, demanded significant processing capabilities. For instance, computing the cross-sections for the Local Fermi Gas (LFG) model required peak memory allocations reaching up to 250~GB of RAM to properly obtain the robust statistical results presented in this dissertation.

\end{document}